\documentclass{article}
\usepackage{amssymb}
\usepackage{geometry}

\input{tcilatex}
\begin{document}

\QTP{Body Math}
$\bigskip $

\QTP{Body Math}
$\bigskip $

\QTP{Body Math}
$\bigskip $

\QTP{Body Math}
$\bigskip $ \ \ \ \ \ \ \ \ \ \ \ \ \ \ \ \ \ \ \ \ \ \ \ \ \ 

\QTP{Body Math}
\ \ \ \ \ \ \ \ \ \ \ \ \ \ \ \ \ \ \ \ \ \ \ {\LARGE \ Topological
character of hydrodynamic }

\QTP{Body Math}
\ \ \ \ \ \ \ \ \ \ \ \ \ \ \ \ \ \ \ \ \ \ \ \ \ {\LARGE screening in
suspensions of hard spheres:}

\QTP{Body Math}
\ \ \ \ \ \ \ \ \ \ \ \ \ \ \ \ \ \ \ \ \ \ \ {\LARGE \ an example of
universal phenomenon}

\QTP{Body Math}
\ \ \ \ \ \ \ \ \ \ \ \ \ \ \ \ \ \ \ \ \ \ \ \ 

\QTP{Body Math}
\ \ \ \ \ \ \ \ \ \ \ \ \ \ \ \ \ \ \ 

\QTP{Body Math}
$\ $

\QTP{Body Math}
$\bigskip $ \ \ \ \ \ \ \ \ \ \ \ \ \ \ \ \ \ \ \ \ \ \ \ \ \ \ \ \ \ \ \ \
\ \ \ \ Ethan E. Ballard\footnote{%
Sections 1-4 of this work comprise part of the thesis of E.Ballard done in
compliance with requirements for the PhD degree in materials science.},
Arkady L. Kholodenko$\footnote{%
Corresponding author
\par
\textsl{E-mail addresses}: string@clemson.edu (A.Kholodenko),
eballar@clemson.edu (E.Ballard)}$

\QTP{Body Math}
\ \ \ \ \ \ \ \ \ \ \ \ \ \ \ \ \ \ \ 375 H.L.Hunter laboratories, Clemson
University, Clemson, SC 29634-0973, USA

\QTP{Body Math}
$\bigskip $\textbf{Abstract}

\QTP{Body Math}
\ \ Although in the case of polymer solutions the existence of hydrodynamic
screening has been theoretically established some time ago, use of the same
methods for suspensions of hard spheres thus far have failed to produce
similar results. In this work we reconsider this problem. Using
superposition of topological and London-style qualitative arguments we prove
the existence of screening in hard sphere suspensions. Even though some of
these arguments were\ employed \ initially for treatments of
superconductivity and superfluidity, we find analogs of these phenomena in
nontraditional \ settings such as in colloidal suspensions, turbulence,
magnetohydrodynamics, etc. In particular, in suspensions we demonstrate that
the hydrodynamic screening is an exact analog of Meissner effect in
superconductors. The extent of screening depends on the volume fraction of
hard spheres. The zero volume fraction limit corresponds to the normal
state. The case of finite volume fractions-to the mixed state typical for
superconductors of the second kind with such a state becoming \ fully
"superconducting" at the critical volume fraction $\varphi ^{\ast }$ for
which the (zero frequency) relative viscosity $\eta (relative)$ diverges.
Brady and, independently, Bicerano \textit{et al} using scaling-type
arguments predicted that for $\varphi $ close to $\varphi ^{\ast }$ the
viscosity $\eta (relative)$ behaves as $C(1-\varphi /\varphi ^{\ast })^{-2}$
with $C$ being some constant. Their prediction is well supported by
experimental data. \ In this work we explain such a behavior of viscosity in
terms of a topological-type transition which\ mathematically can be made
isomorphic to the more familiar Bose-Einstein condensation transition. \
Because of this, the results\ and methods of this work are not limited to
suspensions.\ In the concluding section \ we describe other applications
ranging from turbulence and magnetohydrodynamics to high temperature
superconductors and QCD, etc.

\QTP{Body Math}
\ \qquad

\QTP{Body Math}
$\ PACS:$ 47.57. E-; 47.57.Qk; 47.65.-d; 66.20.Cy; 67.25.dk; 11.25.-w;
12.38.-t; 14.80.Hv; 47.10 Df

\QTP{Body Math}
$\ $

\QTP{Body Math}
$\bigskip $

\QTP{Body Math}
$Keywords:$ Colloidal suspensions; Generalized Stokes-Einsten relation;
Hydrodynamics and hydrodynamic screening; Bose-Einstein condensation; \
London equation; Ginzburg-Landau theory of superconductivity; Topological
field theories; Theory of knots and links; Helicity in hydro and
magnetohydrodynamics; Classical mechanics in the vortex formalism; High T$%
_{c}$ superconductivity; Abelian projection for QCD; String models

\QTP{Body Math}
$\bigskip $

\section{Introduction}

In his 1905-1906 papers on Brownian motion for suspensions of hard spheres,
Einstein obtained now famous relation for the self-diffusion coefficient $%
D_{0}$\ for the noninteracting hard spheres of radius $R$ immersed in a
solvent at temperature $T$ [1]:%
\begin{equation}
D_{0}=\frac{k_{B}T}{\gamma }=\frac{k_{B}T}{6\pi R\eta _{0}}.  \tag{1.1}
\end{equation}%
In this formula $\eta _{0}$ is the viscosity of a pure solvent and $k_{B}$
is the Boltzmann's constant. \ This result is valid only in the infinite
dilution limit. In another paper [2] Einstein took into account the effects
of finite concentrations and obtained the first nonvanishing correction to $%
\eta _{0}$ for small but finite concentrations. It is given by

\begin{equation}
\eta /\eta _{0}=1+2.5\text{ }\varphi +O\left( \varphi ^{2}\right)  \tag{1.2}
\end{equation}%
with $\varphi $\ being the volume fraction $\varphi :=\frac{n}{V}\frac{4}{3}%
\pi R^{3}$. \ In this formula $\mathit{n}$ is the number of monodisperse
hard spheres in the volume $\mathit{V}$. If we formally replace $\eta _{0}$
by $\eta $\ in Eq. (1.1), the obtained result can be cautiously used as a
definition for the cooperative diffusion coefficient \textit{D}, i.e.

\begin{equation}
D=\frac{k_{B}T}{6\pi R\eta }.  \tag{1.3}
\end{equation}%
Below, we use symbols $D_{0}$ for the self-diffusion coefficient and $D$\
for the cooperative diffusion coefficient. \ By combining Eq.s (1.1) -
(1.3), we also obtain:

\begin{equation}
D/D_{0}=1-2.5\varphi +O\left( \varphi ^{2}\right) .  \tag{1.4}
\end{equation}%
Eq.(1.4) compares well with experimental results, e.g. those discussed in
Ref.[3]\footnote{%
The data on page 5 and \ in Table 2 of this reference support our conjecture.%
}. Numerous attempts have been made to obtain results like Eq.(1.4)
systematically. The above results are restricted by the observation that
Stoke's formula for friction $\gamma $ is applicable only for time scales
longer than the characteristic relaxation time $\tau _{r}$ of the solvent, $%
\tau _{r}:=\rho R^{2}/\pi \eta _{0}),$ e.g. see [4]. \ In this formula $\rho 
$ is the density of pure solvent. This requirement provides the typical
cut-off time scale, while the parameter $R$ serves as a typical space
cut-off for the problems we are going to study in this work.

By analogy with the theory of nonideal gases, expansion Eq. (1.2) is
referred to as a "virial". Unlike the theory of nonideal gases, where the
virial coefficients are known \ exactly to a very high order [5], values for
coefficients in the virial expansion for $\eta $ have been an active area of
research to date even in the low concentration regime. \ A considerable
progress was made in obtaining closed form approximations describing the
rheological properties of suspensions of hard spheres in a broad range of
concentrations [6-8]. Similar results for particles of other geometries are
much less complete [7,9]. An extension of these results to solutions of
polymers has taken place in parallel with these developments [10]. A
noticeable advancements have been made in our understanding of rheology of
dilute and semidilute \ polymer solutions for fully flexible polymers and
rigid rods. It should be noted, though, that polymers add further
complexities because the connectivity of the polymer chain backbone plays an
essential role in calculations of rheological properties of polymer
solutions. The effect of chain connectivity on viscoelastic properties of
polymer solutions has been an object of extensive discussion, and many of
theoretical difficulties encountered in describing these solutions are
shared by suspensions of \ hard spheres. In particular, it is known [11],
that particles immersed in a viscous fluid affect the motion of each other
both hydrodynamically and by direct interaction (hard core, etc). Since the
motion of particles in a fluid is correlated, it contributes to the
distribution of local velocities within the fluid. Behavior of many systems
(e.g. those listed in Section 6) other than the hard sphere suspensions
happens to be closely related or even isomorphic to that \ noticed in
suspensions. This observation makes study of suspensions important in many
areas of physics, chemistry and biology. For reasons which will become
apparent upon reading, in this work we shall mention only physical
applications.

In a polymer solutions when the polymer concentration $\varphi $\ increases,
it is believed that the hydrodynamic interactions become unimportant due to
the effects of \ hydrodynamic screening [12].\ To our knowledge, screening
has not been established in the theory of hard sphere suspensions. If it \
would occur in suspensions, the screened particle motion could be affected
only by thermal fluctuations (truly Brownian motion!). Such Brownian hard
spheres can be described by the short range interacting random walk model
[13]. \ For finite concentrations, we expect the longer range hydrodynamic
interactions to be very important. \ That this is indeed the case, is the
central theme of our paper. In what follows, we provide the theoretical
arguments in favor of hydrodynamic \ interparticle interactions and
screening which must be present in solutions at non-vanishing
concentrations. By exploiting analogies between electrodynamics and fluid
mechanics we shall demonstrate that hydrodynamic screening occurs in much
the same way as screening of the magnetic field in superconductors.
Therefore, mathematically, the description of screening in suspensions is
analogous to that for the Meissner effect in superconductors. This
observation \ will allow us to account for a number of interesting
properties of suspensions. For instance, the viscosity of hard sphere
suspensions is known to diverge beyond some critical concentration $\varphi
^{\ast }$. This phenomenon has been observed experimentally and is well
documented, e.g. see Refs. [14-18]. \ All these references are concerned
with changes in rheological\ properties of suspensions occurring with
changes in concentration $\varphi .$ Using scaling-type arguments Brady,
Ref.[19], and, independently, Bicerano \textit{et al}, Ref. [20], had found
that near $\varphi =\varphi ^{\ast }$ the relative viscosity $\eta /\eta
_{0} $ diverges as $\eta /\eta _{0}=C(1-\varphi /\varphi ^{\ast })^{-2}$
with $C$ being some constant. Furthermore, as it is shown by Bicerano 
\textit{et al}, such analytical dependence of relative viscosity on
concentration $\varphi $ actually works extremely well for all
concentrations. In view of (1.3), it is reasonable to expect vanishing of $D$
for $\varphi \rightarrow \varphi ^{\ast }.$ \ This phenomenon was indeed
observed \ in Ref.[21]. .

\ Theoretically, the result for relative viscosity was obtained as result of
a combined \ nontrivial use of topological and combinatorial arguments. Such
arguments can also be used, for instance, for description of the onset of
turbulence in fluids or gases. As described in Ref.[22], such a regime in
these substances is characterized by the sharp increase in the viscosity
(just like in suspensions). According to Chorin, Ref.[22], Section 6.8, one
can think about such an increase as analogous to processes which take place
in superfluid $^{4}$He when one goes in temperatures from below to above $%
\lambda -$transition, that is from the superfluid to normal fluid state.
Such a transition is believed to be associated with uninhibited
proliferation of \ tangled vortices on any scale. In this work we
demonstrate that Chorin's conjecture is indeed correct. This interpretation
is possible only if \ both\ topological and combinatorial arguments are
rigorously \ and carefully taken into account. Surprisingly, when this is
done, the emerging description becomes isomorphic to that known for the
Bose-Einstein condensation transition. Because of this, \ in addition to
turbulence, in concluding section of this \ work we briefly discuss a number
of apparently different physical systems \ whose behavior under certain
conditions resembles that found in colloidal supensions.

The rest of this paper is organized as follows. In section 2, we introduce
notations, discuss \ experimental data with help of previously found
generalized Stokes-Einstein relation [23] and make conjectures about how
these results should be interpreted in the case if hydrodynamic screening
does exist. Some familiarity with Ginzburg-Landau (G-L) theory of
superconductivity is expected for proper understanding of this and the
following sections. In Section 3 we study in detail how many particle
diffusion processes should be affected if hydrodynamic interactions are
taken into account. The major new results of this section are given in
Section 3.3. where we rigorously demonstrate that account of hydrodynamic
interactions causes modification of Fick's laws of diffusion in the same way
as presence of electromagnetic field \ causes modification of the
Schrodinger's equation for charged particles \ The gauge fields emerging in
the modified Fick's equations are of zero curvature implying involvement of
the Chern-Simons topological field theory. The following Section 4 considers
in detail the implications of the results obtained in Section 3. The major
new result of this section is given in Section 4.4. where we adopted the
logic of the ground breaking paper by London and London [24] in order to
demonstrate the existence of hydrodynamic screening. Thus, the phenomenon of
screening in suspensions is analogous to the Meissner effect in
superconductors [25]. In Section 5 we follow the logic of Ginzburg-Landau
paper [26] elaborating the work by London brothers and develop similar
G-L-type theory for suspensions. The major new result of this section is
presented in Section 5.5. in which by using combinatorial and topological
methods \ we reproduce the scaling results by Brady [19] and Bicerano et al,
Ref.[20]. In Section 6 we place the obtained results in a much broader
context. It is done with help of two key concepts: helicity and force-free
fields. They had been in use for some time in areas such as
magnetohydrodynamics, fluid, plasma and gas turbulence, classical mechanics
written in hydrodynamic formalism but not in superconductivity or colloidal
suspensions, etc. \ In this section we mention as well other uses of these
concepts in disciplines such as high temperature superconductivity, quantum
chromodynamics, string theory, non-Abelian fluids, etc. The paper also
contains three appendices which are made sufficiently self contained. They
are not only very helpful in providing details supporting the results of the
main text but also of independentl interest.

\section{Stokes-Einstein Virial Expansions for a Broad Concentration Range}

\subsection{ General Results}

In 1976, Batchelor obtained the following \ general result for the
cooperative diffusion coefficient [27]:

\begin{equation}
D\left( \varphi \right) =\frac{K\left( \varphi \right) }{6\pi \eta r}\frac{%
\varphi }{1-\varphi }\left( \frac{\partial \mu }{\partial \varphi }\right)
_{p,T},  \tag{2.1}
\end{equation}%
where $K\left( \varphi \right) $ is the sedimentation coefficient of the
particles in suspension and $\mu $ is the chemical potential. Batchelor
obtained for $K\left( \varphi \right) $ the following result: 
\begin{equation}
K\left( \varphi \right) =1-6.55\varphi +O\left( \varphi ^{2}\right) 
\tag{2.2}
\end{equation}%
so that (2.1) with thus obtained first order result for $K\left( \varphi
\right) $ can be used only for low concentrations. In Ref.[3] \ an attempt
was made to extend Batchelor's results to higher concentrations. This was
achieved in view of the fact that

\begin{equation}
\frac{\varphi }{1-\varphi }\left( \frac{\partial \mu }{\partial \varphi }%
\right) _{p,T}=\left( \frac{\partial \Pi }{\partial n}\right) _{p,T}, 
\tag{2.3}
\end{equation}%
where $\Pi $ is the osmotic pressure. Use of this result in (2.1) produces:

\begin{equation}
D\left( \varphi \right) =\frac{K\left( \varphi \right) }{6\pi \eta r}\left( 
\frac{\partial \Pi }{\partial n}\right) _{p,T}.  \tag{2.4}
\end{equation}%
The Carnahan-Starling equation of state for hard spheres can be used to
obtain the following result for compressibility

\begin{equation}
\left( \frac{\partial \Pi }{\partial n}\right) _{p,T}=k_{B}T\frac{\left[
\left( 1+2\varphi \right) ^{2}+\left( \varphi -4\right) \varphi ^{3}\right] 
}{\left( 1-\varphi \right) ^{4}},  \tag{2.5}
\end{equation}%
thus converting equation (2.4) into

\begin{equation}
D\left( \varphi \right) =\frac{K\left( \varphi \right) }{6\pi \eta r}k_{B}T%
\frac{\left[ \left( 1+2\varphi \right) ^{2}+\left( \varphi -4\right) \varphi
^{3}\right] }{\left( 1-\varphi \right) ^{4}}.  \tag{2.6}
\end{equation}%
To be in accord with Batchelor's result (2.2) at low concentrations, the
authors [3] suggested replacing of Eq.(2.2) by%
\begin{equation}
K\left( \varphi \right) \approx \left( 1-\varphi \right) ^{6.55}  \tag{2.7}
\end{equation}%
which allows us to rewrite (2.6) in the following final form

\begin{equation}
D/D_{0}=\left( 1-\varphi \right) ^{6.55}\frac{\left[ \left( 1+2\varphi
\right) ^{2}+\left( \varphi -4\right) \varphi ^{3}\right] }{\left( 1-\varphi
\right) ^{4}}  \tag{2.8}
\end{equation}%
convenient for comparison with experimental data.

Such a comparison can be found in Fig.12 of Ref [3] \ where this result is
plotted against author's own experimental data for the cooperative diffusion
coefficient. The experimental data within error margins appears to agree
extremely well with the theoretical curve obtainable from Eq.(2.8). However,
it should be kept in mind that, in fact, originally Eq.(2.2) was determined
only to first order in $\varphi $ (and, therefore, only for the volume
fractions less than about 0.05). Therefore, formally, Eq.(2.8) is in accord
with Eq.(2.2) only for volume fractions of lesser than about 0.03.
Therefore, it is clear from Fig 12 of [3] that to improve the agreement \ in
the whole range of concentrations, a knowledge of a second order in $\varphi 
$ is desirable in (2.2). This problem can be by passed as follows.

From [3 ] the viscosity data from the same experiments were obtained so that
the data can be fit to the following second order expansion:

\begin{equation}
\eta /\eta _{0}=1+2.5\varphi +6.54\varphi ^{2}+O\left( \varphi ^{3}\right) .
\tag{2.9}
\end{equation}%
To obtain this result, the authors constrained the first order coefficient
to 2.5 to comply with Einstein's result (1.2) for viscosity. If one
considers these data without such a constraint, then one obtains,

\begin{equation}
\eta /\eta _{0}=1+2.4\varphi +7.1\varphi ^{2}+O\left( \varphi ^{3}\right) . 
\tag{2.10}
\end{equation}

In the paper by Kholodenko and Douglas [23] the following result for the
cooperative diffusion coefficient was derived (the generalized
Stokes-Einstein relation)

\begin{equation}
D/D_{0}=\frac{1}{\left( \eta /\eta _{0}\right) }\left[ \frac{S\left( \mathbf{%
0},0\right) }{S_{0}\left( \mathbf{0},0\right) }\right] ^{-1/2},  \tag{2.11}
\end{equation}%
where $S\left( \mathbf{0},0\right) $ is the $\mathbf{k}=0,$ zero angle
static scattering form factor. The thermodynamic sum rule for the hard
sphere gas \ produces the following result for this formfactor: 
\begin{equation}
\left[ \frac{S\left( \mathbf{0},0\right) }{S_{0}\left( \mathbf{0},0\right) }%
\right] ^{-1/2}=1+4\varphi +7\varphi ^{2}+O\left( \varphi ^{3}\right) . 
\tag{2.12}
\end{equation}%
By combining Eq.s (2.9)-(2.12) the result for cooperative diffusion is
obtained:

\begin{equation}
D/D_{0}=1+1.6\varphi -3.9\varphi ^{2}+O\left( \varphi ^{3}\right) . 
\tag{2.13}
\end{equation}%
For the sake of comparison with experiment, we made a numerical fit to the
experimental data for higher concentrations obtained in [3]\ by a polynomial
(up to a second order in $\varphi )$ with the result$\footnote{%
The correlation coefficient obtained for this fit is 0.97.}$:

\begin{equation}
D/D_{0}=1+1.5505\varphi -5.3663\varphi ^{2}+O\left( \varphi ^{3}\right) . 
\tag{2.14}
\end{equation}%
Comparison between Eq.s (2.13) and (2.14) shows that the theoretically
obtained result, Eq.(2.13), is in good agreement with the experimental data,
Eq.(2.14), within error margins. Alternatively, we can use the reciprocal of
the empirical expression, Eq.(2.14), in (2.11) to obtain

\begin{equation}
\eta /\eta _{0}=1+2.45\varphi +8.5\varphi ^{2}+O\left( \varphi ^{3}\right) ,
\tag{2.15}
\end{equation}%
which also compares well with the experimental data, Eq.(2.10).

\subsection{The generalized Stokes-Einstein relation and the role of
hydrodynamic screening}

We return now to Eq.(2.11) for further discussion. \ Based on the results of
\ introductory section, especially on\ Eq.s (1.1) and (1.3), we can formally
write:

\begin{equation}
D/D_{0}=\frac{R}{R^{\ast }}\frac{\eta _{0}}{\eta }=1+a_{1}\varphi
+a_{2}\varphi ^{2}+O\left( \varphi ^{3}\right) .  \tag{2.17}
\end{equation}%
The actual values of $a_{i}$ , $i=1,2,..$ can be determined using Eq.(2.11)
written in the following form

\begin{equation}
D/D_{0}=\frac{\left( \xi /\xi _{0}\right) ^{-1}}{\left( \eta /\eta
_{0}\right) }=\frac{1}{\left( \eta /\eta _{0}\right) }\left[ \frac{S\left( 
\mathbf{0},0\right) }{S_{0}\left( \mathbf{0},0\right) }\right] ^{-1/2}, 
\tag{2.18}
\end{equation}%
where $\xi $ is the correlation length, $\xi _{0}$ is the correlation length
in the infinite dilution limit $\xi _{0}\sim R$. To justify such a move we
need to remind to our readers of some facts from the dynamical theory of
linear response. To do so, we\ borrow some results from our previous
work.[23].

\bigskip Generally, both $\mathit{D}$ and $\mathit{D}_{0}$\ are measured by
light scattering experiments. In these experiments the Fourier transform of
the density-density correlator

\begin{equation}
S(\mathbf{R},\tau )=\langle \delta n(\mathbf{r},t)\delta n(\mathbf{r}%
^{\prime },t^{\prime })\rangle  \tag{2.19}
\end{equation}%
is being measured. \ The formfactor, Eq.(2.19), is written with account of
translational invariance, requiring the above correlator to be a function of
relative distance\textbf{\ }$\left\vert \mathbf{r}-\mathbf{r}^{\prime
}\right\vert \equiv \left\vert \mathbf{R}\right\vert $ only. Time
homogeneity, makes it in addition to be a function of $\left\vert
t-t^{\prime }\right\vert $ =$\tau $ only. In this expression, $\langle
...\rangle $ represents an equilibrium thermal average while density
fluctuations are given by $\delta n(\mathbf{r},t)=n(\mathbf{r},t)-\langle
n\rangle $. By definition, the Fourier transform of Eq.(2.19) is given by

\begin{equation}
S(\mathbf{q},\omega )=\int d\mathbf{r}\int d\tau S(\mathbf{R},\tau )e^{i(%
\mathbf{q}\cdot \mathbf{r}-\omega \tau )}.  \tag{2.20}
\end{equation}%
. Using this expression, \ we obtain the initial decay rate $\Gamma _{%
\mathbf{q}}^{(0)}$ as follows [10,23]:

\begin{equation}
\Gamma _{\mathbf{q}}^{(0)}=-\frac{\partial }{\partial \tau }\ln \left[
\int_{-\infty }^{\infty }\frac{d\omega }{2\pi }e^{i\omega \tau }S(\mathbf{q}%
,\omega )\right] _{\tau \rightarrow 0^{+}}.  \tag{2.21}
\end{equation}%
\ With help of this result, the cooperative diffusion coefficient is
obtained as%
\begin{equation}
D=\frac{\partial }{\partial q^{2}}\Gamma _{\mathbf{q}}^{(0)}|_{\mathbf{q}=0}.
\tag{2.22}
\end{equation}%
In the limit of vanishingly low concentrations the self -diffusion
coefficient is known to be [10] 
\begin{equation}
D_{0}=\frac{1}{6}\lim_{t\rightarrow \infty }\frac{1}{t}\left\langle \left\{ 
\mathbf{r}(t)-\mathbf{r}(0)\right\} ^{2}\right\rangle ,  \tag{2.23}
\end{equation}%
where $<...>$ denotes the Gaussian-type average. Following Lovesey [28], it
is convenient to rewrite this result as 
\begin{equation}
D_{0}=\frac{1}{3}\int_{0}^{\infty }dt\left\langle \mathbf{v}\cdot \mathbf{v}%
(t)\right\rangle  \tag{2.24}
\end{equation}%
in view of the fact that if 
\begin{equation}
\mathbf{r}(t)-\mathbf{r}(0)=\int\limits_{0}^{t}\mathbf{v}(\tau )d\tau 
\tag{2.25a}
\end{equation}%
then, 
\begin{equation}
\left\langle \left( \mathbf{r}(t)-\mathbf{r}(0)\right) ^{2}\right\rangle
=\left\langle \left\{ \int\limits_{0}^{t}\mathbf{v}(\tau )d\tau \right\}
^{2}\right\rangle =2\int\limits_{0}^{t}d\tau \int\limits_{0}^{\bar{\tau}}d%
\bar{\tau}\left\langle \mathbf{v}(\tau )\cdot \mathbf{v}(\bar{\tau}%
)\right\rangle =2\int\limits_{0}^{t}d\tau (t-\tau )\left\langle \mathbf{v}%
(\tau )\cdot \mathbf{v}(0)\right\rangle ,  \tag{2.25b}
\end{equation}%
while, by definition, $D_{0}=\frac{1}{6}\lim_{t\rightarrow \infty }\frac{1}{t%
}\left\langle \left( \mathbf{r}(t)-\mathbf{r}(0)\right) ^{2}\right\rangle .$

With these definitions in place and taking into account Eq.(2.18), we would
like now to discuss in more detail the relationship between $D$ and $D_{0}$.
Using Eq.(2.29) of Ref. [23] we obtain%
\begin{equation}
D=\lim_{\tau \rightarrow 0^{+},\mathbf{k}=0}\frac{1}{3}\int\limits_{0}^{%
\infty }dt^{^{\prime \prime }}\frac{<\mathbf{j}(\mathbf{0},t)\cdot \mathbf{j}%
(\mathbf{0},t^{\prime \prime })>}{S(\mathbf{0},0)},  \tag{2.26}
\end{equation}%
where the current $\mathbf{j}$ is given as $\mathbf{j=}\delta n(\mathbf{r},t)%
\mathbf{v}(r,t),$ provided that the non-slip boundary condition%
\begin{equation}
\mathbf{v}_{f}(\mathbf{r},t)=\frac{d\mathbf{r}}{dt}\equiv \mathbf{v}(t) 
\tag{2.27}
\end{equation}%
is applied. Here $\mathbf{v}_{f}(\mathbf{r},t)$ is the velocity of the fluid
and $\mathbf{r}(t)$ is the position of the center of mass of the hard sphere
with respect to the chosen frame of reference. Eq.(2.26) is in agreement
with Eq.(2.24) in view of the fact that in the limit of zero concentration $%
S(\mathbf{0},0)=1$ so that $<\mathbf{j}(\mathbf{0},t)\cdot \mathbf{j}(%
\mathbf{0},t^{\prime \prime })>\rightarrow \left\langle \mathbf{v}\cdot 
\mathbf{v}(t)\right\rangle $ as we would like to demonstrate now. For this
purpose, in view of Eq.(2.22) it is convenient to rewrite the result
Eq.(2.26) is the equivalent form 
\begin{equation}
S(\mathbf{q},0)\Gamma _{\mathbf{q}}^{(0)}=\int\limits_{0}^{\infty
}dt^{\prime \prime }\mathbf{q}\cdot <\mathbf{j}(\mathbf{q},t)\mathbf{j}(-%
\mathbf{q},t^{\prime \prime })>\cdot \mathbf{q}\mid _{\tau \rightarrow 0^{+}}
\tag{2.28}
\end{equation}%
in accord with Eq.(2.15) of Ref.[23]. This relation is very convenient for
theoretical analysis. For instance, it is straightforward to obtain $D$ in
the decoupling approximation as suggested by Ferrell [21]. It is given by 
\begin{equation}
^{{}}\mathbf{q}\cdot <\mathbf{j}(\mathbf{q},t)\mathbf{j}(-\mathbf{q}%
,t^{\prime })>\cdot \mathbf{q}\dot{=}\mathbf{q}\cdot <\mathbf{v}(\mathbf{q}%
^{\prime },t)\mathbf{v}(\mathbf{q-q}^{\prime },t^{\prime })>\cdot \mathbf{q}%
\langle \delta n(\mathbf{q-q}^{\prime },t)\delta n(\mathbf{q}^{\prime
},t^{\prime })\rangle .  \tag{2.29}
\end{equation}%
In Section 5.4. we provide proof \ that the above decoupling is in fact
exact. This provides an explanation why it is working so well in real
experiments.

\ In the meantime, \ we shall consider this decoupling as an approximation.
Once such an approximation is made, the problem of calculation of $D$ is
reduced to the evaluation of correlators defined in Eq.(2.29). For the
velocity-velocity correlator, the following expression was obtained before
(e.g. see Ref.[23], Eq.(2.18)): 
\begin{eqnarray}
&<&v_{i\mathbf{k}}(t)v_{j\mathbf{k}^{\prime }}(t^{\prime })>=(2\pi
)^{3}\delta (\mathbf{k}+\mathbf{k}^{\prime })\{\delta _{ij}-\frac{k_{i}k_{j}%
}{k^{2}}\}\frac{2k_{B}T}{\eta k^{2}}\delta (t-t^{\prime })  \nonumber \\
&\equiv &2k_{B}T\mathcal{H}_{ij}(\mathbf{k})\delta (t-t^{\prime }) 
\TCItag{2.30}
\end{eqnarray}%
with $i,j=1-3.$ This expression defines the Oseen tensor $\mathcal{H}_{ij}(%
\mathbf{k})$ to be discussed in detail in the next section. The presence of
the delta function $\delta (t-t^{\prime })$ in Eq.(2.30) makes it possible
to look only at the equal time density-density correlator in the
decomposition of the $\mathbf{j}-\mathbf{j}$ correlator given by Eq.(2.29).
Such a correlator also was discussed in Ref.[23] where it is shown to be%
\begin{equation}
\langle \delta n(\mathbf{k-k}^{\prime },t)\delta n(\mathbf{k}^{\prime
},t)\rangle =k_{B}T<n>\left[ \frac{\partial }{\partial \Pi }<n>\right] _{T} 
\tag{2.31a}
\end{equation}%
Actually, it is both time and $k-$independent since, as is well known, it is
the thermodynamic sum rule. That is%
\begin{equation}
S(\mathbf{0},0)=k_{B}T<n>\left[ \frac{\partial }{\partial \Pi }<n>\right]
_{T}.  \tag{2.31b}
\end{equation}%
It is convenient to rewrite this result as follows 
\begin{equation}
k_{B}T<n>\left[ \frac{\partial }{\partial \Pi }<n>\right] _{T}=\int d\mathbf{%
R}S\mathbf{(R,}0\mathbf{)\equiv }S(\mathbf{0},0),  \tag{2.31c}
\end{equation}%
implying that 
\begin{equation}
S\mathbf{(R,}0\mathbf{)=}k_{B}T<n>\left[ \frac{\partial }{\partial \Pi }<n>%
\right] _{T}\delta (\mathbf{R}).  \tag{2.32}
\end{equation}%
In view of Eq.(2.12), we notice that in the limit of vanishing
concentrations $S(\mathbf{0},0)=1.$In such an extreme case the decoupling
made in Eq.(2.29) superimposed with the definition, Eq.(2.22), and the fact
that 
\[
\frac{\partial }{\partial q^{2}}\cdot \cdot \cdot =\frac{1}{3}%
tr(\sum\limits_{i,j}\frac{\partial }{\partial q_{i}}\frac{\partial }{%
\partial q_{j}}\cdot \cdot \cdot ) 
\]%
produces the anticipated result, Eq.(2.24), as required. Next, following
Ferrell [21], we regularize the delta function in Eq.(2.32). Using an
identity $1=\int\limits_{0}^{\infty }dxxe^{-x},$ \ the regularized
expression for $S\mathbf{(R,}0\mathbf{)}$ is obtained as follows%
\begin{equation}
S\mathbf{(}r\mathbf{,}0\mathbf{)=}\frac{k_{B}T}{4\pi \xi ^{2}}<n>\left[ 
\frac{\partial }{\partial \Pi }<n>\right] _{T}\frac{1}{r}e^{-\frac{r}{\xi }},
\tag{2.33}
\end{equation}%
where $r=\left\vert \mathbf{R}\right\vert $ and the parameter $\xi $ is
proportional to the \textsl{static} correlation length\footnote{%
For more details, see our work, Ref.[23].}. To use this expression for
calculations of $D,$ by employing Eq.(2.26) we have to transform the
hydrodynamic correlator, Eq.(2.30), into coordinate form as well. Such a
form is given in Eq.(2.33) of Ref.[23] as 
\begin{equation}
<\mathbf{v}(\mathbf{r},t)\cdot \mathbf{v}(\mathbf{r}^{\prime },t^{\prime })>=%
\frac{k_{B}T}{\pi \eta }\frac{1}{\left\vert \mathbf{r}-\mathbf{r}^{\prime
}\right\vert }\delta (t-t^{\prime }).  \tag{2.34}
\end{equation}%
This expression is written with total disregard of possible effects of the
hydrodynamic screening, though. The combined use of Eq.s (2.33) and (2.34)
in Eq.(2.26) produces the anticipated result%
\begin{equation}
D=\frac{k_{B}T}{3\pi \eta \xi }  \tag{2.35}
\end{equation}%
in accord with that obtained by Ferrell, Ref.[21], Eq.(11), provided that we
redefine (still arbitrary) the parameter $\xi $ as $2\check{\xi}$. The
result (2.35) also coincides with Eq.(1.3) if we identify $\check{\xi}$ with 
$R^{\ast }.$ Furthermore, by looking at Eq.(2.18) we realize that in the
infinite dilution limit we have to replace $\check{\xi}$ by $\xi _{0\text{ }%
} $ and, accordingly, $\eta $ by $\eta _{0}$. Such an identification leads
to the generalized Stokes-Einstein relation in the form given by Eq.(2.18)
implying that 
\begin{equation}
\check{\xi}/\xi _{0}=\left[ \frac{S\left( \mathbf{0},0\right) }{S_{0}\left( 
\mathbf{0},0\right) }\right] ^{1/2}.  \tag{2.36a}
\end{equation}%
Since we have noticed before that $S_{0}\left( \mathbf{0},0\right) =1$ this
result can be rewritten as%
\begin{equation}
\check{\xi}=\sqrt{S\left( \mathbf{0},0\right) }\xi _{0}.  \tag{2.36b}
\end{equation}

Suppose now that hydrodynamic interactions are screened to some extent. In
such a case the result Eq.(2.34) should be modified accordingly. Thus, we
obtain 
\begin{equation}
<\mathbf{v}(\mathbf{r},t)\cdot \mathbf{v}(\mathbf{r}^{\prime },t^{\prime })>=%
\frac{k_{B}T}{\pi \eta }\frac{\exp (-\dfrac{r}{\xi _{H}})}{r}\delta
(t-t^{\prime }),  \tag{2.37}
\end{equation}%
where we\ have introduced the \textsl{hydrodynamic} correlation length $\xi
_{H}.$ If, as we shall demonstrate below, the analogy between hydrodynamic
and superconductivity makes sense under some conditions then, using \ this \
assumed analogy we introduce the Ginzburg parameter $\kappa _{G}$ \ for this
problem via known relation [25]: 
\begin{equation}
\xi _{H}=\kappa _{G}\check{\xi}.  \tag{2.38}
\end{equation}%
Using Eq.s (2.33), (2.37) and (2.38) in (2.26), the result for $D$,
Eq.(2.35), acquires the following form:%
\begin{equation}
D=\frac{k_{B}T}{6\pi \Sigma \eta }(1+\frac{1}{\kappa _{G}})^{-1}.  \tag{2.39}
\end{equation}%
Since, the adjustable parameter $\Sigma $ is introduced in Eq.(2.39) quite
arbitrarily, we can, following Ferrell, Ref. [21], take full advantage of
this fact now. To do so, we notice that from the point of view of the
observer, the relation (2.36) holds irrespective of the absence or presence
of hydrodynamic screening. Because of this, we write%
\begin{equation}
\Sigma (1+\frac{1}{\kappa _{G}})=\check{\xi}  \tag{2.40}
\end{equation}%
so that the Eq.(2.36) used in the generalized Stokes-Einstein relation
remains unchanged. By combining Eq.s(2.36b), (2.38) and (2.40) we obtain:%
\begin{equation}
\kappa _{G}=\frac{\xi _{H}}{\xi _{0}}\frac{1}{\sqrt{S\left( \mathbf{0}%
,0\right) }}=\frac{\Sigma }{\xi _{0}}\frac{\kappa _{G}}{\sqrt{S\left( 
\mathbf{0},0\right) }}(1+\frac{1}{\kappa _{G}})  \tag{2.41a}
\end{equation}%
or, equivalently,%
\begin{equation}
\frac{\xi _{0}}{\Sigma }=\frac{1}{\sqrt{S\left( \mathbf{0},0\right) }}(1+%
\frac{1}{\kappa _{G}}).  \tag{2.41b}
\end{equation}%
In this equation the parameter $\Sigma $ is \ still undefined. We can define
this parameter now based on physical arguments. In particular, let us set $%
\Sigma =S\left( \mathbf{0},0\right) \xi _{0}.$ Then, we end up with the
equation 
\begin{equation}
1+\frac{1}{\kappa _{G}}=\frac{1}{\sqrt{S\left( \mathbf{0},0\right) }} 
\tag{2.42}
\end{equation}%
leading to 
\begin{equation}
\kappa _{G}=\frac{1}{\frac{1}{\sqrt{S\left( \mathbf{0},0\right) }}-1}. 
\tag{2.43}
\end{equation}%
To reveal the physical meaning of this equation we use Eq.s (2.36b), (2.38)
and (2.43) in order to obtain%
\begin{equation}
\xi _{H}=\frac{\sqrt{S\left( \mathbf{0},0\right) }\xi _{0}}{\dfrac{1}{\sqrt{%
S\left( \mathbf{0},0\right) }}-1}.  \tag{2.44}
\end{equation}%
From Eq.(2.12) we notice that by considering the infinite dilution limit $%
\varphi \rightarrow 0,$ we obtain: $\xi _{H}\rightarrow \infty ,$ implying
absence of hydrodynamic screening. Consider the opposite case: $\varphi
\rightarrow \infty $ (that is, in practice, $\varphi $ being large). Looking
at Eq.(2.12) we notice that in this case $\xi _{H}\rightarrow 0$ indicating
the complete screening, as expected. Using Eq.(2.42), these results allow us
to rewrite the generalized Stokes -Einstein relation, Eq.(2.18), in the
equivalent form emphasizing the role of hydrodynamic screening. Thus, we
obtain, 
\begin{equation}
D/D_{0}=\frac{1}{\left( \eta /\eta _{0}\right) }(1+\frac{1}{\kappa _{G}}). 
\tag{2.45}
\end{equation}

\bigskip

\section{Diffusion processes in the presence of hydrodynamic interactions}

\subsection{Some facts from the diffusion theory}

If $n$ is the local density, then the flux\textbf{\ $j$}$=n\mathbf{v}$%
\textit{\ }obeys Fick's first law:%
\begin{equation}
\mathbf{j}=-D\mathbf{\nabla }n,  \tag{3.1}
\end{equation}%
where $D$ is the (in general, cooperative) diffusion coefficient, and $%
\mathbf{v}$ is the local velocity. Upon substitution of this expression into
the continuity equation

\begin{equation}
\frac{\partial n}{\partial t}+\mathbf{\nabla }\cdot \mathbf{j}=0  \tag{3.2}
\end{equation}%
we obtain the diffusion equation commonly known as Fick's second law

\begin{equation}
\frac{\partial n}{\partial t}=D\nabla ^{2}n\text{.}  \tag{3.3}
\end{equation}%
In the presence of some external forces, i.e. 
\begin{equation}
\mathbf{F}=-\mathbf{\nabla }U,  \tag{3.4}
\end{equation}%
the diffusion laws must be modified. This is achieved by assuming the
existence of some kind of friction, i.e.by assuming that there exists a
relation 
\begin{equation}
\gamma \mathbf{v}=\mathbf{F}  \tag{3.5}
\end{equation}%
between the local velocity $\mathbf{v}$ and force $\mathbf{F}$\textbf{\ }%
with the coefficient of proportionality $\gamma $ being, for instance (in
the case of hard spheres), of the type given in Eq.(1.1). With such an
assumption, the diffusion current\textbf{, }Eq.(3.1)\textbf{, }is modified
now as follows%
\begin{equation}
\mathbf{j}=-D\mathbf{\nabla }n-\frac{n}{\gamma }\mathbf{\nabla }U.  \tag{3.6}
\end{equation}%
Such a definition makes sense. Indeed, in the case of equilibrium$,$ when
the concentration $n_{eq}$ obeys the Boltzmann's law%
\begin{equation}
n_{eq}=n_{0}\exp (-\frac{U}{k_{B}T}),  \tag{3.7}
\end{equation}%
vanishing of the current in Eq.(3.6) is assured by substitution of Eq.(3.7)
into Eq.(3.6) thus leading to the already cited Einstein result, Eq.(1.1),
for $D_{0}$. As in the case of Eq.(1.3), we shall assume that for finite
concentrations one can still use the Einstein-like result for the diffusion
coefficient. With such an assumption, the current $\mathbf{j}$ in Eq.(3.6)
acquires the following form [12]:%
\begin{equation}
\mathbf{j}=-\frac{n}{\gamma }\mathbf{\nabla }(k_{B}T\ln n+U)\equiv -\frac{n}{%
\gamma }\mathbf{\nabla }\mu \mathbf{,}\text{ }  \tag{3.8}
\end{equation}%
where the last equality defines the nonequilibrium chemical potential $\mu ,$
e.g. like that given in Eq.(2.1). Alternatively, the modified flux velocity $%
\mathbf{v}_{f}$ is given now by $-\frac{1}{\gamma }\mathbf{\nabla }\mu $ so
that the continuity Eq.(3.2) reads as 
\begin{equation}
\frac{\partial n}{\partial t}+\mathbf{\nabla }\cdot (n\mathbf{v}_{f})=0. 
\tag{3.9}
\end{equation}%
Exactly the same equation can be written for the probability density $\Psi $
if we formally replace $n$ by $\Psi $ in the above equation [10]. Such an
interpretation of diffusion is convenient since it allows \ one to talk
about diffusion in terms of the trajectories of Brownian motion of
individual particles whose positions $\mathbf{x}_{n}(t),n=1,2,...$ are
considered to be as random variables. Then, the probability $\Psi $
describes such collective Brownian motion process described by the following
Schrodinger-like equation 
\begin{equation}
\frac{\partial }{\partial t}\Psi =-\sum\limits_{n}\frac{\partial }{\partial 
\mathbf{x}_{n}}(\Psi \mathbf{v}_{fn})  \tag{3.10}
\end{equation}%
in which the velocity $\mathbf{v}_{fn}$ is given by 
\begin{equation}
\mathbf{v}_{fn}=-\sum\limits_{m}L_{nm}\frac{\partial }{\partial \mathbf{x}%
_{m}}(k_{B}T\ln \Psi +U).  \tag{3.11}
\end{equation}%
Thus, we obtain our final result 
\begin{equation}
\frac{\partial }{\partial t}\Psi =\sum\limits_{m,n}\frac{\partial }{\partial 
\mathbf{x}_{n}}L_{nm}(k_{B}T\frac{\partial }{\partial \mathbf{x}_{m}}\Psi +%
\frac{\partial U}{\partial \mathbf{x}_{m}}\Psi )  \tag{3.12}
\end{equation}%
adaptable for hydrodynamic extension. For this purpose, we need to remind
our readers \ of some basic facts from hydrodynamics

\subsection{Hydrodynamic fluctuations and Oseen tensor}

The analog of Newton's equation \ for fluids is the Navier-Stokes equation.
It is given by [29]

\begin{equation}
\frac{\partial }{\partial t}\mathbf{v}+(\mathbf{v}\cdot \mathbf{\nabla })%
\mathbf{v}=-\frac{1}{\rho }\mathbf{\nabla }P+\Gamma \nabla ^{2}\mathbf{v} 
\tag{3.13}
\end{equation}%
where $P$ is the hydrodynamic pressure, $\Gamma =\eta _{0}/\rho _{0}$ is the
kinematic viscosity and $\rho _{0}$ is the density of the the pure solvent.
At low Reynold's numbers, the convective term $($\textbf{$v$}$\cdot \mathbf{%
\nabla })\mathbf{v}$ can be neglected [29], p.63. We shall also assume that
the fluid is incompressible, i.e%
\begin{equation}
\text{div }\mathbf{v}=0.  \tag{3.14}
\end{equation}%
Under such conditions the Fourier transformed Navier-Stokes equation can be
written as

\begin{equation}
\frac{\partial }{\partial t}\mathbf{v}_{\mathbf{k}}=-\Gamma \mathbf{k}^{2}%
\mathbf{v}_{\mathbf{k}}-\frac{i\mathbf{k}}{\rho }P_{\mathbf{k}}.  \tag{3.15}
\end{equation}%
Let us add a fluctuating source term $\mathbf{f}_{\mathbf{k}}$ to the right
hand side of Eq.(3.15). Then, using \ the incompressibility condition,
Eq.(3.14), we obtain: 
\begin{equation}
P_{\mathbf{k}}=-i\rho \frac{\mathbf{k}\cdot \mathbf{f}_{\mathbf{k}}(t)}{k^{2}%
}.  \tag{3.16}
\end{equation}%
Introducing the transverse tensor $T_{ij}(\mathbf{k})=\delta _{ij}-\frac{%
k_{i}k_{j}}{k^{2}}$ and decomposing a random force as

\begin{equation}
f_{i\mathbf{k}}^{T}\left( t\right) =\sum\limits_{j}T_{ij}(\mathbf{k})f_{j%
\mathbf{k}}\left( t\right)  \tag{3.17}
\end{equation}%
eventually replaces the Navier-Stokes equation by the Langevin-type equation
for the transverse velocity fluctuations:%
\begin{equation}
\frac{\partial }{\partial t}\mathbf{v}_{\mathbf{k}}+\mathbf{\Gamma k}^{2}%
\mathbf{v}_{\mathbf{k}}=\mathbf{f}_{\mathbf{k}}^{T}\left( t\right) . 
\tag{3.18}
\end{equation}%
As is usually done for such type of equations, we shall assume that the
random fluctuating forces are Gaussianly distributed. This assumption is
equivalent to the statement that 
\begin{equation}
\left\langle f_{i\mathbf{k}}^{T}(t)f_{j\mathbf{k}^{\prime }}^{T}(t^{^{\prime
}})\right\rangle =T_{ij}(\mathbf{k})(2\pi )^{3}\delta (\mathbf{k}+\mathbf{k}%
^{\prime })\tilde{D}\delta (t-t^{\prime })  \tag{3.19}
\end{equation}%
with parameter $\tilde{D}$ to be determined. A formal solution of the
Langevin-type Eq.(3.18) is given by 
\begin{equation}
\mathbf{v}_{\mathbf{k}}(t)=\mathbf{v}_{\mathbf{k}}(0)e^{-\tilde{\Gamma}%
t}+\int\limits_{0}^{t}dt^{\prime }\mathbf{f}_{\mathbf{k}}^{T}\left(
t^{\prime }\right) e^{-\tilde{\Gamma}(t-t^{\prime })}  \tag{3.20}
\end{equation}%
with $\tilde{\Gamma}=\mathbf{k}^{2}\Gamma .$ Introducing $\mathbf{v}_{%
\mathbf{k}}(t)-\mathbf{v}_{\mathbf{k}}(0)e^{-\tilde{\Gamma}t}=\mathbf{\hat{v}%
}_{\mathbf{k}}(t),$ we obtain 
\begin{equation}
\left\langle \hat{v}_{i\mathbf{k}}(t)\hat{v}_{j\mathbf{k}^{\prime
}}(t^{\prime })\right\rangle =\left\langle \int\limits_{0}^{t}dt^{\prime }%
\mathbf{f}_{\mathbf{k}}^{T}\left( t^{\prime }\right) e^{-\tilde{\Gamma}%
(t-t^{\prime })}\int\limits_{0}^{t^{\prime }}dt^{\prime \prime }\mathbf{f}_{%
\mathbf{k}^{\prime }}^{T}(t^{\prime \prime })e^{-\tilde{\Gamma}(t^{\prime
}-t^{\prime \prime })}\right\rangle .  \tag{3.21}
\end{equation}%
To calculate this correlator, and to determine the parameter $2\tilde{D},$
we consider the equal time correlator first. In such a case the
equipartition theorem produces the following result:%
\begin{equation}
\left\langle \hat{v}_{i\mathbf{k}}(t)\hat{v}_{j\mathbf{k}^{\prime
}}(t)\right\rangle =T_{ij}(\mathbf{k})(2\pi )^{3}\delta (\mathbf{k}+\mathbf{k%
}^{\prime })\frac{k_{B}T}{\rho }.  \tag{3.22}
\end{equation}%
Taking into account Eq.s (3.19) and (3.22) we obtain for the
velocity-velocity correlator, Eq.(3.21), the following result%
\begin{eqnarray}
\left\langle \hat{v}_{i\mathbf{k}}(t)\hat{v}_{j\mathbf{k}^{\prime
}}(t^{\prime })\right\rangle &=&T_{ij}(\mathbf{k})(2\pi )^{3}\delta (\mathbf{%
k}+\mathbf{k}^{\prime })\frac{2k_{B}T}{\rho \tilde{\Gamma}}\frac{\tilde{%
\Gamma}}{2}\exp (-\tilde{\Gamma}\left\vert t-t^{\prime }\right\vert ) 
\nonumber \\
&=&2k_{B}T\mathcal{H}_{ij}(\mathbf{k})\frac{\tilde{\Gamma}}{2}\exp (-\tilde{%
\Gamma}\left\vert t-t^{\prime }\right\vert ).  \TCItag{3.23}
\end{eqnarray}%
In the limit $\tilde{\Gamma}\rightarrow \infty $ the combination $\frac{%
\tilde{\Gamma}}{2}\exp (-\tilde{\Gamma}\left\vert t-t^{\prime }\right\vert )$
can be replaced by $\delta (t-t^{\prime }).$ In this limit the obtained
expression coincides with already cited Eq.(2.30). Furthermore, the constant 
$\tilde{D}$ can be chosen as $\frac{k_{B}T}{\rho }.$ To prove the
correctness of these assumptions, we take a Fourier transform (in time
variable) in order to obtain 
\begin{equation}
\left\langle \hat{v}_{i\mathbf{k}}(\omega )\text{ }\hat{v}_{j(-\mathbf{k)}%
}(-\omega )\right\rangle \dot{=}\frac{2k_{B}T}{\rho }\frac{\tilde{\Gamma}}{%
\omega ^{2}+\tilde{\Gamma}^{2}}T_{ij}(\mathbf{k}).  \tag{3.24}
\end{equation}%
This result coincides with Eq.(89.17) of Ref.[25] as required. Here the sign 
$\dot{=}$ means "up to a delta function prefactor". Incidentally, these
prefactors were preserved in another volume of Landau and Lifshitz, e.g. see
Ref. [30], Eq.(122.12). Since in the limit $\omega \rightarrow 0$ we
reobtain (upon inverse Fourier transform in time) Eq.(2.30), this fact
provides the needed justification for replacement of the factor $\frac{%
\tilde{\Gamma}}{2}\exp (-\tilde{\Gamma}\left\vert t-t^{\prime }\right\vert )$
by $\delta (t-t^{\prime }).$

In polymer physics, Ref.[10], typically only this $\omega \rightarrow 0$
limit is considered, which is equivalent to considering physical processes
at time scales much larger than the characteristic time scale $\tau
_{r}=\rho R^{2}/\pi \eta _{0}$ mentioned in the Introduction. Although this
fact \ could cause some inconsistencies (e.g. see discussion below), we
shall follow the traditional pathway by considering mainly this limit
causing us to drop altogether time-dependence in Eq.(3.15) thus bringing it
to the form considered in the book by Doi and Edwards, Ref.[10], Eq.
(3.III.2). Following these authors, this approximation allows us to specify
a random force $\mathbf{f}(\mathbf{r})$ as%
\begin{equation}
\mathbf{f}(\mathbf{r})=\sum\limits_{n}\mathbf{F}_{n}\delta (\mathbf{r}-%
\mathbf{R}_{n})  \tag{3.25}
\end{equation}%
implying that particle (hard sphere) locations are at the points $\mathbf{R}%
_{n}$ so that the fluctuating component of fluid velocity $\mathbf{v}(%
\mathbf{r})$ at $\mathbf{r}$ is given by 
\begin{equation}
\mathbf{v}(\mathbf{r})=\sum\limits_{n}\mathbf{H}(\mathbf{r}-\mathbf{R}%
_{n})\cdot \mathbf{F}_{n}  \tag{3.26}
\end{equation}%
with the Oseen tensor $\mathbf{H}_{ij}(\mathbf{r})$ in the coordinate
representation given by 
\begin{equation}
\mathbf{H}_{ij}(\mathbf{r})=\frac{1}{8\pi \eta \left\vert \mathbf{r}%
\right\vert }(\delta _{ij}+\hat{r}_{i}\hat{r}_{j}).  \tag{3.27}
\end{equation}%
In this expression $\hat{r}_{i}=\dfrac{r_{i}}{\left\vert \mathbf{r}%
\right\vert }.$In view of Eq.(2.27), we can rewrite Eq.(3.26) in the
following suggestive form%
\begin{equation}
\mathbf{v}(\mathbf{R}_{n})=\sum\limits_{m(m\neq n)}\mathbf{H}(\mathbf{R}_{n}-%
\mathbf{R}_{m})\cdot \mathbf{F}_{m}\text{, }  \tag{3.28}
\end{equation}%
for velocity $\mathbf{v}(\mathbf{R}_{n})$ of the particle located at $%
\mathbf{R}_{n}.$

\subsection{Fick's laws in the presence of hydrodynamic interactions.
Emergence of gauge fields}

\bigskip By comparing Eq.s(3.12) and (3.28) we could write the Fick's first
law explicitly should the Oseen tensor be also defined for $m=n$. \ But it
is not defined in this case. As in electrostatics, self-interactions must be
excluded from consideration. In view of the results of Section 2, the
situation can be repaired if we assume that at some concentrations the
hydrodynamic interactions are totally screened. In such a case only the
usual Brownian motion of individual particles is expected to survive. With
these remarks, Fick's first law for such hydrodynamically interacting
suspensions of spheres can be written now as follows%
\begin{equation}
\mathbf{v}_{f}(\mathbf{R}_{n})=-\sum\limits_{m}\mathbf{\tilde{H}}(\mathbf{R}%
_{n}-\mathbf{R}_{m})\cdot \frac{\partial }{\partial \mathbf{R}_{m}}%
(k_{B}T\ln \Psi +U),  \tag{3.29}
\end{equation}%
where the redefined Oseen tensor $\mathbf{\tilde{H}}_{ij}(\mathbf{R})$ has
the diagonal part $\mathbf{\tilde{H}}_{ii}(\mathbf{R})=\frac{1}{\gamma }$ in
accord with Eq.(1.1). The potential $U$ comes from short-range non-
hydrodynamic interactions between particles, which are always present. Using
this result and Eq.(3.10), we finally arrive at the Fick's second law%
\begin{equation}
\frac{\partial \Psi }{\partial t}=\sum\limits_{n,m}\frac{\partial }{\partial 
\mathbf{R}_{n}}\cdot \mathbf{\tilde{H}}(\mathbf{R}_{n}-\mathbf{R}_{m})\cdot
(k_{B}T\frac{\partial \Psi }{\partial \mathbf{R}_{m}}+\frac{\partial U}{%
\partial \mathbf{R}_{m}}\Psi )  \tag{3.30}
\end{equation}%
in accord with Eq.(3.110) of Ref.[10]. Since this equation contains both
diagonal and nondiagonal terms the question arises about its mathematical
meaning. That is, we should inquire: under what conditions does the solution
to this equation exist? \ The solution will exist if and only if the above
equation can be brought to the diagonal form. To do so, as it is usually
done in mathematics, we have to find generalized coordinates in which the
above equation will acquire the diagonal form. Although the attempts to do
so were made by several authors, most notably, by Kirkwood, e.g. see
Ref.[10], chr-3 and references therein, in this work we would like to extend
their results to account for effects of gauge invariance.

We begin with the following auxiliary problem: since $\nabla ^{2}=$div$\cdot 
\mathbf{\nabla }$ $\equiv \mathbf{\nabla }\cdot \mathbf{\nabla ,}$ we are
interested in finding how this result changes if we transform it from the
flat Euclidean space to the space described in terms of generalized
coordinates. This task is easy if we take into account that in the Euclidean
space%
\begin{equation}
\mathbf{\nabla }\cdot \mathbf{\nabla =}\sum\limits_{i,j}\frac{\partial }{%
\partial x_{i}}h^{ij}\frac{\partial }{\partial x_{j}},  \tag{3.31}
\end{equation}%
with $h^{ij}$ being a diagonal matrix with unit entries. We notice that the
above expression is a scalar and, hence, it is covariant. This means, that
we can replace the usual derivatives by covariant derivatives, the metric
tensor $h^{ij}$ by the metric tensor $g^{ij}$ in the curved space so that in
this, the most general case, we obtain%
\begin{equation}
D_{i}g^{ij}D_{j}f(\mathbf{x})=\frac{\partial }{\partial x_{i}}g^{ij}\frac{%
\partial }{\partial x_{j}}f(\mathbf{x})+g^{kj}\Gamma _{ik}^{i}\frac{\partial 
}{\partial x_{j}}f(\mathbf{x}),  \tag{3.32}
\end{equation}%
where summation over repeated indices is assumed, as usual. The covariant
derivative $\mathbf{D}_{i}$ is defined for a scalar $f$ as $D_{i}f=\frac{%
\partial }{\partial x_{i}}f$ and for contravariant vector X$^{i}$ as%
\begin{equation}
D_{j}X^{i}=\frac{\partial X^{i}}{\partial x_{j}}+\Gamma _{jk}^{i}X^{i} 
\tag{3.33}
\end{equation}%
with Christoffel symbol $\Gamma _{jk}^{i}$ defined in a usual way of
Riemannian geometry. A precise definition of this symbol is going to be
given below. Since $\Gamma _{ik}^{i}=\frac{\partial }{\partial x_{k}}\ln 
\sqrt{g}$ , we can rewrite Eq.(3.32) in the following alternative final form%
\begin{equation}
\nabla ^{2}f=D_{i}g^{ij}D_{j}f(\mathbf{x})=\frac{1}{\sqrt{g}}\frac{\partial 
}{\partial x_{i}}[g^{ij}\sqrt{g}\frac{\partial }{\partial x_{i}}f] 
\tag{3.34}
\end{equation}%
so that in Eq.(3.3) the operator$\ \nabla ^{2}$ is replaced \ now by that
given by Eq.(3.34). To make this presentation complete, we have to include
the relation 
\begin{equation}
g_{ij}=\frac{\partial r^{k}}{\partial q^{i}}\frac{\partial r^{l}}{\partial
q^{j}}h_{kl}.  \tag{3.35}
\end{equation}%
In the simplest case, when we are dealing with 3 dimensional vectors, so
that $\mathbf{r}=\mathbf{r}(q_{1},q_{2},q_{3}),$ sometimes it is convenient
to introduce vectors%
\begin{equation}
\mathbf{e}_{i}=\frac{\partial \mathbf{r}}{\partial q^{i}}  \tag{3.36}
\end{equation}%
and the metric tensor%
\begin{equation}
g_{ij}=\mathbf{e}_{i}\cdot \mathbf{e}_{j}  \tag{3.37}
\end{equation}%
with "$\cdot "$ being the usual Euclidean scalar product sign. Definitions
Eq.(3.35) and (3.37) are obviously equivalent in the present case. Because
of this, it is clear that upon transformation to the curvilinear coordinates
the Riemann curvature tensor \ written in terms of $g_{ij}$ is \ still zero
since it is obviously zero for the $h_{kl}.$ The curvature tensor will be
introduced and discussed below. Before doing so, using the example we have
just described, we need to rewrite Eq.(3.30) in terms of generalized
coordinates. In the present case, we must have $3N$ generalized coordinates
and the tensor $h_{kl}$ is not a unit tensor anymore. Our arguments will not
change if we replace Eq.(3.30) by that in which the potential $U=0$.
Furthermore, we shall adsorb the factor $k_{B}T$ into the tensor $\mathbf{%
\tilde{H}}$\textbf{\ }and this redefined tensor we shall use instead of $%
h_{kl}.$ Evidently, the final result for the Laplacian, Eq.(3.34), will
remain unchanged. The question arises: if in the first example the
Riemannian curvature tensor remains flat \textsl{after} \ a coordinate
transformation (since the tensor $h_{kl}$ is the tensor describing the flat
Euclidean space), what can be said about the Riemann tensor in the present
case? To answer this question consider once again Eq.(3.35), this time with
the tensor \textbf{H}$_{mn}$ instead of h$_{ij}.$ For the sake of argument,
let us ignore for a moment the fact that each of \ generalized coordinates
is 3-dimensional. Then, we obtain, 
\begin{equation}
g_{\alpha \beta }=\frac{\partial R^{k}}{\partial q^{\alpha }}\text{\~{H}}%
_{kl}\frac{\partial R^{l}}{\partial q^{\beta }},  \tag{3.38a}
\end{equation}%
where we introduced a set of new generalized coordinates $%
\{Q\}=\{q_{1},...,q_{N}\}$ so that $R_{l}=R_{l}(\{Q\})$. We shall use Greek
indices for new coordinates and Latin for old. In the case of 3 dimensions
the above result becomes 
\begin{equation}
g_{\mathbf{\alpha \beta }}=\frac{\partial \mathbf{R}^{\mathbf{k}}}{\partial 
\mathbf{q}^{\mathbf{\alpha }}}\cdot \text{\~{H}}_{\mathbf{kl}}\cdot \frac{%
\partial \mathbf{R}^{\mathbf{l}}}{\partial \mathbf{q}^{\mathbf{\beta }}} 
\tag{3.38b}
\end{equation}%
with "$\cdot "$ being the Euclidean scalar product sign as before. The
indices \textbf{k}, \textbf{l,} $\mathbf{\alpha }$ and $\mathbf{\beta }$ now
have 3 components each. We are interested in generalized coordinates which
make the metric tensor $g_{\mathbf{\alpha \beta }}$ diagonal. By analogy
with Eq.(3.36), \ we introduce now a scalar product 
\begin{equation}
<\mathbf{R}\cdot \mathbf{R}>\equiv \mathbf{R}^{\mathbf{k}}\cdot \text{\~{H}}%
_{\mathbf{kl}}\cdot \mathbf{R}^{\mathbf{l}}  \tag{3.39}
\end{equation}%
so that instead of the vectors $\mathbf{e}_{i}$ we obtain now%
\begin{equation}
\mathbf{e}_{\mathbf{\alpha }}=\frac{\partial \mathbf{R}}{\partial \mathbf{q}%
^{\mathbf{\alpha }}}  \tag{3.40}
\end{equation}%
and, accordingly,%
\begin{equation}
g_{\mathbf{\alpha \beta }}=<\mathbf{e}_{\mathbf{\alpha }}\cdot \mathbf{e}_{%
\mathbf{\beta }}>.  \tag{3.41}
\end{equation}%
The Christoffel symbol can be defined now as 
\begin{equation}
\frac{\partial \mathbf{e}_{\mathbf{\alpha }}}{\partial \mathbf{q}^{\mathbf{%
\beta }}}=\Gamma _{\mathbf{\alpha \beta }}^{\mathbf{\gamma }}\mathbf{e}_{%
\mathbf{\gamma }}.  \tag{3.42}
\end{equation}%
To find \ the needed generalized coordinates, we impose an additional
constraint%
\begin{equation}
\frac{\partial \mathbf{e}_{\mathbf{\alpha }}}{\partial \mathbf{q}^{\mathbf{%
\beta }}}=\frac{\partial \mathbf{e}_{\mathbf{\beta }}}{\partial \mathbf{q}^{%
\mathbf{\alpha }}}  \tag{3.43}
\end{equation}%
compatible with the symmetry of the tensor \~{H}$_{\mathbf{kl}}.$ By
combining Eq.s(3.42) and (3.43) we obtain,%
\begin{equation}
\Gamma _{\mathbf{\alpha \beta }}^{\mathbf{\gamma }}\mathbf{e}_{\mathbf{%
\gamma }}=\Gamma _{\mathbf{\beta \alpha }}^{\mathbf{\gamma }}\mathbf{e}_{%
\mathbf{\gamma }}  \tag{3.44}
\end{equation}%
implying that $\Gamma _{\mathbf{\alpha \beta }}^{\mathbf{\gamma }}=\Gamma _{%
\mathbf{\beta \alpha }}^{\mathbf{\gamma }}$. That is, the imposition of the
constraint, Eq.(3.43), is equivalent to requiring that our new generalized
space is Riemannian (that is, without torsion). In such a space we would
like to consider the following combination 
\begin{equation}
R_{\mathbf{\alpha \beta }}\equiv \frac{\partial ^{2}\mathbf{e}_{\mathbf{%
\alpha }}}{\partial \mathbf{q}^{\mathbf{\alpha }}\partial \mathbf{q}^{%
\mathbf{\beta }}}-\frac{\partial ^{2}\mathbf{e}_{\mathbf{\beta }}}{\partial 
\mathbf{q}^{\mathbf{\beta }}\partial \mathbf{q}^{\mathbf{\alpha }}}. 
\tag{3.45}
\end{equation}%
Again, using Eq.(3.42) we obtain 
\begin{equation}
\frac{\partial }{\partial \mathbf{q}^{\mathbf{\alpha }}}(\Gamma _{\mathbf{%
\alpha \beta }}^{\mathbf{\gamma }}\mathbf{e}_{\mathbf{\gamma }})=\left( 
\frac{\partial }{\partial \mathbf{q}^{\mathbf{\alpha }}}\Gamma _{\mathbf{%
\alpha \beta }}^{\mathbf{\gamma }}\right) \mathbf{e}_{\mathbf{\gamma }%
}+\left( \frac{\partial }{\partial \mathbf{q}^{\mathbf{\alpha }}}\mathbf{e}_{%
\mathbf{\gamma }}\right) \Gamma _{\mathbf{\alpha \beta }}^{\mathbf{\gamma }}.
\tag{3.46}
\end{equation}%
Analogously, we obtain%
\begin{equation}
\frac{\partial }{\partial \mathbf{q}^{\mathbf{\beta }}}(\Gamma _{\mathbf{%
\beta \alpha }}^{\mathbf{\gamma }}\mathbf{e}_{\mathbf{\gamma }})=\left( 
\frac{\partial }{\partial \mathbf{q}^{\mathbf{\beta }}}\Gamma _{\mathbf{%
\beta \alpha }}^{\mathbf{\gamma }}\right) \mathbf{e}_{\mathbf{\gamma }%
}+\left( \frac{\partial }{\partial \mathbf{q}^{\mathbf{\beta }}}\mathbf{e}_{%
\mathbf{\gamma }}\right) \Gamma _{\mathbf{\beta \alpha }}^{\mathbf{\gamma }}.
\tag{3.47}
\end{equation}%
Finally, we use Eq.(3.42) in Eq.(3.46) and (3.47) in order to obtain the
following result for $R_{\mathbf{\alpha \beta }}$%
\begin{eqnarray}
R_{\mathbf{\alpha \beta }} &=&\left( \frac{\partial }{\partial \mathbf{q}^{%
\mathbf{\alpha }}}\Gamma _{\mathbf{\alpha \beta }}^{\mathbf{\gamma }}\right) 
\mathbf{e}_{\mathbf{\gamma }}-\left( \frac{\partial }{\partial \mathbf{q}^{%
\mathbf{\beta }}}\Gamma _{\mathbf{\beta \alpha }}^{\mathbf{\gamma }}\right) 
\mathbf{e}_{\mathbf{\gamma }}+\Gamma _{\mathbf{\alpha \beta }}^{\mathbf{%
\omega }}\Gamma _{\mathbf{\omega \alpha }}^{\mathbf{\gamma }}\mathbf{e}_{%
\mathbf{\gamma }}-\Gamma _{\mathbf{\omega \beta }}^{\mathbf{\gamma }}\Gamma
_{\mathbf{\beta \alpha }}^{\mathbf{\omega }}\mathbf{e}_{\mathbf{\gamma }} 
\nonumber \\
&\equiv &R_{\mathbf{\alpha \alpha \beta }}^{\gamma }\mathbf{e}_{\mathbf{%
\gamma }}  \TCItag{3.48}
\end{eqnarray}%
The second line defines the Riemann curvature tensor. In the most general
case it is given by $R_{\mathbf{\alpha \delta \beta }}^{\mathbf{\gamma }}.$
By combining Eq.s(3.40), (3.43), (3.45) and (3.48) we conclude that 
\begin{equation}
\frac{\partial ^{2}\mathbf{e}_{\mathbf{\alpha }}}{\partial \mathbf{q}^{%
\mathbf{\alpha }}\partial \mathbf{q}^{\mathbf{\beta }}}=\frac{\partial ^{2}%
\mathbf{e}_{\mathbf{\beta }}}{\partial \mathbf{q}^{\mathbf{\beta }}\partial 
\mathbf{q}^{\mathbf{\alpha }}}  \tag{3.49}
\end{equation}%
implying that the Riemann tensor is zero so that the connection $\Gamma _{%
\mathbf{\alpha \beta }}^{\mathbf{\gamma }}$ is flat. For such a case we can
replace the covariant derivative $D_{i}$ by $\nabla _{i}$ +$A_{i}$ [31]. The
vector field $A_{i}$ is defined as follows. Introduce a 1-form $A$ via $%
A=A_{i}dx^{i}$, $A_{i}=A_{i}^{\alpha }T^{\alpha },$ where in the non
-Abelian case $T^{\alpha }$ is one of infinitesimal generators of some Lie
group $G$ obeying the commutation relations $[T^{\alpha },T^{\beta
}]=if^{\alpha \beta \gamma }T^{\gamma }$ of the associated with it Lie
algebra. In addition, tr[$T^{\alpha }T^{\beta }]=\frac{1}{2}\delta ^{\alpha
\beta }$. The Chern-Simons (C-S) functional $CS(A)$ \ producing upon
minimization the needed flat connections is given by[31,32] 
\begin{eqnarray}
CS(\mathbf{A}) &=&\frac{k}{4\pi }\int\limits_{M}tr(\mathbf{A}\wedge d\mathbf{%
A}+\frac{2}{3}\mathbf{A}\wedge \mathbf{A}\wedge \mathbf{A})  \nonumber \\
&=&\frac{k}{8\pi }\int\limits_{M}\varepsilon ^{ijk}tr(A_{i}(\partial
_{j}A_{k}-\partial _{k}A_{j})+\frac{2}{3}A_{i}[A_{j},A_{k}])  \TCItag{3.50}
\end{eqnarray}%
with $k$ being some integer. Minimization of this functional produces an
equation for the flat connections. Indeed, we have%
\begin{eqnarray}
\frac{8\pi }{k}CS(\mathbf{A}+\mathbf{B}) &=&\int\limits_{M}tr(\mathbf{B}%
\wedge d\mathbf{A}+\mathbf{A}\wedge \mathbf{B}+2\mathbf{B}\wedge \mathbf{A}%
\wedge \mathbf{A})  \nonumber \\
&=&2\int\limits_{M}tr(\mathbf{B}\wedge (d\mathbf{A}+\mathbf{A}\wedge \mathbf{%
A}),  \TCItag{3.51}
\end{eqnarray}%
where we took into account that 
\[
\int\limits_{M}tr(A_{i}dx^{i}\wedge \frac{\partial B_{k}}{\partial dx^{j}}%
dx^{j}\wedge dx^{k})=\int\limits_{M}tr(B_{k}dx^{k}\wedge \frac{\partial A_{i}%
}{\partial dx^{j}}dx^{j}\wedge dx^{i}). 
\]%
From here, by requiring 
\begin{equation}
\frac{\delta }{\delta B}CS(\mathbf{A}+\mathbf{B})=0  \tag{3.52}
\end{equation}%
we obtain our final result: 
\begin{equation}
d\mathbf{A}+\mathbf{A}\wedge \mathbf{A}\equiv (\frac{\partial A_{i}}{%
\partial x_{j}}-\frac{\partial A_{j}}{\partial x_{i}}+[A_{i},A_{j}])dx^{i}%
\wedge dx^{j}\equiv F(\mathbf{A})dx^{i}\wedge dx^{j}=0.  \tag{3.53}
\end{equation}%
In the last equality we have taken into account that both in the C-S and
Yang-Mills theory $F(\mathbf{A})$ is the curvature associated with
connection $\mathbf{A}$. Vanishing of curvature produces Eq.(3.53) for the
field $\mathbf{A}$. Irrespective to the explicit form of the field $\mathbf{A%
}$, we have just demonstrated that, at least in the case when the potential $%
U$ in Eq.(3.30) is zero, this equation can be brought into diagonal form
provided that the operator $\nabla _{i}$ is replaced by $\nabla _{i}$ +$%
A_{i} $ with the field $A_{i}$ to be specified below, in the next section.

\section{An interplay between topology and randomness: connections with the
vortex model of superfluid $^{4}$He}

\subsection{General comments}

\bigskip The C-S functional, Eq.(3.50), whose minimization produces
Eq.(3.53) for the field $\mathbf{A}$ was introduced into physics by Witten
[32] and was discussed in the context of polymer physics in our previous
works summarized in Ref.[33]. Since polymer physics of fully flexible
polymer chains involves diffusion-type equations [10], the connections
between polymer and colloidal physics are apparent. For this reason, we
follow \ Ref.[32] in our exposition and use it as general source of
information.

Specifically, as explained by Witten [32], theories based on the C-S
functional are known as topological field theories. The averages in these
theories produce all kinds of topological invariants (depending upon the
generators T$^{\alpha }$ in the non Abelian case) which are observables for
such theories. In the present case the question arises: should we use the
non Abelian version of the C-S field theory or is it sufficient to use only
its Abelian version, to be defined shortly? \ Since both versions of C-S
theory were discussed in the context of polymer physics in Ref.[33], we
would like to argue that, for the purposes of this work, the Abelian version
of the C-S theory is sufficient. We shall provide the proof of this fact in
this section.

The action functional for the abelian C-S field theory is given by\footnote{%
E.g. see Eq.(4.12) of Ref.[33].} 
\begin{equation}
S_{C-S}^{A}[\mathbf{A}]=\frac{k}{8\pi }\int\limits_{M}d^{3}x\varepsilon
^{\mu \nu \rho }A_{\mu }\partial _{\nu }A_{\rho }  \tag{4.1}
\end{equation}%
With such defined functional one calculates the (topological) averages with
help of the C-S probability measure%
\begin{equation}
<\cdot \cdot \cdot >_{C-S}\equiv \hat{N}\int D[\mathbf{A}]\exp
\{iS_{C-S}^{A}[\mathbf{A}]\}\cdot \cdot \cdot .  \tag{4.2}
\end{equation}%
The random objects which are subject to averaging are the Abelian Wilson
loops $W(C)$ defined by 
\begin{equation}
W(\text{C})=\exp \{ie\oint\limits_{\text{C}}d\mathbf{r}\cdot \mathbf{A\},} 
\tag{4.3}
\end{equation}%
where C is some closed contour in 3 dimensional space (normally, without
self-intersections), and $e$ is some constant ("charge") whose exact value
is of no interest to us at this moment. The averages of products of Wilson's
loops (perhaps, forming a link $L$) 
\begin{equation}
W(L)=\prod\limits_{i=1}^{n}W(\text{C}_{i})  \tag{4.4}
\end{equation}%
are the main objects of study in such a topological field theory.
Substitution of $W(L)$ into Eq.(4.2) produces the following result [32]%
\begin{equation}
<W(L)>_{C-S}=\exp \{i\left( \frac{2\pi }{k}\right)
\sum\limits_{i,j}e_{i}e_{j}lk(i,j)\}  \tag{4.5}
\end{equation}%
with the (Gauss) linking number $lk(i,j)$ defined as 
\begin{equation}
lk(i,j)=\frac{1}{4\pi }\oint\limits_{\text{C}_{i}}\oint\limits_{\text{C}_{j}}%
\left[ d\mathbf{r}_{i}\times d\mathbf{r}_{j}\right] \cdot \dfrac{\left( 
\mathbf{r}_{i}-\mathbf{r}_{j}\right) }{\left\vert \mathbf{r}_{i}-\mathbf{r}%
_{j}\right\vert ^{3}}\equiv \frac{1}{4\pi }\int\limits_{0}^{\text{T}%
_{i}}\int\limits_{0}^{\text{T}_{j}}ds_{i}ds_{j}[\mathbf{v}(s_{i})\times 
\mathbf{v}(s_{j})]\cdot \dfrac{\left( \mathbf{r}(s_{i})-\mathbf{r}%
(s_{j})\right) }{\left\vert \mathbf{r}(s_{i})-\mathbf{r}(s_{j})\right\vert
^{3}}.  \tag{4.6}
\end{equation}%
Here T$_{i}$ and T$_{j}$ are respectively the contour lengths of contours C$%
_{i}$ and C$_{j}$ and $\mathbf{v}(s)=\frac{d}{ds}\mathbf{r}(s).$ With the
Gauss linking number defined in such a way, in view\ of Eq.(4.5), it should
be clear that we must to consider as well self-linking numbers $lk(i,i).$ \
Such a technicality requires us to think about the so called framing
operation discussed in some detail in both Ref.s [32] and [33]. We shall
ignore this technicality until Section 6 for reasons which will become
apparent.

\subsection{An interplay between the topology and randomness in hydrodynamics%
}

Following Tanaka and Ferrari, Refs [34, 35], we rewrite the Gauss linking
number in a more physically suggestive form. For this purpose, we introduce
the "magnetic" field $\mathbf{B}(\mathbf{r})$ via 
\begin{equation}
\mathbf{B}(\mathbf{r})=\frac{1}{4\pi }\oint\limits_{\text{C}_{j}}d\mathbf{r}%
_{j}\times \dfrac{\left( \mathbf{r}-\mathbf{r}(s_{j})\right) }{\left\vert 
\mathbf{r}-\mathbf{r}(s_{j})\right\vert ^{3}}  \tag{4.7}
\end{equation}%
allowing us to rewrite the linking number $lk(i,j)$ as 
\begin{equation}
lk(i,j)=\oint\limits_{\text{Ci}}d\mathbf{r}_{i}\cdot \mathbf{B}(\mathbf{r}%
_{i}).  \tag{4.8}
\end{equation}%
Eq.(4.7) for the field $\mathbf{B}(\mathbf{r})$ is known in magnetostatics
as Biot-Savart law, e.g. see Ref.[36], Eq.(30.14). Because of this, we
recognize that%
\begin{equation}
\mathbf{\nabla }\cdot \mathbf{B=}0  \tag{4.9}
\end{equation}%
and 
\begin{equation}
\mathbf{\nabla }\times \mathbf{B=j,}  \tag{4.10}
\end{equation}%
where $\mathbf{j}(\mathbf{r})=\oint\limits_{C}ds\mathbf{v(}s\mathbf{)}\delta 
\mathbf{(r-r(}s\mathbf{)).}$To connect these results with hydrodynamics, we
introduce the vector potential $\mathbf{A}$\textbf{\ }in such a way that%
\textbf{\ }%
\begin{equation}
\mathbf{\nabla }\times \mathbf{A=B.}  \tag{4.11}
\end{equation}%
Using this result in Eq.(4.10) we obtain as well%
\begin{equation}
\nabla ^{2}\mathbf{A}=-\mathbf{j}  \tag{4.12}
\end{equation}%
in view of the fact that $\mathbf{\nabla }\cdot \mathbf{A=}0.$ In
hydrodynamics we can represent the local fluid velocity following Ref.[37],
page 86, as 
\begin{equation}
\mathbf{v}=\mathbf{\nabla }\times \mathbf{A}  \tag{4.13}
\end{equation}%
and define the vorticity $\mathbf{\vec{\omega}}$ as 
\begin{equation}
\mathbf{\vec{\omega}=\nabla }\times \mathbf{v.}  \tag{4.14}
\end{equation}%
By analagy with Eq.s(4.10) and (4.12) we now obtain 
\begin{equation}
\nabla ^{2}\mathbf{A}=-\mathbf{\vec{\omega}.}  \tag{4.15}
\end{equation}%
Hence, to apply previous results to hydrodynamics the following
identification should be made: 
\begin{equation}
\mathbf{\vec{\omega}\rightleftarrows j.;v\rightleftarrows B.}  \tag{4.16}
\end{equation}%
The kinetic energy $\mathcal{E}$ of a fluid in a volume $M$ is given by 
\begin{equation}
\mathcal{E=}\frac{\rho }{2}\int\limits_{M}\mathbf{v}^{2}d^{3}\mathbf{r}. 
\tag{4.17}
\end{equation}%
We would like now to explain how this energy is related to the above defined
linking numbers. For this purpose, we introduce the following auxiliary
functional:%
\begin{equation}
\mathcal{F}[\mathbf{A}]_{i}=\oint\limits_{\text{C}_{i}}d\mathbf{r}_{i}\cdot 
\mathbf{A}(\mathbf{r}_{i}).  \tag{4.18}
\end{equation}%
Use of the theorem by Stokes produces%
\begin{equation}
\mathcal{F}[\mathbf{A}]_{i}=\oint\limits_{\text{C}_{i}}d\mathbf{r}_{i}\cdot 
\mathbf{A}(\mathbf{r}_{i})=\iint\limits_{S_{i}}d\mathbf{S}_{i}\cdot \left( 
\mathbf{\nabla }\times \mathbf{A}\right) =\iint\limits_{S_{i}}d\mathbf{S}%
_{i}\cdot \mathbf{v=}\iint\limits_{S_{i}}d\mathbf{S}_{i}\oint\limits_{\text{C%
}_{j}}ds_{j}\mathbf{v(}s_{j}\mathbf{)\delta (r}_{i}\mathbf{-r(}s_{j}\mathbf{%
)).}  \tag{4.19}
\end{equation}%
At the same time, for the linking number, Eq.(4.8), an analogous procedure
leads to the following chain of equalities%
\begin{equation}
lk(i,j)=\oint\limits_{\text{Ci}}d\mathbf{r}_{i}\cdot \mathbf{B}(\mathbf{r}%
_{j})=\oint\limits_{\text{Ci}}d\mathbf{r}_{i}\cdot \mathbf{v}(\mathbf{r}%
_{j})=\iint\limits_{S_{i}}d\mathbf{S}_{i}\cdot \left( \mathbf{\nabla }\times 
\mathbf{v}\right) =\iint\limits_{S_{i}}d\mathbf{S}_{i}\cdot \mathbf{\vec{%
\omega}.}  \tag{4.20a}
\end{equation}%
Since the same vector potential was used in both Eq.s(4.11) and (4.13) we
notice that Eq.s (4.12) and (4.15) also imply that%
\begin{equation}
\mathbf{\vec{\omega}=}e\mathbf{j,}  \tag{4.20b}
\end{equation}%
where $e$ is some constant to be determined. Because of this, we obtain 
\begin{equation}
e\mathcal{F}[\mathbf{A}]_{i}=lk(i,j)=e\iint\limits_{S_{i}}d\mathbf{S}%
_{i}\oint\limits_{\text{C}_{j}}ds_{j}\mathbf{v(}s_{j}\mathbf{)}\delta 
\mathbf{(r}_{i}\mathbf{-r(}s_{j}\mathbf{)).}  \tag{4.21}
\end{equation}%
Since the obtained equivalence is of central importance for the entire work,
we would like to discuss a few additional details of immediate relevance. In
particular, from Eq.(4.20b), which we shall call from now on, the London
equation (e.g. see the Subsection 4.4 below), it should be clear that the as
yet unknown constant $e$ must have dimensionality of inverse length L$^{-1}. 
$ This fact should be taken into account when we consider the following
dimensionless\footnote{%
In view of the fact that the dimensionality of $e$ is fixed we have
introduced a factor $f$ which makes the functional $\mathcal{W[}\mathbf{A}]$
dimensionless. This factor will be determined shortly below.} functional

\begin{equation}
\mathcal{W[}\mathbf{A}]=\frac{\rho }{2k_{B}T}\int\limits_{M}d^{3}\mathbf{%
r(\nabla \times A)}^{2}+i\frac{e}{f}\sum\limits_{j}\oint\limits_{\text{Cj}}d%
\mathbf{r}_{j}\cdot \mathbf{A}(\mathbf{r}_{j})  \tag{4.22}
\end{equation}%
and the path integral associated with it, i.e. 
\begin{equation}
\check{N}\int D[\mathbf{A}]\delta (\mathbf{\nabla }\cdot \mathbf{A})\exp \{-%
\mathcal{W[}\mathbf{A}]\}\equiv <W(L)>_{T}  \tag{4.23a}
\end{equation}%
to be compared with Eq.s(4.2) and (4.5). Here the thermal average $<\cdot
\cdot \cdot >_{T}$ is defined by 
\begin{equation}
<\cdot \cdot \cdot >_{T}=\check{N}\int D[\mathbf{A}]\delta (\mathbf{\nabla }%
\cdot \mathbf{A})\exp \{-\frac{\rho }{2k_{B}T}\int\limits_{M}d^{3}\mathbf{%
r(\nabla \times A)}^{2}\}\cdot \cdot \cdot .  \tag{4.23b}
\end{equation}%
Calculation of this Gaussian path integral is complicated by the presence of
a delta constraint (Coulomb gauge) in the path integral measure.
Fortunately, this path integral can be found in the paper by Brereton and
Shah [38]. Without providing the details, these authors presented the
following final result in notations adapted to this work:%
\begin{equation}
<W(L)>_{T}=\exp \{-\frac{1}{2\rho }\left( \frac{e}{f}\right)
^{2}\sum\nolimits_{i,j=1}^{^{\prime
}}\int\limits_{0}^{t}\int\limits_{0}^{t}ds_{i}ds_{j}\mathbf{\dot{r}}%
(s_{i})\cdot \mathbf{\tilde{H}}[\mathbf{r}(s_{i})-\mathbf{r}(s_{j})]\cdot 
\mathbf{\dot{r}}(s_{j})\}.  \tag{4.24}
\end{equation}%
The Oseen tensor $\mathbf{\tilde{H}}(\mathbf{R})$ in this expression was
previously defined in Eq.(3.27) and the prime on the summation sign means
that the diagonal part of this tensor should be excluded. Even though
calculations \ leading to this result are not given in Ref.[38], they can be
easily understood field-theoretically. For this purpose, we have to
regularize the delta function constraint in the path integral measure in
Eq.(4.23) very much the same way as Ferrell, Ref.[21], did it in the case of
hydrodynamics as we discussed in Section 2. Specifically, we write%
\begin{eqnarray}
&&\check{N}\int D[\sigma (\mathbf{r})]\exp (-\frac{1}{2\xi }\int d^{3}%
\mathbf{r}\sigma ^{2}(\mathbf{r}))\int D[\mathbf{A}]\delta (\mathbf{\nabla }%
\cdot \mathbf{A-\sigma (r)})\exp \{-\frac{\rho }{2k_{B}T}\int\limits_{M}d^{3}%
\mathbf{r(\nabla \times A)}^{2}\}\cdot \cdot \cdot  \nonumber \\
&=&\check{N}\int D[\mathbf{A}]\exp \{-\frac{\rho }{2k_{B}T}%
\int\limits_{M}d^{3}\mathbf{r(\nabla \times A)}^{2}-\frac{1}{2\xi }\int d^{3}%
\mathbf{r(\nabla }\cdot \mathbf{A)}^{2}\}\cdot \cdot \cdot  \nonumber \\
&=&\check{N}\int D[\mathbf{A}]\exp \{-\frac{\rho }{2k_{B}T}%
\int\limits_{M}d^{3}\mathbf{r}A_{\mu }[-\delta _{\mu \nu }\mathbf{\nabla }%
^{2}-(1-\frac{1}{\tilde{\xi}})\partial _{\mu }\partial _{\nu }]A_{\nu
}\}\cdot \cdot \cdot .  \TCItag{4.25}
\end{eqnarray}%
with some adjustable regularizing parameter $\tilde{\xi}.$Also, for the
quadratic form (in $\mathbf{A}$\textbf{)} in the exponent of the last
expression we obtain 
\begin{equation}
\int\limits_{M}d^{3}\mathbf{r}A_{\mu }[-\delta _{\mu \nu }\mathbf{\nabla }%
^{2}-(1-\frac{1}{\tilde{\xi}})\partial _{\mu }\partial _{\nu }]A_{\nu }=\int
d^{3}\mathbf{k}A_{\mu }(\mathbf{k})\mathbf{[\delta }_{\mu \nu }\mathbf{k}%
^{2}-(1-\frac{1}{\tilde{\xi}})k_{\mu }k_{\nu }]A_{\nu }  \tag{4.26}
\end{equation}%
The inverse of the matrix $\mathbf{[\delta }_{\mu \nu }\mathbf{k}^{2}-(1-%
\frac{1}{\tilde{\xi}})k_{\mu }k_{\nu }]$ is easy to find following Ramond
[39]. Indeed, we write 
\begin{equation}
\mathbf{[\delta }_{\mu \nu }\mathbf{k}^{2}-(1-\frac{1}{\tilde{\xi}})k_{\mu
}k_{\nu }][X(\mathbf{k})\delta _{\nu \rho }+Y(\mathbf{k})k_{\nu }k_{\rho
}]=\delta _{\mu \rho }.  \tag{4.27}
\end{equation}%
From here the unknown functions $X(\mathbf{k})$ and $Y(\mathbf{k})$ can be
determined so the inverse matrix is given explicitly by 
\begin{equation}
\lbrack X(\mathbf{k})\delta _{\nu \rho }+Y(\mathbf{k})k_{\nu }k_{\rho }]=%
\frac{1}{\mathbf{k}^{2}}[\delta _{\nu \rho }-(1-\tilde{\xi})\frac{k_{\nu
}k_{\rho }}{\mathbf{k}^{2}}].  \tag{4.28}
\end{equation}%
In the limit $\tilde{\xi}\rightarrow 0$ we recover the Oseen tensor (up to a
constant $1/\eta )$ in the $k$-space representation in accord with
Ref.[10].These results explain why in the average, Eq.(4.24), there are no
diagonal terms. Now we are ready to determine the constant $e$ introduced in
Eq.(4.22).

\subsection{Reparametrization invariance and vortex-vortex interactions}

\bigskip The important result for $<W(L)>_{T}$ contains random velocities $%
\mathbf{\dot{r}}(s)$ and thus, seemingly, additional averaging is required.
\ The task now lies in finding the explicit form of this averaging. To do
so, several steps are required. To begin, we notice that in the absence of
hydrodynamic interactions Eq.(3.30) acquires the following form%
\begin{equation}
\frac{\partial \Psi }{\partial t}=D_{0}\sum\limits_{n}\frac{\partial ^{2}}{%
\partial \mathbf{R}_{n}^{2}}\Psi  \tag{4.29}
\end{equation}%
with diffusion coefficient $D_{0}$ defined in Eq.(1.1). \ If Eq.(4.29) we
treat $\Psi $ as Green's function (e.g. see Appendix A for details), then it
can be formally represented in the path integral form as 
\begin{equation}
\Psi (t;\mathbf{R}_{1,...,}\mathbf{R}_{\text{N}})=\int D[\{\mathbf{R}(\tau
)\}]\exp (-\frac{1}{4D_{0}}\int\limits_{0}^{t}\sum\nolimits_{i-1}^{\text{N}}%
\left[ \mathbf{\dot{r}}(\tau _{i})\right] ^{2}d\tau _{i}).  \tag{4.30}
\end{equation}%
In this expression we have suppressed the explicit $\mathbf{R}$-dependence
of the path integral to avoid excessive notation. Hydrodynamic interactions
can now be accounted for as follows 
\begin{eqnarray}
\mathcal{F} &=&\check{N}\int D[\mathbf{A}]\exp \{-\frac{\rho }{2k_{B}T}%
\int\limits_{M}d^{3}\mathbf{r}A_{\mu }[-\delta _{\mu \nu }\mathbf{\nabla }%
^{2}-(1-\frac{1}{\tilde{\xi}})\partial _{\mu }\partial _{\nu }]A_{\nu }\} 
\nonumber \\
&&\times \int D[\{\mathbf{r}(\tau )\}]\exp (-\frac{1}{4D_{0}}%
\int\limits_{0}^{t}(\sum\nolimits_{j-1}^{\text{N}}\left[ \mathbf{\dot{r}}%
(\tau _{j})\right] ^{2}d\tau _{j})\exp \{i\frac{e}{f}\int\limits_{0}^{t}\sum%
\nolimits_{j-1}^{\text{N}}\left[ \mathbf{\dot{r}}(\tau _{j})\right] \cdot 
\mathbf{A[r(}\tau _{j}\mathbf{)]}d\tau _{j}\mathbf{\}}  \nonumber \\
&\equiv &<\prod\limits_{j=1}^{\text{N}}\int D[\{\mathbf{r}(\tau _{j})\}]\exp
(-\frac{1}{4D_{0}}\int\limits_{0}^{t}\left[ \mathbf{\dot{r}}(\tau _{j})%
\right] ^{2}d\tau _{j})\exp \{i\frac{e}{f}\int\limits_{0}^{t}\mathbf{\dot{r}}%
(\tau _{j})\cdot \mathbf{A[r(}\tau _{j}\mathbf{)]}d\tau _{j}\mathbf{\}>}_{T}.
\TCItag{4.31}
\end{eqnarray}%
Perturbative calculation of path integrals of the type 
\begin{equation}
\text{I}[\mathbf{A};t]=\int D[\{\mathbf{r}(\tau _{j})\}]\exp (-\frac{1}{%
4D_{0}}\int\limits_{0}^{t}\left[ \mathbf{\dot{r}}(\tau _{j})\right]
^{2}d\tau _{j}\exp \{i\frac{e}{f}\int\limits_{0}^{t}\left[ \mathbf{\dot{r}}%
(\tau _{j})\right] \cdot \mathbf{A[r(}\tau _{j}\mathbf{)]}d\tau _{j}\mathbf{%
\}}  \tag{4.32}
\end{equation}%
was considered by Feynman long \ ago, Ref.[40]. From this paper it follows
that the most obvious way to do such a calculation is to write the usual Schr%
\"{o}dinger-like equation%
\begin{equation}
\left( \frac{\partial }{\partial t}-D_{0}(\mathbf{P}-ie\mathbf{A}%
)^{2}\right) G(t,\mathbf{r};t^{\prime }\mathbf{r}^{\prime })=0\text{, }%
\mathbf{r}\neq \mathbf{r}^{\prime }\text{ }  \tag{4.33}
\end{equation}%
and to take into account that $(\mathbf{P}-ie\mathbf{A})^{2}=\mathbf{P}%
^{2}-ie\mathbf{A}\cdot \mathbf{P-}ie\mathbf{P}\cdot \mathbf{A}-e^{2}\mathbf{A%
}^{2}\simeq \mathbf{P}^{2}-ie\mathbf{A}\cdot \mathbf{P+}O\mathbf{(A}^{2})$ \
(since $\mathbf{P}\cdot \mathbf{A=}0).$ This result is useful to compare
with Eq.(3.32) in order to recognize that the field $\mathbf{A}$ is indeed a
connection.

To use these results, we would like to rewrite Eq.(3.30) in the alternative
form which (for $U=0$) is given by%
\begin{equation}
\frac{\partial \Psi }{\partial t}=D_{0}\sum\limits_{n}\text{ }\frac{\partial
^{2}}{\partial \mathbf{R}_{n}^{2}}\Psi \text{\ +\ }k_{B}T\sum%
\nolimits_{m,n,i,j}^{\prime }\text{\ }\mathbf{\tilde{H}}_{ij}(\mathbf{R}_{n}-%
\mathbf{R}_{m})\text{\ }\frac{\partial }{\partial R_{in}}\text{\ }\frac{%
\partial }{\partial R_{jm}}\text{\ }\Psi \text{.}  \tag{4.34}
\end{equation}%
In arriving at this equation we took into account Eq.(3.14). Consider such
an equation for $n=2$. In this case, we rewrite Eq.(4.34) in the style of
quantum mechanics, i.e.%
\begin{equation}
\left( \frac{\partial }{\partial t}-H_{1}-H_{2}-V_{12}\right) \Psi =0 
\tag{4.35}
\end{equation}%
in which, as in quantum mechanics, we shall treat $V_{12}$ as a
perturbation. The best way of dealing with such problems is to use the
method of Green's functions. For our reader's convenience we present some
facts about this method in Appendix A. Eq.(A.10) of this Appendix provides
an equation for the effective potential $\mathcal{V}.$ A similar type of
equation was obtained in the book by Doi and Edwards, Ref.[10], in Section
5.7.3., who used methods of the effective medium theory. Using this theory
they were able to prove the existence of screening for the case of polymer
solutions. \ We shall reach an analogous conclusion about screening in
colloidal suspension using different arguments to be discussed in the next
subsection. In the meantime, we would like to provide arguments justifying
our previously made approximation: $(\mathbf{P}-ie\mathbf{A})^{2}\simeq 
\mathbf{P}^{2}-ie\mathbf{A}\cdot \mathbf{P+}O\mathbf{(A}^{2}).$ Using
results of Appendix A, we introduce the one-particle Green's function $G_{0}$
as a solution to the equation 
\begin{equation}
\left( \frac{\partial }{\partial t}-D_{0}\frac{\partial ^{2}}{\partial 
\mathbf{R}^{2}}\right) G_{0}(\mathbf{R},t;\mathbf{R},t^{\prime })=\delta (%
\mathbf{R}-\mathbf{R}^{\prime })\delta (t-t^{\prime }).  \tag{4.36}
\end{equation}%
Having in mind the determination of previously introduced factor $f$ (in
Eq.(4.22))$,$ it is convenient to rescale the variables in this equation to
convert it into dimensionless form. Evidently, the most convenient choice is 
$t=\tau /(D_{0}/R_{0}^{2})$ and $\mathbf{R}=R_{0}\mathbf{\tilde{R}}$ with $%
R_{0}$ is the hard sphere radius introduced in Eq.(1.1) and $\tau $ and $%
\mathbf{\tilde{R}}$ being dimensionless time and space coordinates. Below,
we shall avoid \ the use of tildas for $\mathbf{\tilde{R}}$ and shall still
write $t$ instead of $\tau \mathbf{.}$The original symbols can be restored
whenever they are required. Having this in mind, next we consider the
two-particle Green's function $G_{0}.$In the absence of interactions, it is
just the product of two Green's functions of the type given by Eq.(4.36). As
a result, the Dyson-type equation for the full Green's function for
Eq.(4.34) ($n=2$ case) is given by 
\begin{eqnarray}
G(\mathbf{R}_{1},\mathbf{R}_{2},t;\mathbf{R}_{1}^{^{\prime \prime }},\mathbf{%
R}_{2}^{^{\prime \prime }},t^{\prime \prime }) &=&G_{0}(\mathbf{R}_{1},t;%
\mathbf{R}_{1}^{\prime },t^{\prime })G_{0}(\mathbf{R}_{2},t;\mathbf{R}%
_{2}^{\prime },t^{\prime })  \nonumber \\
&&+\int G_{0}(\mathbf{R}_{1},t;\mathbf{R}_{1}^{\prime },t_{1}^{\prime
})G_{0}(\mathbf{R}_{2},t;\mathbf{R}_{2}^{\prime },t_{1}^{\prime })V(\mathbf{R%
}_{1}^{\prime },\mathbf{R}_{2}^{\prime })G(\mathbf{R}_{1}^{\prime },\mathbf{R%
}_{2}^{\prime },t_{1}^{\prime };\mathbf{R}_{1}^{^{\prime \prime }},\mathbf{R}%
_{2}^{^{\prime \prime }},t^{\prime \prime })d\mathbf{R}_{1}^{\prime }d%
\mathbf{R}_{2}^{\prime }dt_{1}^{\prime }  \nonumber \\
&&  \TCItag{4.37}
\end{eqnarray}%
in which the potential $V(\mathbf{R}_{1}^{\prime },\mathbf{R}_{2}^{\prime
})= $\ $k_{B}T\mathbf{\tilde{H}}_{ij}(\mathbf{R}_{1}-\mathbf{R}_{2})$\ $%
\frac{\partial }{\partial R_{1i}}$\ $\frac{\partial }{\partial R_{2i}}.$ As
before, summation over repeated indices is assumed. \ Using results of
Appendix A and Eq.(4.37) it is possible to write now the equation for the
effective potential. In view of the results to be discussed in the next
subsection, this is actually unnecessary. Hence, we proceed with other tasks
at this point. Specifically, taking into account Eq.(3.27) in which the
explicit form of the Oseen tensor is given, we conclude that the nondiagonal
part of this tensor can be discarded in the Dyson Eq.(4.37). This is so
because of the following obvious identity: $\left[ (\mathbf{r}_{1}-\mathbf{r}%
_{2})\cdot \mathbf{r}_{1}\right] \left[ (\mathbf{r}_{1}-\mathbf{r}_{2})\cdot 
\mathbf{r}_{2}\right] +\left[ (\mathbf{r}_{2}-\mathbf{r}_{1})\cdot \mathbf{r}%
_{1}\right] \left[ (\mathbf{r}_{2}-\mathbf{r}_{1})\cdot \mathbf{r}_{2}\right]
=0$ associated with the scalar products of unit vectors in Eq.(3.27).
Evidently, it is always possible to select a coordinate system centered,
say, at $\mathbf{r}_{1}$Alternatively, this result can be easily proven in $%
k $-space taking into account the incompressibility constraint. Furthermore,
these observations cause us to write the potential $V(\mathbf{R}_{1},\mathbf{%
R}_{2})$ in the following \ dimensionful form\footnote{%
Using dimensional analysis performed for Eq.(4.36) the result, Eq.(4.38),
can be easily rewritten also in dimensionless form}%
\begin{equation}
V(\mathbf{R}_{1}^{\prime },\mathbf{R}_{2}^{\prime })=\ \frac{k_{B}T}{4\pi
\eta }\frac{1}{\left\vert \mathbf{R}_{1}-\mathbf{R}_{2}\right\vert }\ \frac{%
\partial }{\partial \mathbf{R}_{1}}\cdot \ \frac{\partial }{\partial \mathbf{%
R}_{2}}.  \tag{4.38a}
\end{equation}%
Using dimensional analysis of Eq.(4.36), this result can be easily rewritten
also in dimensionless form. Explicitly, it is given by 
\begin{equation}
V(\mathbf{R}_{1}^{\prime },\mathbf{R}_{2}^{\prime })=\ \frac{k_{B}T}{4\pi
\eta R_{0}^{2}D_{0}}\frac{1}{\left\vert \mathbf{R}_{1}-\mathbf{R}%
_{2}\right\vert }\ \frac{\partial }{\partial \mathbf{R}_{1}}\cdot \ \frac{%
\partial }{\partial \mathbf{R}_{2}}  \tag{4.38b}
\end{equation}%
in which the scalar product can be of any sign. This fact is of importance
because of the following.

Using Eq.(4.31) and proceeding with calculations of the path integral
following Feynman's prescriptions [40], we obtain exactly the same equation
as that given by Eq.(4.37). This observation allows us to determine the
constants $e$ and $f$ explicitly. In view of the results just obtained, the
constant $e$ can be determined only with accuracy up to a sign. \ Taking
this into account, the value of $e$ is determined as $e=\pm \dfrac{1}{R_{0}}%
\sqrt{\dfrac{D_{0}\rho }{4\pi \eta }},$ while the constant $f$ is given by $%
D_{0}^{{}}$ in view of the fact that the field $\mathbf{A}$ in Eq.(4.22) has
dimensionality $L^{2}/t$ , \ i.e. that of the diffusion coefficient, while
the dimensionality of $e$ is fixed by the Eq.(4.20b), so that the
combination $eds\mathbf{\dot{r}}(s)$ is dimensionless.

Using these results and Eq.(4.38), we can rewrite $<W(L)>_{T}$ defined by
Eq.(4.24) in the following \ manifestly dimensionless physically suggestive
form%
\begin{equation}
<W(L)>_{T}=\exp (-\frac{k_{B}T}{D_{0}8\pi \eta }\sum\nolimits_{i,j=1}^{^{%
\prime }}\frac{s_{i}}{R_{0}}\frac{s_{j}}{R_{0}}\oint \oint \frac{\left\vert d%
\mathbf{r}(\tau _{i})\cdot d\mathbf{r}(\tau _{j})\right\vert }{\left\vert 
\mathbf{r}(\tau _{i})-\mathbf{r}(\tau _{j})\right\vert })  \tag{4.39}
\end{equation}%
where we have introduced the dimensionless Ising spin-like variables $s_{i}$
playing the role of charges accounting for the sign of the product $\frac{%
\partial }{\partial \mathbf{R}_{1}}\cdot \ \frac{\partial }{\partial \mathbf{%
R}_{2}}.$ Since the whole system must be "electrically neutral", at this
point it is possible to develop the Debye-H\"{u}ckel-type theory of
hydrodynamic screening by analogy with that developed for Coulombic systems,
e.g. see Ref.[41]. \ Nevertheless, below we choose another, more elegant
pathway to arrive at the same conclusions.

Before doing so, we notice that there is an important difference between the
double integral, Eq.(4.39), and $\frac{1}{4D_{0}}\int\limits_{0}^{t}\left[ 
\mathbf{\dot{r}}(\tau _{j})\right] ^{2}d\tau _{j}$ present in the exponent
in Eq.(4.31). While the double integral, Eq.(4.39), is manifestly
reparametrization invariant, the diffusion integral is not. This means that
we can always reparametrize time in this diffusion integral so that the
coefficient $\left( 4D\right) ^{-1}$ can be made equal to any preassigned
nonnegative integer. This was effectively done already when we introduced
the dimensionless variables in Eq.(4.36). Such inequivalence between these
two types of integrals can be eliminated if we replace this diffusion -type
integral by that which is manifestly reparametrization- invariant. In such a
case the total action is given by 
\begin{equation}
S=m_{0}\sum\limits_{i}\oint d\tau _{i}\sqrt{\mathbf{r}^{2}(\tau _{i})}+\frac{%
k_{B}T}{D_{0}8\pi \eta }\sum\nolimits_{i,j=1}^{^{\prime }}s_{i}s_{j}\oint
\oint \frac{\left\vert d\mathbf{r}(\tau _{i})\cdot d\mathbf{r}(\tau
_{j})\right\vert }{\left\vert \mathbf{r}(\tau _{i})-\mathbf{r}(\tau
_{j})\right\vert }).  \tag{4.40}
\end{equation}%
It should be noted that use of a symbol $\oint $ instead of $\int $ in
Eq.(4.40) is a delicate matter. In [33] we demonstrated that in the limit of
long times (that is in the limit $\omega \rightarrow 0$ used in this work)
all random walks are asymptotically closed (that is, the Brownian trajectory
in this limit becomes very much the same as known for ring polymers)%
\footnote{%
Additional mathematical results on this property are discussed in Section
6.2.}. Since the result, Eq.(4.40), is manifestly reparametrization
invariant, such a replacement is permissible. Additional explanations are
given in Appendix B which we recommend to read only after reading of Section
5.

The constant $m_{0}$ in Eq.(4.40) will be determined in the next section.
The form of the action given by Eq.(4.40) is almost identical to that for
the action for the superfluid liquid $^{4}$He as discussed in the book by
Kleinert [42], page 300. From the same book, it also follows that the
Ginzburg-Landau theory of superconductivity also can be recast in the same
form. We said "almost identical to" meaning that in these two theories (of
superfluidity and superconductivity) the self-interaction of vortices is
also allowed so that if the above expression \ would represent the dual
(vortex) description of colloidal suspension dynamics (e.g. see Appendix B),
then the prime in the double summation above can be removed since the
vortices are allowed to intersect with themselves.

In the direct case, when the focus of attention is on particles, removal of
the prime in the double summation in Eq.(4.40) would imply that the Oseen
tensor is defined for particles hydrodynamically interacting with
themselves. This assumption is not present in the original Doi-Edwards
formulation, Ref.[10]. As we noticed already in Eq.(3.29), the diagonal part
of the Oseen tensor is associated with self-diffusion. The question
therefore arises: can this "almost equivalence" be converted into full
equivalence? \ The main feature of superconductors is the existence of the
Meissner (for hard spheres)\ and dual (for vortices) Meissner effect. In the
present case such an effect is equivalent to the existence of hydrodynamic
screening. Hence, to prove such an equivalence requires us to prove the
existence of hydrodynamic screening for suspensions. Evidently, we cannot
immediately use Eq.(4.40) for such a proof. Therefore, in the next
subsection we use London-style arguments to arrive at the desired conclusion.

\subsection{ London-style theory of hydrodynamic screening}

\bigskip We begin our proof by taking into account the non-slip boundary
condition, Eq.(2.27): 
\begin{equation}
\mathbf{v}(\mathbf{r},t)=\frac{d\mathbf{r}}{dt}=\mathbf{v}(t).  \tag{2.27}
\end{equation}%
Within the approximations made, we also have to impose the incompressibility
requirement 
\begin{equation}
\mathbf{\nabla }\cdot \mathbf{v}(\mathbf{r},t)=0.  \tag{3.14}
\end{equation}%
Because of this requirement, the current $\mathbf{j}=\rho \mathbf{v}$
becomes $\mathbf{j}=n_{0}\mathbf{v}$ with the density $n_{0}$ being a
constant. Since $\mathbf{j}$\textbf{\ }is a vector, we can always represent
it as 
\begin{equation}
\mathbf{j}=\alpha \mathbf{\nabla }\psi  \tag{4.41}
\end{equation}%
with suitably chosen scalar $\psi $ and some proportionality constant $%
\alpha .$ To choose such a scalar we take into account that in the present
case 
\begin{equation}
\mathbf{\nabla }\cdot \mathbf{j=}0  \tag{4.42}
\end{equation}%
implying 
\begin{equation}
\nabla ^{2}\psi =0.  \tag{4.43}
\end{equation}%
The vector $\mathbf{j}$\textbf{\ } given by Eq\textbf{.(}4.41\textbf{) }is
not uniquely defined. It\textbf{\ }will still obey the Eq.(4.42) if we write%
\begin{equation}
\mathbf{j}=\alpha \mathbf{\nabla }\psi \pm g\mathbf{A}  \tag{4.44}
\end{equation}%
for a vector \textbf{A} such that $\mathbf{\nabla }\cdot \mathbf{A=}0.$
Evidently$,$ a vector obeying Eq.(4.13) by construction possess this
property. The choice of the sign "+" or "-" in the above equation can be
determined based on the following arguments. Since $\mathbf{j}=n_{0}\mathbf{v%
}$ and since $n_{0}$ is constant, we can replace Eq.(4.44) by 
\begin{equation}
\mathbf{v}=\alpha \mathbf{\nabla }\psi \pm g\mathbf{A}  \tag{4.45}
\end{equation}%
by suitably redefinig constants $\alpha $ and $g$. Next, we assume that $%
\mathbf{v}$ is a random variable so that on average $<\mathbf{v}>=0$ thus
implying 
\begin{equation}
<\alpha \mathbf{\nabla }\psi >\pm g<\mathbf{A>=}0.  \tag{4.46}
\end{equation}%
This equation causes us to choose the sign "-". After this, we can write for
the correlator 
\begin{equation}
<\mathbf{v}\cdot \mathbf{v}>=\alpha ^{2}<\mathbf{\nabla }\psi \cdot \mathbf{%
\nabla }\psi >+g^{2}<\mathbf{A}\cdot \mathbf{A}>=2g^{2}<\mathbf{A}\cdot 
\mathbf{A}>.  \tag{4.47}
\end{equation}%
In view of our choice of \textbf{A}, the $<\mathbf{A}\cdot \mathbf{A}>$
correlator coincides with that given in the exponent of Eq.(4.24). Now we
take into account Eq.(4.20b) where, of course, we replace $\mathbf{j}$ by $%
\mathbf{v}$ so that using the dictionary, Eq.(4.16), we arrive at 
\begin{equation}
\mathbf{\vec{\omega}}=e\mathbf{v}\text{ \ \ \ \ \ \ \ \ \ \ (London equation)%
}  \tag{4.48}
\end{equation}%
supplemented with 
\begin{equation}
\mathbf{\vec{\omega}=\nabla }\times \mathbf{v}\text{ \ \ \ \ \ (Maxwell
equation).}  \tag{4.14}
\end{equation}%
Such an identification becomes apparent because of the following arguments.
Let us use Eq.(4.45) in Eq.(4.48) in order to obtain%
\begin{equation}
\mathbf{\vec{\omega}}=e\mathbf{(}\alpha \mathbf{\nabla }\psi -e\mathbf{A).} 
\tag{4.49}
\end{equation}%
In this equation we replaced the constant $g$ by $e$. Furthermore, since
Eq.(4.49) formally looks like the Fick's first law, we can as well rewrite
this result as 
\begin{equation}
\mathbf{\vec{\omega}}=e\mathbf{(}\frac{D_{0}}{2\pi }\mathbf{\nabla }\psi -e%
\mathbf{A).}  \tag{4.50}
\end{equation}%
By applying to both sides of this equation the curl operator and taking into
account Eq.(4.13), we obtain%
\begin{equation}
\mathbf{\nabla }\times \mathbf{\vec{\omega}=-}e^{2}\mathbf{v.}  \tag{4.51}
\end{equation}%
Taking into account the Maxwell's Eq.(4.14) and using it in Eq.(4.51) we
obtain as well 
\begin{equation}
\nabla ^{2}\mathbf{v}=e^{2}\mathbf{v.}  \tag{4.52a}
\end{equation}%
Equivalently, we obtain, 
\begin{equation}
\nabla ^{2}\mathbf{A=}e^{2}\mathbf{A.}  \tag{4.51b}
\end{equation}%
Using Eq.s(4.47), (4.52a) and following the same steps as in the Appendix A
of our previous work, Ref.[23], we obtain 
\begin{equation}
\left\langle \mathbf{v(r)\cdot v(0)}\right\rangle =\frac{const}{r}\exp (-%
\frac{\mathbf{r}}{\xi _{H}}),  \tag{4.53}
\end{equation}%
where $\xi _{H}=e^{-1}=\left( \dfrac{1}{R_{0}}\sqrt{\dfrac{D_{0}\rho }{4\pi
\eta }}\right) ^{-1}$ and the constant in Eq.(4.53) can be obtained from
comparison between this equation and Eq.(2.37). The analogous result is also
obtained for the $<\mathbf{A}\cdot \mathbf{A}>$ correlator.

In accord with Eq.(2.44) we obtain the\ result of central importance $\xi
_{H}\rightarrow \infty $ when $\rho \rightarrow 0,$ implying absence of
screening in the infinite dilution limit. Our derivation explains the
rationale behind the identification of Eq.s (4.14) and (4.48) with the
Maxwell and London equations in the theory of superconductivity, Ref.[25],
pages 174, 175. Evidently, such an identification becomes possible only in
view of the topological nature of the London equation, Eq.(4.48), coming
from identification of Eq.(4.19) with (4.20a).

\section{Exotic superconductivity of colloidal suspensions}

\subsection{General Remarks}

In the previous section we developed a theory of hydrodynamic screening
following ideas of the London brothers, Ref.[24]. As is well known, their
seminal work found its most notable application in the theory of ordinary
superconductors [25]. At the same time, Eq.(4.40) \ was originally used in
the theory of superfluid $^{4}$He. In the book by Kleinert [42] it is shown
that Eq.(4.40) can be rewritten in such a way that it will acquire the same
form as used in the phenomenological Ginzburg-Landau (G-L) theory of
superconductivity [25]. We would like to arrive at the same conclusions
differently. In doing so we also would like to determine both the physical
and mathematical meaning of the parameter $m_{0}$ which was left
undetermined in Eq.(4.40). We shall develop our arguments mainly following
the original G-L pathway. \ 

It should be said, though, that in the present case the connections with
superconductivity are only in the structure of equations to be derived. The
underlying physics is similar but not identical to that for superconductors.
Indeed, in the case of superconductors one typically is talking about the
supeconducting-to-normal transition controlled by temperature. Also, one is
talking about the temperature-dependent "critical" magnetic field (the upper
and the lower critical magnetic fields in the case of superconductors of the
second kind) which destroys the superconductivity. In the present case of
colloidal suspensions there is no explicit temperature dependence: \textsl{%
the same} phenomena can take place at \textsl{any} temperature at which the
solvent is not frozen. If we account for short range forces, then, of
course, one can study a situation in which such a colloidal suspension is
undergoing a temperature-controlled phase transition. Such a case requires a
separate treatment and will not be considered in this work. \ In the present
case the phase diagram can be qualitatively described as follows. The
infinite dilution limit corresponds to the \textsl{normal} state. The regime
of finite concentrations corresponds to a \textsl{mixed} state, typical for
superconductors of the \textsl{second kind}, and the dramatic jump in
viscosity discussed in the Introduction and in Section 2 corresponds to the
transition to the "fully superconducting" state. Such a difference from the
usual superconductors brings some new physics into play which may be useful,
in other disciplines, e.g. in the high energy physics or turbulence, etc.%
\footnote{%
E.g. see Section 6.}

\subsection{ G-L style derivation of equations of superconductivity for
colloidal suspensions}

\bigskip We begin with the one of Maxwell's equations in its conventional
form, e.g. as given in Ref.[25], page 181,%
\begin{equation}
\mathbf{\nabla }\times \mathbf{B}=\frac{4\pi }{c}\mathbf{j.}  \tag{5.1}
\end{equation}%
In the G-L theory we have for the current $\mathbf{j}$\textbf{\ }the
following result\textbf{:}%
\begin{equation}
\mathbf{j}=-\frac{i\tilde{e}\hbar }{2m}(\varphi ^{\ast }\nabla \varphi
-\varphi \nabla \varphi ^{\ast })-\frac{2\tilde{e}^{2}}{mc}\left\vert
\varphi \right\vert ^{2}\mathbf{A.}  \tag{5.2}
\end{equation}%
Both equations can be obtained by minimization of the following (\textsl{%
truncated}) G-L functional\footnote{%
This truncation is known in literature as the "\textsl{London limit}".}%
\begin{equation}
\mathcal{F[}\mathbf{A},\varphi ]=\int d^{3}r\{\frac{\left( \mathbf{\nabla }%
\times \mathbf{A}\right) ^{2}}{8\pi }+\frac{\hbar ^{2}}{4m}\left\vert (%
\mathbf{\nabla }-\frac{2i\tilde{e}}{\hbar c}\mathbf{A)}\varphi \right\vert
^{2}\}  \tag{5.3}
\end{equation}%
with respect to $\mathbf{A}$. Substitution of the ansatz $\varphi =\dfrac{%
\sqrt{n_{s}}}{2}\exp (i\psi )$ into Eq.(5.2) leads to the current%
\begin{equation}
\mathbf{j}=\frac{\tilde{e}\hbar }{2m}n_{s}(\mathbf{\nabla }\psi -\frac{2%
\tilde{e}}{\hbar c}\mathbf{A})  \tag{5.4a}
\end{equation}%
to be compared with our Eq.(4.50). Evidently, this result is equivalent to
the postulated London equation for superconductors%
\begin{equation}
\mathbf{\nabla }\times \mathbf{j=-}\dfrac{en_{s}}{mc}\mathbf{B.}  \tag{5.4b}
\end{equation}%
\ At the same time, a comparison of Eq.(5.4a) with Eq.(4.50) leads to the
following \ chain of identifications: $\dfrac{\tilde{e}\hbar }{2m}%
n_{s}\rightleftarrows eD_{0}$ and $\dfrac{\tilde{e}^{2}}{mc}%
n_{s}\rightleftarrows e^{2}.$ Consequently, we obtain as well: $\dfrac{\hbar
^{2}}{2m}\rightleftarrows D_{0},\dfrac{\tilde{e}n_{s}}{\hbar }\rightarrow e$%
; $\dfrac{\tilde{e}^{2}n_{s}}{m}\rightarrow e^{2},c\rightleftarrows 4\pi
\rightarrow \dfrac{1}{2\pi },\dfrac{2\tilde{e}}{\hbar }\rightleftarrows 
\dfrac{e}{D_{0}}\rightleftarrows \dfrac{2e}{n_{s}}.$ Using these
identifications, we can rewrite the functional $\mathcal{F[}\mathbf{A}%
,\varphi ]$ as follows%
\begin{equation}
\mathcal{F[}\mathbf{A},\varphi ]=\frac{\rho }{2}\int d^{3}r\{\left( \mathbf{%
\nabla }\times \mathbf{A}\right) ^{2}+D_{0}\left\vert (\mathbf{\nabla }-i%
\frac{2\pi e}{D_{0}}\mathbf{A)}\varphi \right\vert ^{2}\}.  \tag{5.5}
\end{equation}%
In the traditional setting, the \ superconducting density $n_{s}$ is
determined from the full G-L functional 
\begin{equation}
\mathcal{F[}\mathbf{A},\varphi ]=\int d^{3}r\{\frac{\left( \mathbf{\nabla }%
\times \mathbf{A}\right) ^{2}}{8\pi }+\frac{\hbar ^{2}}{4m}\left\vert (%
\mathbf{\nabla }-\frac{2i\tilde{e}}{\hbar c}\mathbf{A)}\varphi \right\vert
^{2}+a\left\vert \varphi \right\vert ^{2}+\frac{b}{2}[\left\vert \varphi
\right\vert ^{2}]^{2}\},  \tag{5.6}
\end{equation}%
e.g. by minimization with respect to $\varphi ^{\ast }$. In fact, to obtain $%
n_{s}$ it is formally sufficient to treat only the case when $\mathbf{A}=0$.
Indeed, under this condition we obtain%
\begin{equation}
a\varphi _{c}+b\left\vert \varphi _{c}\right\vert ^{2}\varphi _{c}=0, 
\tag{5.7}
\end{equation}%
which has a nontrivial solution only for $a<0.$ In this case we get $n_{s}=%
\frac{\left\vert a\right\vert }{b},$ provided that $b>0,$ as usual. If we
use this result back in Eq.(5.6), that is we use $\varphi _{c}=\dfrac{\sqrt{%
n_{s}}}{2}\exp (i\psi )$ in Eq.(5.6) then, the polynomial (in $\varphi )$
part of the functional becomes a constant. This constant is divergent when
the volume of the system goes to infinity. To prevent this from happening
another constant term is typically added to the functional $\mathcal{F[}%
\mathbf{A},\varphi ]$ so that it acquires the following canonical form%
\begin{equation}
\mathcal{F[}\mathbf{A},\varphi ]=\int d^{3}\mathbf{r}\{\frac{\left( \mathbf{%
\nabla }\times \mathbf{A}\right) ^{2}}{8\pi }+\frac{\hbar ^{2}}{4m}%
\left\vert (\mathbf{\nabla }-\frac{2i\tilde{e}}{\hbar c}\mathbf{A)}\varphi
\right\vert ^{2}+\frac{b}{2}(\left\vert \varphi \right\vert
^{2}-n_{s})^{2}\}.  \tag{5.8}
\end{equation}%
Then, when $\varphi _{c}=\dfrac{\sqrt{n_{s}}}{2}\exp (i\psi ),$ the
polynomial (in $\varphi )$ part of the functional vanishes and, accordingly,
in this limit we require 
\begin{equation}
\int d^{3}r\left\vert (\mathbf{\nabla }-\frac{2i\tilde{e}}{\hbar c}\mathbf{A)%
}\varphi _{c}\right\vert ^{2}\rightarrow 0  \tag{5.9}
\end{equation}%
as well. This leads us to the equation 
\begin{equation}
\frac{\hbar c}{i2\tilde{e}}\frac{1}{\varphi _{c}}\mathbf{\nabla }\varphi
_{c}=\mathbf{A}  \tag{5.10a}
\end{equation}%
or to%
\begin{equation}
\frac{\hbar c}{2\tilde{e}}\mathbf{\nabla }\psi =\mathbf{A.}  \tag{5.10b}
\end{equation}%
This equation coincides(on average) with the previously obtained Eq.(4.46)
(with redefinitions described above) as required and will be treated further
in Section 5.4.

It should be noted though that originally, in London's theory, Ref.[24], the 
$n_{s}$ was left as an adjustable parameter and, hence, microscopically
undefined. This is important in our case since the phenomenon of
supercoductivity can be looked upon (as in thermodynamics) without any
reference to spontaneous symmetry breaking, Higgs effect, etc. \ At the
level of G-L theory, the London equations are reproduced with help of the
truncated G-L functional. Hence, in principle, in the present case use of
the truncated functional, Eq.(5.5), is also sufficient. At the macroscopic
mean field level the presence of \ polynomial terms in the full-G-L
functional, Eq.s(5.6) and (5.8) seems somewhat artificial. They do not
reveal their microscopic origin and are introduced just to fit the data. \
We would like to use some known facts from the path integral treatments of
superconductivity/superfluidity in order to reveal their physical meaning.
Such information is also useful for development of \ the hydrodynamic theory
of colloidal suspensions.

\subsection{Path integrals associated with the G-L functional}

\bigskip In view of Eq.(4.40), we begin our discussion with the simplest
case of the path integral for a single "relativistic" scalar particle.

Following Polyakov, Ref.[43], the Euclidean version of propagator for such a
(Klein-Gordon) particle is given by 
\begin{equation}
G(x,x^{\prime })=\int \left( \frac{\mathfrak{D}\mathbf{x}(\tau )}{\mathfrak{D%
}f(\tau )}\right) \exp (-m_{0}\int\limits_{0}^{1}d\tau \sqrt{\mathbf{\dot{x}}%
^{2}(\tau )}),  \tag{5.11a}
\end{equation}%
where in the most general case 
\begin{equation}
\mathbf{\dot{x}}^{2}(\tau )=g_{\mu \nu }(\mathbf{x})\frac{dx^{\mu }}{d\tau }%
\frac{dx^{\nu }}{d\tau }.  \tag{5.11b}
\end{equation}%
This propagator is of interest in string theory since it represents a
reduced form of the propagator for the bosonic string. As in the case of a
string, the action of this path integral is manifesttly
reparametrization-invariant, i.e. invariant under changes of the type $%
\mathbf{x}$($\tau )\rightarrow \mathbf{x}(f(\tau ))$ (with $f(\tau )$ being
some nonnegative monotonically increasing function). The path integral
measure is designed to absorb this redundancy. The full account of this
absorption is cumbersome. Because of this, instead of copying Polyakov's
treatment of such a path integral, we shall adopt a simplified treatment
allowing us to recover Polyakov's final results. We begin with an obvious
well-known identity%
\begin{equation}
\left( \frac{1}{4\pi t}\right) ^{\frac{d}{2}}\exp (-\frac{1}{4}\frac{\mathbf{%
x}^{2}}{t})=\int\limits_{\mathbf{x}(0)=0}^{\mathbf{x}(t)=\mathbf{x}}%
\mathfrak{D}[\mathbf{x}(\tau )]\exp \{-\frac{1}{4}\int\limits_{0}^{t}d\tau
\left( \frac{d\mathbf{x}}{d\tau }\right) ^{2}\}.  \tag{5.12}
\end{equation}%
This identity is used below as follows. Consider the propagator for the
Klein-Gordon (K-G) field given by 
\begin{equation}
G(\mathbf{x})=\int \frac{d^{d}\mathbf{k}}{\left( 2\pi \right) ^{d}}\frac{%
\exp (i\mathbf{k\cdot x})}{\mathbf{k}^{2}+m^{2}}.  \tag{5.13}
\end{equation}%
By employing the identity%
\begin{equation}
\frac{1}{a}=\int\limits_{0}^{\infty }dx\exp (-ax)  \tag{5.14}
\end{equation}%
Eq.(5.13) can be rewritten as follows%
\begin{eqnarray}
G(\mathbf{x}) &=&\int\limits_{0}^{\infty }dt\exp (-tm^{2})\int \frac{d^{d}%
\mathbf{k}}{\left( 2\pi \right) ^{d}}\exp (i\mathbf{k\cdot x-}t\mathbf{k}%
^{2})  \nonumber \\
&=&\frac{1}{\mathcal{E}}\int\limits_{0}^{\infty }dt\exp (-\mathcal{E}%
tm^{2})\int \frac{d^{d}\mathbf{k}}{\left( 2\pi \right) ^{d}}\exp (i\mathbf{%
k\cdot x-}\mathcal{E}t\mathbf{k}^{2})  \nonumber \\
&=&\frac{1}{\mathcal{E}}\int\limits_{0}^{\infty }dt\exp (-t\mathcal{E}\text{%
\textit{m}}^{2})\left( \frac{1}{4\pi \mathcal{E}t}\right) ^{\frac{d}{2}}\exp
(-\frac{1}{4}\frac{\mathbf{x}^{2}}{\mathcal{E}t})  \nonumber \\
&=&\frac{1}{\mathcal{E}}\int\limits_{0}^{\infty }dt\exp (-t\mathfrak{m}%
^{2})\int\limits_{\mathbf{x}(0)=0}^{\mathbf{x}(t)=\mathbf{x}}\mathfrak{D}[%
\mathbf{x}(\tau )]\exp \{-\frac{1}{4\mathcal{E}}\int\limits_{0}^{t}d\tau
\left( \frac{d\mathbf{x}}{d\tau }\right) ^{2}\}  \TCItag{5.15}
\end{eqnarray}%
where we used the identity, Eq.(5.12), to obtain the last line and
introduced an arbitrary nonnegative parameter $\mathcal{E}$ for comparison
with results by Polyakov. Specifically, using page 163 of the book by
Polyakov (and comparing our 3rd line above with the 3rd line of his
Eq.(9.63)) we can make the following identifications: $\mathcal{%
E\rightleftarrows \varepsilon }$, $m^{2}\rightleftarrows \mu .$ Since,
according to Polyakov, $\mu =\varepsilon ^{-1}(m_{0}-\dfrac{c}{\sqrt{%
\varepsilon }})$ with $c$ being some constant, we obtain: $m_{0}=\mathcal{E}%
m^{2}+\dfrac{c}{\sqrt{\varepsilon }}$. That is, the physical mass $m^{2}$
entering the K-G equation is obtained as the limit of the expression ($%
\varepsilon \rightarrow 0)$ 
\begin{equation}
m=\lim_{m_{0}\rightarrow m_{cr}}\varepsilon ^{-\frac{1}{2}}(m_{0}-m_{cr})^{%
\frac{1}{2}}.  \tag{5.16}
\end{equation}%
Clearly, such an expression is nonnegative by construction. From the last
line of Eq.(5.15) it follows that the propagator for the K-G field is just
the direct Laplace transform of the nonrelativistic "diffusion" propagator,
Eq.(5.12), with the Laplace variable $\mathit{m}$ playing a role of a mass
for such a field. In the Euclidean version of the K-G propagator this mass
cannot be negative since in such a case the identity Eq.(5.14) cannot be
used so that the connection between the nonrelativistic and the K-G
propagators is lost. However, Eq.(5.2) seemingly is for the \textsl{quantum}
current while the propagator in Eq.(5.12) is describing Brownian motion, not
quantum diffusion. To fix the problem we have to replace time $t$ in
Eq.(5.12) by $it$ and, accordingly, to make changes in Eq.(5.15). This then
converts the Laplace transform into the Fourier transform, provided that the
nonrelativistic propagator describes the \textsl{retarded} Green's function.
To use the full strength of the apparatus of quantum field theory one needs
to use the causal Green's functions. This is required by the relativistic
covariance treating space and time coordinates on the same footing. Once all
of these requirements are met, it becomes possible to treat the case of a
negative mass.

It should be emphasized at this point that the London-style derivation given
in the previous section formally \textsl{does not} require such quantum
mechanical analogy. Because of this, the following problem emerges: is it
possible to reproduce the functional integral $\mathcal{F}$ defined by
Eq.(4.31) using the truncated G-L functional for superconductivity in the
exponent of the associated path integral? \ We would like to provide an
affirmative answer to this question now.

We begin with the partition function $Z$ for the two-component scalar
K-G-type field%
\begin{equation}
\ln Z=-\ln [\det (-\nabla ^{2}+m^{2})]  \tag{5.17}
\end{equation}%
Since 
\begin{equation}
\ln [\det (-\nabla ^{2}+m^{2})]=tr\left[ \ln (-\nabla ^{2}+m^{2})\right] 
\tag{5.18}
\end{equation}%
and 
\begin{equation}
tr\left[ \ln (-\nabla ^{2}+m^{2})\right] =\int \frac{d^{d}\mathbf{k}}{\left(
2\pi \right) ^{d}}\ln (\mathbf{k}^{2}+m^{2}),  \tag{5.19}
\end{equation}%
we can use the results of our previous work, Ref.[44], for evaluation of the
last expression. \ Thus, we obtain, 
\begin{eqnarray}
tr\left[ \ln \frac{(-\nabla ^{2}+m^{2})}{(-\nabla ^{2})}\right]
&=&\int\limits_{0}^{m^{2}}dy\frac{d}{dy}\int \frac{d^{d}\mathbf{k}}{\left(
2\pi \right) ^{d}}\ln (\mathbf{k}^{2}+y)  \nonumber \\
&=&\int\limits_{0}^{m^{2}}dy\int \frac{d^{d}\mathbf{k}}{\left( 2\pi \right)
^{d}}\frac{1}{\mathbf{k}^{2}+y}=\int\limits_{0}^{m^{2}}dyG(\mathbf{0};y) 
\nonumber \\
&=&\int\limits_{0}^{\infty }dt\int\limits_{0}^{m^{2}}dy\exp
(-ty)\int\limits_{\mathbf{x}(0)=\mathbf{0}}^{\mathbf{x}(t)=\mathbf{0}}%
\mathfrak{D}[\mathbf{x}(\tau )]\exp \{-\frac{1}{4}\int\limits_{0}^{t}d\tau
\left( \frac{d\mathbf{x}}{d\tau }\right) ^{2}\}  \nonumber \\
&=&\int\limits_{0}^{\infty }\frac{dt}{t}(1-\exp (-m^{2}t))\int\limits_{%
\mathbf{x}(0)=\mathbf{0}}^{\mathbf{x}(t)=\mathbf{0}}\mathfrak{D}[\mathbf{x}%
(\tau )]\exp \{-\frac{1}{4}\int\limits_{0}^{t}d\tau \left( \frac{d\mathbf{x}%
}{d\tau }\right) ^{2}\}.  \TCItag{5.20}
\end{eqnarray}%
Following the usual practice, we shall write $\oint $ instead of $%
\int\limits_{\mathbf{x}(0)=\mathbf{0}}^{\mathbf{x}(t)=\mathbf{0}}$ in the
path integral and consider a formal (that is diverging!) expression for the
free energy $\mathcal{F}_{0}$ 
\begin{equation}
\exp \left( -\mathcal{F}_{0}\right) =\ln Z_{0}=-\ln [\det (-\nabla
^{2}+m^{2})]=\int\limits_{0}^{\infty }\frac{dt}{t}\exp (-m^{2}t)\oint 
\mathfrak{D}[\mathbf{x}(\tau )]\exp \{-\frac{1}{4}\int\limits_{0}^{t}d\tau
\left( \frac{d\mathbf{x}}{d\tau }\right) ^{2}\}  \tag{5.21}
\end{equation}%
by keeping in mind that this result makes sense mathematically only when the
same expression with $m^{2}=0$ is subtracted from it as required by
Eq.(5.20). Inclusion of the electromagnetic field into this scheme can be
readily accomplished now. For this purpose we replace the $\mathbf{\nabla }$
operator by its covariant derivative:$\mathbf{\nabla \rightarrow D\equiv
\nabla -}ie\mathbf{A}$\textbf{\ }\ (we put $D_{0}=1$ in view of developments
presented in Eq.(5.15)) \ Using $\mathbf{D}$ instead of $\mathbf{\nabla }$
in Eq.(5.20) we have to evaluate the following path integral\footnote{%
For $m^{2}=0$ this is just part of the truncated G-L functional.} 
\begin{equation}
\lbrack \det^{{}}(-\mathbf{D}^{2}+m^{2})]^{-1}=\int D[\bar{\varphi},\varphi
]\exp (-\frac{1}{2}\int d^{3}r\{\bar{\varphi}(-\mathbf{D}^{2}+m^{2})\varphi
\}).  \tag{5.22}
\end{equation}%
For $\mathbf{A}=0$ we did this already while for $\mathbf{A}\neq 0$ we can
treat terms containing $\mathbf{A}$ as perturbation. We can do the same for
the path integral in Eq.(4.32). This is easy to understand if we realize
that 
\begin{equation}
\int\limits_{0}^{\infty }dt\exp (-m^{2}t)\text{I}[\mathbf{A};t]\mid _{%
\mathbf{A}=0}=-\frac{d}{dm^{2}}\int\limits_{0}^{\infty }\frac{dt}{t}\exp
(-m^{2}t)\oint \mathfrak{D}[\mathbf{x}(\tau )]\exp \{-\frac{1}{4}%
\int\limits_{0}^{t}d\tau \left( \frac{d\mathbf{x}}{d\tau }\right) ^{2}\} 
\tag{5.23}
\end{equation}%
Therefore, the final answer reads as follows%
\begin{eqnarray}
\exp \left( -\mathcal{F}\right) &=&\ln Z=-\ln [\det (-\mathbf{D}%
^{2}+m^{2})]=\int\limits_{0}^{\infty }\frac{dt}{t}\exp (-m^{2}t)\oint 
\mathfrak{D}[\mathbf{x}(\tau )]\exp \{-\int\limits_{0}^{t}d\tau \lbrack 
\frac{1}{4}\left( \frac{d\mathbf{x}}{d\tau }\right) ^{2}+i\frac{e}{f}\mathbf{%
\dot{x}}\cdot \mathbf{A}[\mathbf{x}(\tau )]]\}  \nonumber \\
&=&\int\limits_{0}^{\infty }\frac{dt}{t}\exp (-m^{2}t)\oint \mathfrak{D}[%
\mathbf{x}(\tau )]\exp \{-\frac{1}{4}\int\limits_{0}^{t}d\tau \left( \frac{d%
\mathbf{x}}{d\tau }\right) ^{2}\}\exp \{-i\frac{e}{f}\oint d\mathbf{r}\cdot 
\mathbf{A}[\mathbf{x}(\tau )]\}.  \TCItag{5.24}
\end{eqnarray}%
This result demonstrates that applying the operator $\int\limits_{0}^{\infty
}\frac{dt}{t}\exp (-m^{2}t)$ to I$[\mathbf{A};t]$ \ defined in Eq.(4.32)
makes it equivalent to the "matter" part of truncated G-L functional for
superconductivity as needed. This raises a question about comparison of the
full G-L functional with the "diffusion" path integrals of \ Section 4.
Evidently, this can be done only if in the original diffusion Eq.(3.30) we
do not discard the potential $U$. If we do not discard the potential and if,
instead, we ignore the hydrodynamic interactions completely, we would end up
with the following path integral for interacting Brownian particles in the
canonical ensemble%
\begin{equation}
\Xi =\int \prod\limits_{l=1}^{N}\mathcal{D}[\mathbf{x}(\tau _{l})]\exp \{-%
\frac{1}{4D_{0}}\sum\limits_{i=1}^{N}\int\limits_{0}^{t}d\tau _{i}\left( 
\frac{d\mathbf{x}}{d\tau _{i}}\right)
^{2}-\sum\limits_{i<j}^{N}\int\limits_{0}^{t}d\tau
_{i}\int\limits_{0}^{t}d\tau _{j}V[\mathbf{x}(\tau _{i})-\mathbf{x}(\tau
_{j})]\}.  \tag{5.25}
\end{equation}%
It is essential that this expression \textsl{does not} contain
self-interactions typical for problems involving polymer chains with
excluded volume-type interactions. The situation here resembles that
encountered when, following Doi and Edwards, Ref.[10], we redefined the
Oseen tensor in Eq.s (3.29\ ) and (3.30) so that \ it acquired the diagonal
part. In the present case we must require the diagonal part to be zero at
the end of calculations. These results, correct for colloidal particles, may
become incorrect in the present case for the following reason. From looking
either at Eq.(5.24) or Eq.(4.39), we recognize that in these cases we are
dealing with assemblies of loops (vortices) which are in one-to one
correspondence with diffusing particles. While this topic is studied in
detail in the next subsection and Appendix B, here we notice that if
Eq.(5.25) is written for such loops, then the excluded volume requirement
becomes essential, even for a single loop. Indeed, the existence of such a
loop is possible only if the field $\mathbf{A}$\textbf{\ }associated with
these loops\textbf{\ }is uniquely defined. This is possible only if the loop
contour does not have self-interactions. This is the origin of the excluded
volume constraint requirement. With this restriction imposed, we introduce a
density%
\begin{equation}
\rho (\mathbf{r})=\sum\limits_{i=1}^{N}\int\limits_{0}^{t}d\tau _{i}\delta (%
\mathbf{x}-\mathbf{x}(\tau _{i}))  \tag{5.26}
\end{equation}%
so that the binary potential in Eq.(5.25) can be written as 
\begin{equation}
\frac{1}{2}\sum\limits_{i,j}^{N}\int\limits_{0}^{t}d\tau
_{i}\int\limits_{0}^{t}d\tau _{j}V[\mathbf{x}(\tau _{i})-\mathbf{x}(\tau
_{j})]=\frac{1}{2}\int d\mathbf{r}\int d\mathbf{r}^{\prime }\rho (\mathbf{r}%
)V[\mathbf{r}-\mathbf{r}^{\prime }]\rho (\mathbf{r}^{\prime }).  \tag{5.27}
\end{equation}%
Then, using the Hubbard-Stratonovich (H-S) identity we obtain, 
\begin{equation}
\exp (-\frac{1}{2}\int d\mathbf{r}\int d\mathbf{r}^{\prime }\rho (\mathbf{r}%
)V[\mathbf{r}-\mathbf{r}^{\prime }]\rho (\mathbf{r}^{\prime }))=\mathfrak{N}%
\int D[\psi (\mathbf{r})]\exp (-\frac{1}{2}\int d\mathbf{r}\int d\mathbf{r}%
^{\prime }\psi (\mathbf{r})V^{-1}[\mathbf{r}-\mathbf{r}^{\prime }]\psi (%
\mathbf{r}^{\prime }))\exp (i\int dr\psi (\mathbf{r})\rho (\mathbf{r})) 
\tag{5.28}
\end{equation}%
with $\mathfrak{N}$ being a normalization constant (bringing the above
identity to the statement $1=1$ for $\rho =0).$ Use of this result in
Eq.(5.25) in which self exclusion is allowed converts this partition
function into the following form (written for the loop ensemble)%
\begin{equation}
\Xi =\mathfrak{N}\int D[\psi (\mathbf{r})]\exp (-\frac{1}{2}\int d\mathbf{r}%
\int d\mathbf{r}^{\prime }\psi (\mathbf{r})V^{-1}[\mathbf{r}-\mathbf{r}%
^{\prime }]\psi (\mathbf{r}^{\prime }))\prod\limits_{i=1}^{N}G_{i}(0;t\mid
\psi )  \tag{5.29}
\end{equation}%
where 
\begin{equation}
G_{i}(0;t\mid \psi )=\oint \mathfrak{D}[\mathbf{x}(\tau _{i})]\exp
\{-\int\limits_{0}^{t}d\tau _{i}[\frac{1}{4}\left( \frac{d\mathbf{x}}{d\tau
_{i}}\right) ^{2}+ie\psi \lbrack \mathbf{x}(\tau _{i})]]\}.  \tag{5.30}
\end{equation}%
In the case of polymers, typically, one uses the delta function-type
potential for description of the interactions. This observation is helpful
in the present case as well because of the following. Consider the G-L
functional, Eq.(5.6), and use the H-S identity for the interaction term%
\begin{equation}
\exp (-\frac{b}{2}[\left\vert \varphi \right\vert ^{2}]^{2})=\mathfrak{N}%
\int D[\psi (\mathbf{r})]\exp (-\frac{1}{2b}\int d\mathbf{r}\int d\mathbf{r}%
^{\prime }\psi (\mathbf{r})\psi (\mathbf{r}^{\prime }))\exp (i\int dr\psi (%
\mathbf{r}))\left\vert \varphi \right\vert ^{2}).  \tag{5.31}
\end{equation}%
This allows us to replace the determinant, Eq.(5.22), by the following (more
general) determinant%
\begin{equation}
\lbrack \det^{{}}(-\mathbf{D}^{2}+m^{2}+i\psi )]^{-1}=\int D[\bar{\varphi}%
,\varphi ]\exp (-\frac{1}{2}\int d^{3}r\{\bar{\varphi}(-\mathbf{D}%
^{2}+m^{2}+i\psi \lbrack \mathbf{r}])\varphi \})  \tag{5.32}
\end{equation}%
which, in view of Eq.(5.24), can be equivalently rewritten as 
\begin{eqnarray}
\exp \left( -\mathcal{F}\right) &=&\ln Z=-\ln [\det (-\mathbf{D}%
^{2}+m^{2}+i\psi )]=\mathfrak{N}\int D[\psi (\mathbf{r})]\exp (-\frac{1}{2b}%
\int d\mathbf{r}\int d\mathbf{r}^{\prime }\psi (\mathbf{r})\psi (\mathbf{r}%
^{\prime }))  \nonumber \\
&&\times \int\limits_{0}^{\infty }\frac{dt}{t}\exp (-m^{2}t)\oint \mathfrak{D%
}[\mathbf{x}(\tau )]\exp \{-\frac{1}{4}\int\limits_{0}^{t}d\tau \left( \frac{%
d\mathbf{x}}{d\tau }\right) ^{2}\}\exp \{-ie\oint d\mathbf{x}\cdot \mathbf{A}%
[\mathbf{x}(\tau )]\}+i\oint d\tau \psi \lbrack \mathbf{x}(\tau )]\} 
\nonumber \\
&=&\int\limits_{0}^{\infty }\frac{dt}{t}\exp (-m^{2}t)\oint \mathfrak{D}[%
\mathbf{x}(\tau )]\exp \{-\frac{1}{4}\int\limits_{0}^{t}d\tau \left( \frac{d%
\mathbf{x}}{d\tau }\right) ^{2}\}  \nonumber \\
&&\times \exp \{-i\frac{e}{f}\oint d\mathbf{x}\cdot \mathbf{A}[\mathbf{x}%
(\tau )]-\frac{b}{2}\oint d\tau \oint d\tau ^{\prime }\delta (\mathbf{x}%
(\tau )-\mathbf{x}(\tau ^{\prime })\}.  \TCItag{5.33}
\end{eqnarray}%
Alternatively, this result can be rewritten as a grand canonical ensemble of
selfavoiding loops 
\begin{eqnarray}
Z[\mathbf{A;}b]-1 &=&\sum\limits_{n=1}^{\infty }\frac{1}{n!}%
\prod\limits_{l=1}^{n}[\int\limits_{0}^{\infty }\frac{dt_{l}}{t_{l}}\exp
(-m^{2}t_{l})\oint \mathfrak{D}[\mathbf{x}(\tau _{l})]\exp \{-\frac{1}{4}%
\sum\limits_{l=1}^{n}\int\limits_{0}^{t}d\tau _{l}\left( \frac{d\mathbf{x}}{%
d\tau _{l}}\right) ^{2}\}  \nonumber \\
&&\times \exp \{-i\frac{e}{f}\oint d\mathbf{x}\cdot \mathbf{A}[\mathbf{x}%
(\tau _{l})]\}-\frac{b}{2}\sum\limits_{l,m=1}^{n}\oint d\tau _{l}\oint d\tau
_{m}^{\prime }\delta (\mathbf{x}(\tau _{l})-\mathbf{x}(\tau _{m}^{\prime
})\}.  \TCItag{5.34}
\end{eqnarray}%
This result is useful to compare with Eq.(4.31). From such a comparison it
is evident that Eq.(5.34) is compatible with that obtained previously. It
accounts for the effects of non hydrodynamic-type interactions which can be
incorporated, in principle, in the diffusion Eq.(3.30) in which the
potential $U$ must be specified. Clearly, \ the use of path integrals makes
such a task much simpler. However, even though the above derivation is
intuitively appealing, strictly speaking, it cannot be used for a number of
reasons. Unlike the G-L functional, Eq.(5.8), which is convenient for
studying of topological and nonperturbative effects, Eq.(5.34) makes sense
only in perturbative calculations. This means that phenomena such as
screening (caused by the Higgs effect) cannot be captured with such a
formalism alone. These observations explain why screening effects were found
in solutions of polymers but not in colloidal suspensions, Ref. [10,12].
Furthermore, Eq.(5.34) contains a mixture of reparametrization-invariant and
non invariant terms. This is questionable mathematically. It would be more
logical to have the entire action reparametrization-invariant. We study
these issues in some detail in the next subsection.

\subsection{Reparametrization-invariance and its consequences. London-style
analysis}

Since path integrals mathematically can seldom be defined rigorously, we
would like in this subsection to extend the analysis of Sections 4.2.-4.4
avoiding the use of path integrals. We start with a discussion of the
result, Eq.(5.10b), which is the superconducting analog of Eq.(4.46) for
suspensions. Since $\mathbf{B}=\mathbf{\nabla }\times \mathbf{A,}$ (or $%
\mathbf{v}$ $=\mathbf{\nabla }\times \mathbf{A}$ in the case of
suspensions)\ we conclude that Eq.(5.10b) causes the first term in the G-L
functional Eq.(5.6) (or (5.5)) to vanish in the bulk. Nevertheless, it is
perfectly permissible to write 
\begin{equation}
\frac{\hbar c}{2\tilde{e}}\oint d\mathbf{r}\cdot \mathbf{\nabla }\psi =\oint
d\mathbf{r}\cdot \mathbf{A.}  \tag{5.35}
\end{equation}%
and to use Stokes' theorem 
\begin{equation}
\oint\limits_{C}d\mathbf{r}\cdot \mathbf{A=}\iint d\mathbf{S\cdot (\nabla }%
\times \mathbf{A)=}\iint d\mathbf{S\cdot B=}n\frac{hc}{2\tilde{e}} 
\tag{5.36}
\end{equation}%
with $n=0,\pm 1,\pm 2,...$ \ An analogous result for suspensions reads as
follows%
\begin{equation}
\oint\limits_{C}d\mathbf{r}\cdot \mathbf{A=}\iint d\mathbf{S\cdot (\nabla }%
\times \mathbf{A)=}\iint d\mathbf{S\cdot v=}n\frac{D_{0}}{e}.  \tag{5.37a}
\end{equation}%
In view of Eq.s(4.19) and (4.20), this result leads also to 
\begin{equation}
\frac{1}{D_{0}}\iint d\mathbf{S\cdot \tilde{\omega}=}n=\frac{e}{D_{0}}%
\oint\limits_{C}d\mathbf{r}\cdot \mathbf{A}  \tag{5.37b}
\end{equation}%
which is the same as Eq.(4.21).

These results can be interpreted in a number of ways. For the sake of
argument, we would like to explore the more established case of
superconductivity first. Following Lund and Regge, Ref.[45], we suppose that
the vector potential $\mathbf{A}$ can be presented as follows%
\begin{equation}
\mathbf{A(r)=}\frac{k}{4\pi }\oint\limits_{C}\frac{1}{\left\vert \mathbf{r}-%
\mathbf{r}(\sigma )\right\vert }\left( \frac{\partial \mathbf{r}}{\partial
\sigma }\right) d\sigma \equiv \frac{k}{4\pi }\oint\limits_{C}\frac{1}{%
\left\vert \mathbf{r}-\mathbf{r}(\sigma )\right\vert }\mathbf{v}(\sigma
)d\sigma  \tag{5.38}
\end{equation}%
with the appropriately chosen constant $k.$ This result easily follows from
Eq.(4.15) under the assumption that $\tilde{\omega}(\mathbf{r}%
)=k\oint\limits_{C}d\sigma \mathbf{v}(\sigma )\delta (\mathbf{r}-\mathbf{r}%
(\sigma ))$ (which is the same as our Eq.(4.20b)). Substitution of this
result into Eq.(5.36) produces%
\begin{equation}
\oint\limits_{C_{1}}d\mathbf{r}\cdot \mathbf{A=}\frac{k}{4\pi }%
\oint\limits_{C_{1}}\oint\limits_{C_{2}}d\mathbf{\sigma }d\mathbf{\sigma }%
^{\prime }\frac{\mathbf{v}(\sigma )\cdot \mathbf{v}(\sigma ^{\prime })}{%
\left\vert \mathbf{r(\sigma )}-\mathbf{r}(\sigma ^{\prime })\right\vert }=n%
\frac{hc}{2\tilde{e}}.  \tag{5.39}
\end{equation}%
The obtained result allows us to determine the constant $k.$ To do so we
need to demonstrate that the above double integral is a linking number, e.g.
see Eq.(4.6). The proof of this result depends upon correctness of the
following statement 
\begin{eqnarray}
\oint\limits_{C}d\mathbf{r}\cdot \mathbf{A} &\mathbf{=}&\varkappa
\oint\limits_{C}d\mathbf{r}\cdot \mathbf{B=}\varkappa \oint\limits_{C}d%
\mathbf{r}\cdot \mathbf{\nabla }\times \mathbf{A=}  \nonumber \\
&=&\frac{\varkappa k}{4\pi }\oint\limits_{C_{1}}d\sigma \mathbf{v}(\sigma
)\cdot \oint\limits_{C_{2}}d\mathbf{\sigma }^{\prime }\mathbf{v}(\sigma
^{\prime })\times \frac{(\mathbf{r(\sigma )}-\mathbf{r}(\sigma ^{\prime }))}{%
\left\vert \mathbf{r(\sigma )}-\mathbf{r}(\sigma ^{\prime })\right\vert ^{3}}
\nonumber \\
&=&\frac{\varkappa k}{4\pi }\oint\limits_{C_{1}}\oint\limits_{C_{2}}d\sigma
d\sigma ^{\prime }\left[ \mathbf{v}(\sigma )\times \mathbf{v}(\sigma
^{\prime })\right] \cdot \frac{(\mathbf{r(\sigma )}-\mathbf{r}(\sigma
^{\prime }))}{\left\vert \mathbf{r(\sigma )}-\mathbf{r}(\sigma ^{\prime
})\right\vert ^{3}}  \nonumber \\
&=&\varkappa klk(1,2)  \TCItag{5.40}
\end{eqnarray}%
with linking number $lk(1,2)$ defined in Eq.(4.6). If the above result is
correct and the constant $\varkappa $ can be found then, the constant $k$
can be determined from Eq.(5.39). Hence, the task lies in demonstrating that
the nonzero constant $\varkappa $ does exist. To do so we shall use the
standard London analysis. Thus, we write 
\begin{equation}
\mathbf{\nabla }\times \mathbf{B}=\frac{4\pi }{c}\mathbf{j}\text{ \ \ \ \
(Maxwell equation) }  \tag{5.1}
\end{equation}%
and 
\begin{equation}
\mathbf{\nabla }\times \mathbf{j=-}\dfrac{en_{s}}{mc}\mathbf{B}\text{ \ \
(London \ equation).}  \tag{5.4b}
\end{equation}%
Since, $\mathbf{B}=\mathbf{\nabla }\times \mathbf{A,}$ and $\mathbf{\nabla }%
\cdot \mathbf{B}=\mathbf{\nabla }\cdot \mathbf{A=}0,$ we have $\varkappa 
\mathbf{B=A}$ so that we obtain 
\begin{equation}
\mathbf{\nabla }\times \mathbf{A=}\varkappa \frac{4\pi }{c}\mathbf{j} 
\tag{5.41}
\end{equation}%
and, from here%
\begin{equation}
-\mathbf{\nabla }^{2}\mathbf{A}=\mathbf{\nabla }\times \mathbf{\nabla }%
\times \mathbf{A=}\varkappa \frac{4\pi }{c}\left( \mathbf{\nabla }\times 
\mathbf{j}\right) =-\varkappa \dfrac{4\pi en_{s}}{mc^{2}}\mathbf{B=-}%
\varkappa ^{2}\dfrac{4\pi en_{s}}{mc^{2}}\mathbf{A}  \tag{5.42}
\end{equation}%
which is the familiar screening-type equation, e.g. see Eq.(4.51b). Since,
in the conventional setting the penetration depth $\delta ^{2}$ is known to
be $\delta ^{2}=\left( \dfrac{4\pi en_{s}}{mc^{2}}\right) ^{-1},$ we can
chose $\varkappa ^{2}\mathbf{=}1$ implying that $k=\dfrac{hc}{2\tilde{e}}.$
The choice $\varkappa \mathbf{=}1$ does not mean of course that the constant 
$\varkappa $ is dimensionless. Because of this, we obtain 
\begin{equation}
\frac{1}{4\pi \varkappa }\oint\limits_{C_{1}}\oint\limits_{C_{2}}d\mathbf{%
\sigma }d\mathbf{\sigma }^{\prime }\frac{\mathbf{v}(\sigma )\cdot \mathbf{v}%
(\sigma ^{\prime })}{\left\vert \mathbf{r(\sigma )}-\mathbf{r}(\sigma
^{\prime })\right\vert }=lk(1,2)  \tag{5.43}
\end{equation}%
in accord with Eq.(4.21). Next, if we take into account screening effects,
the conclusions we've reached will remain the same due to reparametrization
invariance of both sides of Eq.(5.43). Indeed, consider one loop, say $%
C_{1}, $ going from -$\infty $ to +$\infty $ in the z-direction. If we
compactify $\mathbf{R}^{3}$ by adding one point at infinity so that \textbf{R%
}$^{3}$ becomes \textbf{S}$^{3}$, then such a loop will be closed. Another
loop can stay mainly in the x-y plane so that the linking number becomes the
winding number, e.g. see Ref.[46], page 134. Under these conditions the
screening factor $exp(-\dfrac{r}{\delta })$ \footnote{%
Emergence of such a screening factor can be easily understood if we replace
Eq.(4.15) by Eq.(5.42) with the right hand side given by $\tilde{\omega}(%
\mathbf{r})=k\oint\limits_{C}\mathbf{v}(\sigma )\delta (\mathbf{r}-\mathbf{r}%
(\sigma ))$ in accord with Eq.(5.38).}under the integral of the left hand
side of Eq.(5.43) is unimportant since we can always arrange our windings in
such a way that $r\ll \delta $ for any preassigned $\func{nonzero}\delta $
so that the screening factor becomes unimportant.

The above analysis can be extended to the case of colloidal suspensions in
view of the results of Sections 4.2 and 4.4. implying that in both
superconductivity and colloidal suspensions the phase transition is
topological in nature (e.g. in the colloidal case Eq.(4.39) is a topological
invariant to be considered in the next subsection). Evidently, such a
conclusion cannot be reached by perturbatively calculating the Green's
function in Eq.(4.37).

In Section 5.1 we discussed similarities and differences between
superconductors and colloidal suspensions. It is appropriate now to add a
few additional details to the emerging picture. In the case of
superconductivity correctness of the topological picture depends upon the
existence of nontrivial solutions of Eq.(5.42). These are possible only when
the parameter $n_{s}$ is nonzero. When it becomes zero the above picture
breaks down. In the case of suspensions the role of the parameter $\varkappa
^{-1}$ is played by the density-dependent parameter $e$. This can be easily
seen if we take into account that dimensional analysis requires us to
replace Eq.(5.38) by%
\begin{equation}
\mathbf{A(r)=}\frac{D_{0}}{4\pi }\oint\limits_{C}\frac{1}{\left\vert \mathbf{%
r}-\mathbf{r}(\sigma )\right\vert }\mathbf{v}(\sigma )d\sigma  \tag{5.44}
\end{equation}%
so that by employing Eq.(5.37b) we obtain, 
\begin{equation}
\frac{e}{4\pi }\oint\limits_{C_{1}}\oint\limits_{C_{2}}d\mathbf{\sigma }d%
\mathbf{\sigma }^{\prime }\frac{\mathbf{v}(\sigma )\cdot \mathbf{v}(\sigma
^{\prime })}{\left\vert \mathbf{r(\sigma )}-\mathbf{r}(\sigma ^{\prime
})\right\vert }=n=lk(1,2)  \tag{5.45}
\end{equation}%
as expected.

\subsection{Bose-Einstein-type transition in a system of linked loops}

In the Introduction we noted that Chorin, Ref.[22], conjectured that the
superfluid-to normal transition in $^{4}$He is associated with vortices
causing a sharp increase in viscosity. In this subsection we wold like to
demonstrate that, at least for colloidal suspensions, his conjecture is
correct: the sharp increase in viscosity is associated with the lambda-type
transition. Instead of treating this problem in full generality, i.e. for
the nonideal Bose gas, we simplify matters and consider a Bose condensation
type transition typical for the ideal Bose gas. It should be noted though
that our simplified treatment is motivated only by the fact that it happens
to be sufficient for comparison with experimental data. In other cases, such
a restriction can be lifted.

To develop such a theory we use the information obtained in the previous
subsection augmented by some additional facts needed for completion of our
task. In particular, we are interested in \ the expression for the kinetic
energy. Up to a constant it is given by 
\begin{equation}
E\dot{=}\frac{1}{2}\int d^{3}\mathbf{r(\nabla \times A)}^{2}  \tag{5.46}
\end{equation}%
and is manifestly nonnegative. Using known facts from vector analysis this
expression can be rewritten as follows 
\begin{equation}
E\dot{=}\frac{1}{2}\int d^{3}\mathbf{r(\nabla \times A)}\cdot \mathbf{v=}%
\frac{1}{2}\int d^{3}\mathbf{r[A\cdot \tilde{\omega}+}div\mathbf{[A\times
v]]=}\frac{1}{2}\int d^{3}\mathbf{rA\cdot \tilde{\omega}}  \tag{5.47}
\end{equation}%
In view of Eq.(5.38) we can rewrite this result as 
\begin{equation}
E\dot{=}\frac{k^{2}}{2}\oint\limits_{C_{1}}\oint\limits_{C_{2}}d\mathbf{%
\sigma }d\mathbf{\sigma }^{\prime }\frac{\mathbf{v}(\sigma )\cdot \mathbf{v}%
(\sigma ^{\prime })}{\left\vert \mathbf{r(\sigma )}-\mathbf{r}(\sigma
^{\prime })\right\vert }  \tag{5.48}
\end{equation}%
to be compared with Eq.(5.45). Using such a comparison we arrive at an
apparent contradiction: while \ an expression for $E$ should be nonnegative,
the linking number $lk(1,2)$ can be both positive or negative. If we make
the replacement $\mathbf{r}\rightarrow -\mathbf{r}$ in Eq.(5.48) nothing
changes but if we do the same for $lk(1,2)$ it changes the sign. Thus, if we
want to use $lk(1,2)$ in Eq.(5.48) we have to use $\left\vert
lk(1,2)\right\vert $. This number was introduced by Arnold and is known in
literature as \textsl{entanglement complexity}\footnote{%
For more deatails about this number and its many applications can be found
in our works, Refs.[47,48].}. \ Evidently, in view of this remark, $n$ in
Eq.(5.45) can be only nonnegative. If we require our system to be invariant
with respect to rotations of \ the coordinate frame, Eq.(4.39) should be
rewritten according to the procedure developed in our work, Ref.[49]. This
means, that we introduce a set of linking numbers: $%
n_{1},n_{2},...,n_{i},... $ so that \ for a given $n\footnote{%
E.g. see Eq.(4.4).},$ the set of $\frac{1}{2}n(n-1)\equiv N$ possible
linking numbers can be characterized by the total linking number $L$, i.e.
we have%
\begin{equation}
\sum\limits_{i=1}^{N}n_{i}=L.  \tag{5.49}
\end{equation}%
This result can be rewritten alternatively as follows. Let $C_{1}$\ be the
number of links with linking number $\QTR{sl}{1}$, $C_{2}$ the number of
links with linking number 2 and so on. Then, we obtain 
\begin{equation}
\sum\limits_{i=1}^{L}iC_{i}=L.  \tag{5.50}
\end{equation}%
Furthermore, we also must require 
\begin{equation}
\sum\limits_{i=1}^{L}C_{i}=N  \tag{5.51}
\end{equation}%
Define the Stirling-type number $\tilde{S}(L,N)$ via the following
generating function\footnote{%
The true Stirling number of the first kind $S(L,N)$ is defined as follows: $%
S(L,N):=(-1)^{L-N}\tilde{S}(L,N).$} 
\begin{equation}
\sum\limits_{N=0}^{L}\tilde{S}(L,N)x^{N}=x(x+1)\cdot \cdot \cdot (x+L-1). 
\tag{5.52}
\end{equation}%
Set in this definition $x=1$. This then allows us to introduce the
probability $p(L,N)=\tilde{S}(L,N)/L!$ The number $\tilde{S}(L,N)$ can be
easily obtained\footnote{%
E.g. see Ref. [49].} with the result given by 
\begin{equation}
\tilde{S}(L,N)=\prod\limits_{i=1}^{N}\frac{L!}{i^{C_{i}}C_{i}!}.  \tag{5.53}
\end{equation}%
With thus obtained results, we are now ready to return to Eq.(4.39) in which
we make a rescaling: $\mathbf{r}(\tau )\rightarrow R_{0}\mathbf{\tilde{r}}%
(\tau )$, with $\mathbf{\tilde{r}}(\tau )$ being dimensionless. After which,
using Eq.(5.45) we can rewrite Eq.(4.39) as follows 
\begin{equation}
<W(L)>_{T}=\exp (-\frac{3\eta _{0}}{\eta }L)  \tag{5.54}
\end{equation}%
Evidently, the numerical factor of 3 in the exponent is non-essential and
can be safely dropped upon rescaling of $L$. To use this expression we
combine it with Eq.(5.34) in which we have to make some adjustments
following Feynman, Ref.[50], pages 62-64. On these pages Feynman discusses a
partition function for the ideal Bose gas written in the path integral form.
We would like to rewrite his result in the notation of our paper. For this
purpose we use Eq.(4.30) in which the path integral is written for a loop
and is in discrete form. We obtain, 
\begin{eqnarray}
h(\nu ) &=&\left( \frac{1}{4\pi D_{0}}\right) ^{\dfrac{3\nu }{2}}\int
\prod\limits_{i=1}^{\nu }d^{3}\mathbf{r}_{i}\exp \{-\frac{1}{4D_{0}}[(%
\mathbf{r}_{1}-\mathbf{r}_{2})^{2}+(\mathbf{r}_{2}-\mathbf{r}_{3})^{2}+...+(%
\mathbf{r}_{\nu -1}-\mathbf{r}_{\nu })^{2}+(\mathbf{r}_{\nu }-\mathbf{r}%
_{1})^{2}\}  \nonumber \\
&=&V\left( \frac{1}{4\pi \nu D_{0}}\right) ^{\dfrac{3\nu }{2}}  \TCItag{5.55}
\end{eqnarray}%
with $V=\int d^{3}\mathbf{r}_{1}.$Under such circumstances the Brownian ring
is made out of $\nu $ links(segments) so that we can identify its length
with $\nu .$ In the present case each such ring is linked with another ring
thus forming a link with a linking number $iC_{i}$, $i=0,1,2,...$ Since the
linking number is independent of the lengths of rings from which it is made,
we can take advantage of this fact by identifying the index $i$ with $\nu .$
By combining Eq.s (5.50)-(5.55) and repeating the same steps as given in
Feynman's lectures we assemble the following \ dimensionless grand partition
function $\mathcal{F}$ 
\begin{eqnarray}
e^{-\mathcal{F}} &=&\sum\limits_{C_{1},...,C_{q},....}\prod\limits_{\nu }%
\frac{h(\nu )^{C_{\nu }}}{C_{\nu }!\nu ^{C_{\nu }}}\exp (-\frac{\eta _{0}}{%
\eta }\nu C_{\nu })=\sum\limits_{C_{1},...,C_{q},....}\prod\limits_{\nu }%
\frac{1}{C_{\nu }!}(h(\nu )\frac{z^{\nu }}{\nu })^{C_{\nu }}  \nonumber \\
&=&\prod\limits_{\nu }\sum\limits_{C_{\nu }=0}^{\infty }\frac{1}{C_{\nu }!}%
(h(\nu )\frac{z^{\nu }}{\nu })^{C_{\nu }}=\exp (\sum\limits_{\nu }h(\nu )%
\frac{z^{\nu }}{\nu }).  \TCItag{5.56}
\end{eqnarray}%
Here the "chemical" potential $z=exp(-\frac{\eta _{0}}{\eta }).$ Taking the
logarithm of both sides of the above equation we obtain the partition
function for the ideal Bose gas. Written per unit volume it reads%
\begin{equation}
\mathcal{F=-}\left( \frac{1}{4\pi D_{0}}\right) ^{\dfrac{3}{2}}\zeta
_{5/2}(z).  \tag{5.57}
\end{equation}%
In this expression $\zeta _{\alpha }(z)$ is Riemann's zeta function 
\begin{equation}
\zeta _{\alpha }(z)=\sum\limits_{n=1}^{\infty }\frac{z^{n}}{n^{\alpha }}. 
\tag{5.58}
\end{equation}%
This function is well defined for $z<1$, i.e. for $\dfrac{\eta }{\eta _{0}}%
<\infty $ and is divergent for $z>1,$ thus indicating a Bose condensation
whose onset is determined by the value $z=1$ (i.e. $\eta =\infty )$ for
which $\zeta _{5/2}(1)=1.341.$ If we follow standard treatments, then we
obtain for the critical density $\rho _{c}$%
\begin{equation}
\rho _{c}=\left( \frac{1}{4\pi D_{0}}\right) ^{\dfrac{3}{2}}2.612. 
\tag{5.59}
\end{equation}%
In view of Eq.(5.55), the obtained result for density has the correct
dimensionality. From here the critical volume fraction is: $\varphi
_{c}=\rho _{c}\frac{4}{3}\pi R_{0}^{3}$. The number $2.612$ is just the
value of $\zeta _{3/2}(1).$ This means, that we can write in the general
case 
\begin{equation}
\rho (z)=\left( \frac{1}{4\pi D_{0}}\right) ^{\dfrac{3}{2}}\zeta _{3/2}(z) 
\tag{5.60}
\end{equation}%
thus giving us the equation 
\begin{equation}
\frac{\rho _{c}-\rho }{\rho _{c}}=1-\frac{\zeta _{3/2}(z)}{\zeta _{3/2}(1)}.
\tag{5.61}
\end{equation}%
In the book by London, Ref.[51], we found the following expansion for $\zeta
_{3/2}(z)$ in the vicinity of $z=1$ $(z<1):$ 
\begin{equation}
\zeta _{3/2}(z)=-3.545\alpha ^{\frac{1}{2}}+2.612+1.460\alpha -0.104\alpha
^{2}+....,  \tag{5.62}
\end{equation}%
where $\alpha =-\ln z.$ Use of this result in Eq.(5.61) produces the
following result: 
\begin{equation}
\frac{\eta }{\eta _{0}}=\left( \frac{1}{3.545}\right) ^{2}(1-\frac{\rho }{%
\rho _{c}})^{-2}  \tag{5.63}
\end{equation}%
in accord with scaling predictions by Brady, Ref. [19], and \ Bicerano 
\textit{et al.}Ref.[20]. It should be noted though that in view of Eq.(5.54)
the actual value of the constant prefactor in Eq.(5.63) is quite arbitrary
and can be adjusted with help of experimental data. For instance, by making
this prefactor of order unity, Bicerano \textit{et all} obtained a very good
agreement with experimental data in the whole range of concentrations, e.g.
see Ref. [20], Fig.4.

\section{Discussion and outlook}

\subsection{\protect\bigskip General comments}

\ With the exception of the work by De Gennes [52] on phase transition in
smectics A, \ the superconductivity and superfluidity phenomena are
typically associated with the domain of low temperature condensed matter
physics\footnote{%
Lately, however, these ideas have began to be popular in color
supercoductivity dealing with quark matter [53].}.This fact remains true
even with account of cuprate superconductors, Ref.[54]. The results obtained
in this work cause us to look at these phenomena differently. For instance,
the previously mentioned relation $\tilde{\omega}(\mathbf{r}%
)=k\oint\limits_{C}\mathbf{v}(\sigma )\delta (\mathbf{r}-\mathbf{r}(\sigma
)) $ used in the work by Lund and Regge, Ref.[45], for fluids, coincides
with our Eq.(4.20b) for colloids. The work of Lund and Regge is based on
previous work by Rasetti and Regge, Ref.[55], on superfluid He and,
therefore, their results are apparently valid only in the domain of low
temperatures.This conclusion is incorrect however as shown in the series of
papers by Berdichevsky, Ref.s [56,57]. Any ideal (that is Euler-type)
incompressible fluid can be treated this way. Furthermore, as results by
Chorin, Ref.[22],\ indicate, the same methods should be applicable for
description of the onset of fluid/gas turbulence. In our work the fluid is
manifestly nonideal. Nevertheless, in the long time (zero frequencies) limit
it can still be treated as if it is ideal. \ 

The most spectacular departure from traditional view on the results by Lund
and Regge was recently made in a series of papers by Schief and
collaborators, Refs.[58,59]. The latest results elaborating on \ his work
can be found in Ref.[60]. Schief demonstrated that the results of\ Lund and
Regge work well in the case of magnetohydrodynamics, that is, ultimately in
the plasma installations designed for controlled thermonuclear synthesis.

The basic underlying physics of all these phenomena can be summarized as
follows. In every system which supports knotted structures, the existence of
a decoupling of topological properties from the conformational (statistical)
properties of flux tubes from which these knots/links are made should be
possible. Since this statement is not restricted to a simple Abelian C-S
field theory describing knots/links existing in G-L theory, in full
generality the theory should include the G-L theory as a special case (as
demonstrated above). Accordingly, the minimization of the corresponding
truncated G-L functional may or may not lead to London-type equations. We
would like to illustrate these general statements by \ specific examples.
This is accomplished below.

\subsection{Helicity and force-free fields imply knoting and linking but not
nesesssarily superconductivity via London mechanism}

The concept of helicity has its origin in theory of neutrino, Ref.[61]. An
expression $\mathbf{\sigma }\cdot \mathbf{p/}\left\vert \mathbf{p}%
\right\vert $ is\ called \textsl{helicity.} Here $\mathbf{\sigma }\cdot 
\mathbf{p=\sigma }_{x}p_{x}+\mathbf{\sigma }_{y}p_{y}+\mathbf{\sigma }%
_{z}p_{z}$, and $p_{i}$ and $\sigma _{i}$ , $i=1-3,$ are being respectively
the components of the momentum and Pauli matrices. The eigenvalue equation 
\begin{equation}
\left[ \mathbf{\sigma }\cdot \mathbf{p/}\left\vert \mathbf{p}\right\vert %
\right] \Psi =\lambda \Psi  \tag{6.1}
\end{equation}%
produces eigenvalues $\lambda $ which can be only $\pm 1.$ Moffat, Ref.[62],
designed a classical analog of the helicity operator. He proposed to use the
product $\mathbf{v}\cdot \mathbf{\nabla }\times \mathbf{v\equiv v}\cdot 
\mathbf{\tilde{\omega}}$ \ for this classical analog. In it, as before, e.g.
see Eq.(4.14), the vorticity field $\mathbf{\tilde{\omega}}$ is used. Moffat
constructed an integral (over the volume $M$) 
\begin{equation}
I=\int\limits_{M}\mathbf{v}\cdot \mathbf{\tilde{\omega}}dV  \tag{6.2}
\end{equation}%
along with two other integrals: the kinematic kinetic energy%
\begin{equation}
\frac{2T}{\rho }E=\int\limits_{M}\mathbf{v}^{2}dV  \tag{6.3}
\end{equation}%
and the rotational kinetic energy 
\begin{equation}
\Omega =\int\limits_{M}\mathbf{\tilde{\omega}}^{2}dV.  \tag{6.4}
\end{equation}%
Then, he used the Schwarz inequality 
\begin{equation}
I^{2}\leq E\Omega \text{ \ or }\Omega \geq \frac{I^{2}}{E}  \tag{6.5}
\end{equation}%
in order to demonstrate that the equality is achieved only if $\mathbf{%
\tilde{\omega}=\alpha v}$ where $\alpha $ is a constant. \ Since this
requirement coincides exactly with our Eq.(4.20b), it is of interest to sudy
this condition further. In particular, under this condition we obtain $%
\alpha I=E$ which would coincide with our Eq.(5.43) (see also 5.48)) should $%
I$ be associated with the linking number. Fortunately, this is indeed the
case. The proof was given by Arnold and is outlined in Ref.[63], pages
141-146. In view of its physical significance, we would like to discuss it
in some detail.

Before doing so, we notice that the condition $\mathbf{\tilde{\omega}=\alpha
v}$ is known in literature as the \textsl{force-free} \textsl{condition }for
the following reason. In electrodynamics, the motion of an electron in a
magnetic field is given by (in the system of units in which $m=c=e=1$) 
\begin{equation}
\frac{d\mathbf{v}}{dt}=\mathbf{v}\times \mathbf{B}  \tag{6.6a}
\end{equation}%
while the use of the Maxwell's equation, our Eq.(4.10), produces as well 
\begin{equation}
\mathbf{v}=\mathbf{\nabla }\times \mathbf{B}=\alpha \mathbf{B}  \tag{6.6b}
\end{equation}%
Using previously established equivalence $\mathbf{v}\rightleftarrows \mathbf{%
B}$ \ and substitution of Eq.(6.6b) into Eq.(6.6a) explains why the
force-free condition is given by $\mathbf{\tilde{\omega}=\alpha v.}$ This
equation can be looked upon as an eigenvalue equation for the operator $%
\nabla \times (\cdot \cdot \cdot ).$ From this point of view the force-free
equation is totally analogous to its quantum counterpart, Eq.(6.1). Details
can be found in Ref.[64].

Going back to Arnold's proof, we note that according to Moffatt, Ref.[62],
page 119, 
\begin{equation}
I=\int\limits_{V}\mathbf{v}\cdot \mathbf{\tilde{\omega}}dV=\frac{1}{4\pi }%
\int\limits_{V(1)}\int\limits_{V(2)}\frac{\mathbf{R}_{12}\cdot \lbrack 
\tilde{\omega}(1)\times \tilde{\omega}(2)]}{\left\Vert \mathbf{R}%
_{12}\right\Vert ^{3}}dV(1)dV(2).  \tag{6.7}
\end{equation}%
Clearly, if as is done by Moffatt and others in physics literature (e.g.
Lund and Regge, etc.), we assume that the vector potential $\mathbf{A}$ can
be given in the form of Eq.(5.44), then $I$ indeed becomes the linking
number, Eq.(4.6). If, however, we do not make such an assumption, then much
more sophisticated methods are required for the proof of this result. Use of
these methods is not of academic interest only, as we would like to explain
now. According to Kozlov, Ref.[65],the force-free case $\mathbf{\tilde{\omega%
}=\alpha v}$ belongs to the category of so called vortex motion in the 
\textsl{weak sense. }There are many other vortex motions for which $\mathbf{v%
}\times \nabla \times \mathbf{v\neq 0.}$ These are vortex motions in the%
\textsl{\ strong sense}. Evidently, any relation with superconductivity or
superfluidity (which is actually only hinted at this stage in view of
results obtained in previous sections) is lost in this (strong) case. But
even with the vorticity present in the weak sense this connection is not
immediately clear.This is so because of multitude of solutions of the
force-free equation as discussed, for example, in Refs.[66, 67]. We would
like to discuss only those solutions which are suitable for use in Arnold's
theorem. These solutions can be obtained as follows. Taking the curl of the
equation 
\begin{equation}
\mathbf{\nabla }\times \mathbf{B}=\alpha \mathbf{B,}  \tag{6.8}
\end{equation}%
provided that $\mathbf{\nabla }\cdot \mathbf{B}=0,$ produces 
\begin{equation}
(\nabla ^{2}+\alpha ^{2})\mathbf{B}=0,  \tag{6.9}
\end{equation}%
to be compared with our result, Eq.(4.52a). Unlike our case,\ which is
motivated by analogies with superconductivity and superfluidity, in the
present case there are many solutions of this equation. We choose only the
solution which illustrates the\ theorem by Arnold. It is given by $\mathbf{v}%
=(Asinz+Ccosy,Bsinx+Acosz,Csiny+Bcosx),$ where $ABC\neq 0$ and $A,B,C\in 
\mathbf{R}$ This solution is obtained for $\alpha =1.$

Following Arnold, we introduce the asymptotic linking number $\Lambda
(x_{1},x_{2})$ via 
\begin{equation}
\Lambda (x_{1},x_{2})=\lim_{T_{1},T_{2}\rightarrow \infty }\frac{1}{4\pi
T_{1}T_{2}}\int\limits_{0}^{T_{1}}\int\limits_{0}^{T_{2}}dt_{1}dt_{2}\frac{(%
\mathbf{\dot{x}}_{1}(t_{1})\times \mathbf{\dot{x}}_{2}(t_{2}))\cdot (\mathbf{%
x}_{1}(t_{1})-\mathbf{x}_{2}(t_{2}))}{\left\Vert \mathbf{x}_{1}(t_{1})-%
\mathbf{x}_{2}(t_{2})\right\Vert ^{3}}.  \tag{6.10a}
\end{equation}%
The theorem proven by Arnold states that if the motion described by
trajectories $\mathbf{x}_{1}(t_{1})$ and $\mathbf{x}_{2}(t_{2})$ is ergodic,
then 
\begin{equation}
\frac{1}{4\pi }\int\limits_{V(1)}\int\limits_{V(2)}\frac{\mathbf{R}%
_{12}\cdot \lbrack \tilde{\omega}(1)\times \tilde{\omega}(2)]}{\left\Vert 
\mathbf{R}_{12}\right\Vert ^{3}}dV(1)dV(2)=\frac{1}{V^{2}}%
\int\limits_{V(1)}\int\limits_{V(2)}\Lambda (x_{1},x_{2})dV(1)dV(2)=lk(1,2).
\tag{6.10b}
\end{equation}%
That is the function $\Lambda (x_{1},x_{2})$ on ergodic trajectories is
almost everywhere constant. This theorem as such does not imply that this
constant is an integer. For us it is important to realize that \textsl{both }
Eq.(4.52a) and Eq.(6.9) can produce trajectories minimizing the Schwarz
inequality thus leading to the condition $\alpha I=E$ with $I$ being either
linking (in the case of suspensions) or self-linking number (depending upon
the problem in question) or a conbination of both. \ Because both Eq.(4.52a)
and (6.9) cause formation of links, the choice between them should be made
on a case-by-case basis. In particular, existence of the Messner effect in
superconductors leaves us with no freedom of choice between these two
equations. \ In the case of magnetohydrodynamics/plasma physics the
situation is less obvious. In the next subsection we shall argue in favour
of superconducting/superfluid choice between these equations. To our
knowledge, such a choice was left unused in plasma physics literature.

\subsection{Ideal magnetohydrodynamics and superfluidity/superconductivity}

\bigskip In order to discuss the work by Schief, Ref.[58], we would like to
remind to our readers of some facts from the work by Lund and Regge
(originally meant to describe superfluid $^{4}$He) since these fact nicely
supplement those presented in previous sections. We already mentioned that
Berdichevsky adopted these results for normal fluids, including those which
are turbulent. Lund and Regge assumed that the vortex has a finite thickness
so that the non-slip boundary condition, Eq.(2.27), should be now amended to
account for finite thickness. The amended equation is given by 
\begin{equation}
v_{i}(t)=\frac{\partial x_{i}}{\partial t}+\frac{\partial x_{i}}{\partial
\sigma }\frac{\partial \sigma }{\partial t},  \tag{6.11}
\end{equation}%
where $\sigma $ parametrizes the coordinate along the vortex line. Eq.(5.38)
taken from work by Lund and Regge then implies:%
\begin{equation}
\varepsilon _{ijk}\frac{\partial x_{j}}{\partial \sigma }(\frac{\partial
x_{k}}{\partial t}-v_{k})=0.  \tag{6.12}
\end{equation}%
This equation is treated as an equation of motion by Lund and Regge obtained
with help of the following Lagrangian 
\begin{equation}
\mathcal{L}=\frac{k\rho }{3}\int\limits_{C}\varepsilon _{ijk}x_{i}\frac{%
\partial x_{j}}{\partial \sigma }\frac{\partial x_{k}}{\partial t}d\sigma -%
\frac{\rho }{2}\int\limits_{V}\mathbf{v}^{2}d^{3}V.  \tag{6.13}
\end{equation}%
\ \ Since the zero thickness limit of the action for this Lagrangian is
given by our Eq.(4.22), which upon integration of the A-field leads to the
result, Eq.(4.24), the same can be done in the present case and,
accordingly, by analogy with the action, Eq.(4.22), which was extended, e.g.
see Eq.(4.40), in the present case it can be extended as well so that the
final result for the action of the Nambu-Goto bosonic string interacting
with electromagnetic-type field reads (using the same signature of
space-time as used in Ref.[45]) 
\begin{equation}
S=-m\int d\sigma d\tau \sqrt{-g}+f\int A_{\mu \nu }\frac{\partial x_{\mu }}{%
\partial \sigma }\frac{\partial x_{\nu }}{\partial t}d\sigma d\tau -\frac{1}{%
4}\int \mathbf{F}^{2}dvol  \tag{6.14}
\end{equation}%
with 
\begin{equation}
\sqrt{-g}=[-\left( \frac{\partial x^{\nu }}{\partial \sigma }\frac{\partial
x_{\nu }}{\partial \sigma }\right) \cdot \left( \frac{\partial x^{\mu }}{%
\partial \tau }\frac{\partial x_{\mu }}{\partial \tau }\right) +\left( \frac{%
\partial x^{\nu }}{\partial \sigma }\frac{\partial x_{\nu }}{\partial \tau }%
\right) ^{2}]^{\frac{1}{2}}.  \tag{6.15}
\end{equation}%
and $m$ and $f$ being some coupling constants. \ The metric of the surface
enclosing the vortex can be always brought to diagonal form by some
conformal transformation\footnote{%
For more details, please see Ref.[68].}. In such coordinates, variation of
the action $S$ produces the following set of equations%
\begin{equation}
m(\frac{\partial ^{2}}{\partial \tau ^{2}}-\frac{\partial ^{2}}{\partial
\sigma ^{2}})x_{\mu }=f\varepsilon ^{\mu \nu \lambda \rho }F_{\nu }\frac{%
\partial x_{\rho }}{\partial \tau }\frac{\partial x_{\lambda }}{\partial
\sigma }  \tag{6.16a}
\end{equation}%
and%
\begin{equation}
\partial ^{\mu }\partial _{\mu }A^{\alpha \beta }=-2f\int d\sigma d\tau (%
\frac{\partial x^{\alpha }}{\partial \sigma }\frac{\partial x^{\beta }}{%
\partial \tau }-\frac{\partial x^{\alpha }}{\partial \tau }\frac{\partial
x^{\beta }}{\partial \sigma })\delta ^{\left( 4\right) }(x(\sigma ,\tau )-y)
\tag{6.16b}
\end{equation}%
provided that $\partial _{\mu }A^{\mu \nu }=0.$ Since the last equation is
just the wave equation with an external source, the equation of motion for
the vortex is Eq.(6.16a). In such a form it was obtained in Ref.[58]
describing vortices in ideal magnetohydrodynamics. Under some physically
plausible condition it was reduced in the same reference to the equation of
motion for the one-dimensional Heisenberg ferromagnet. This result will be
discussed further below from a somewhat different perspective.

It should be noted though that Eq.(6.16a) emerges in Ref.[58] under somewhat
broader conditions than those allowed by the force-free equation. In view of
the content of the next subsection, we would like to reproduce this, more
general case, now. \ For this purpose, we recall that the Euler's equation
for fluids can be written in the form, Ref.[29], 
\begin{equation}
\frac{\partial }{\partial t}\mathbf{\tilde{\omega}=\nabla \times (v\times 
\tilde{\omega}).}  \tag{6.17}
\end{equation}%
In the case when $\mathbf{\tilde{\omega}}$ is time-independent, it is
sufficient to require only that 
\begin{equation}
\mathbf{v\times \tilde{\omega}=\nabla }\Phi  \tag{6.18}
\end{equation}%
with $\Phi $ being some (potential) scalar function. In the case of
hydrodynamics the equation $\Phi =const$ is the famous Bernoulli equation.
Thus, the force-free condition in this case is equivalent to the Bernoulli
condition/equation. In magnetohydrodynamics there is an analog of the
Bernoulli equation as explained in Ref.[69]. So, again, the equation $\Phi
=const$ is equivalent to the force-free condition. \ In the case of
magnetohydrodynamics the vortex Eq.(6.16.a) is obtained under the condition $%
\Phi =const.$ Since Eq.(6.16a) describes the vortex filament, the helicity
integral, Eq.(6.7), describes either linking, self-linking or both. In the
case of self-linking it is known, e.g. see Ref.s[48,63], that $%
lk(1,1)=Tw+Wr. $ Analytically, the writhe $Wr$ term is expressible as in
Eq.(4.6) but with $C_{1}$ and $C_{2}$ now representing the same closed
curve. The need for $Tw$ disappears if the closed curve can be considered to
have zero thickness. More accurately, the closed curve should be a ribbon in
order to have a nonzero $Tw$. This is explained in Ref.[63]. With the
exception of Appendix C, in this work we have ignored such complications.

\subsection{Classical mechanics in the vortex formalism, inertial dynamics
of nonrigid bodies and G-L theory of high temperature superconductors}

\bigskip Euler's Eq.(6.17) can be rewritten in the equivalent form: 
\begin{equation}
\frac{\partial }{\partial t}\mathbf{v}=\mathbf{v}\times \mathbf{\tilde{\omega%
}-\nabla }\Phi .  \tag{6.19}
\end{equation}%
Following Kozlov, Ref.[65], in the case of Hamiltonian mechanics it is
convenient to consider a very similar (Lamb) equation given by 
\begin{equation}
\frac{\partial }{\partial t}\mathbf{u+}\left( \mathbf{\nabla \times u}%
\right) \cdot \mathbf{v=-\nabla }\Phi ,  \tag{6.20}
\end{equation}%
in which the vector $\mathbf{u}$\textbf{\ }is such that $\nabla \cdot 
\mathbf{u}=0.$ It can be demonstrated that Hamiltonian dynamics is
isomorphic to the dynamics described by the above Lamb equation, provided
that we make the following identifications. Let $\Sigma _{t}^{n}$ be a
manifold in phase space $P=T^{\ast }M$ admitting a single-valued projection
onto a configurational space $M$. In canonical coordinates $x$ and $y$ this
manifold is defined by the equation%
\begin{equation}
\mathbf{y}=\mathbf{u}(\mathbf{x},t).  \tag{6.21}
\end{equation}%
It is not difficult to demonstrate that the manifold $\Sigma _{t}^{n}$ is an
invariant manifold for a canonical Hamiltonian $H(\mathbf{x},\mathbf{y},t)$
if and only if the field $\mathbf{y}$ satisfies the Lamb's Eq.(6.20) and
that $\Phi (\mathbf{x},t)=H(\mathbf{x},\mathbf{y}(\mathbf{x},t),t)$ is a
function on $M$ parametrized by time $t$ in such a way that 
\begin{equation}
\mathbf{v}=\frac{\partial H}{\partial \mathbf{y}}\mid _{\mathbf{y}=\mathbf{u}%
}\text{ }  \tag{6.22}
\end{equation}%
and%
\begin{equation}
\mathbf{\dot{y}=-}\frac{\partial H}{\partial \mathbf{x}}\mid _{\mathbf{y}=%
\mathbf{u}}=\frac{\partial \mathbf{u}}{\partial t}+\frac{\partial \mathbf{u}%
}{\partial \mathbf{x}}\cdot \mathbf{v.}  \tag{6.23}
\end{equation}%
Relevance of these results to our discussion can be seen when Eq.(6.23) is
compared with Eq.(6.11) of Lund and Regge. This comparison shows their near
equivalence. In view of this, we would like to exploit this equivalence
further by employing it for analysis of the truncated G-L functional
analogous to our Eq.(5.3) typically used for phenomenological description of
high temperature superconductors [54]. In this case the functional $\mathcal{%
F[}\mathbf{A},\varphi ]$ should be replaced by 
\begin{equation}
\mathcal{\tilde{F}[}\mathbf{A},\varphi ]=\int d^{3}r\{\frac{\left( \mathbf{%
\nabla }\times \mathbf{A}\right) ^{2}}{8\pi }+\frac{\hbar ^{2}}{4m_{\perp }}%
\left\vert (\mathbf{\nabla }_{\perp }-\frac{2i\tilde{e}}{\hbar c}\mathbf{A}%
_{\perp }\mathbf{)}\varphi \right\vert ^{2}+\frac{\hbar ^{2}}{4m_{\parallel }%
}\left\vert (\mathbf{\nabla }_{\parallel }-\frac{2i\tilde{e}}{\hbar c}%
\mathbf{A}_{\parallel }\mathbf{)}\varphi \right\vert ^{2}\}  \tag{6.24}
\end{equation}%
with its components lying in the x-y (cuprate) plane and z-plane
perpendicular to it. By varying this functional with respect to $\mathbf{A}%
_{\perp }$ and $\mathbf{A}_{\parallel }$ separately we obtain respectively
the following components for the Maxwell's equation 
\begin{equation}
\mathbf{\nabla }\times \mathbf{B}_{i}=\frac{4\pi }{c}\mathbf{j}_{i}\text{ (}%
i=\perp \text{and}\parallel \text{)},  \tag{6.25}
\end{equation}%
where 
\begin{equation}
\mathbf{j}_{\perp }=-\frac{ie\hbar }{2m_{\perp }}(\varphi ^{\ast }\mathbf{%
\nabla }_{\perp }\varphi -\varphi \mathbf{\nabla }_{\perp }\varphi ^{\ast })-%
\frac{2\tilde{e}^{2}}{m_{\perp }c}\left\vert \varphi \right\vert ^{2}\mathbf{%
A}_{\perp }\mathbf{.}\text{and }\mathbf{j}_{\parallel }=-\frac{ie\hbar }{%
2m_{\parallel }}(\varphi ^{\ast }\frac{d}{dz}\varphi -\varphi \frac{d}{dz}%
\varphi ^{\ast })-\frac{2\tilde{e}^{2}}{m_{\parallel }c}\left\vert \varphi
\right\vert ^{2}\mathbf{A}_{\parallel }.  \tag{6.26}
\end{equation}%
From here we obtain the phenomenological London-type equations%
\begin{equation}
\mathbf{\nabla }\times \mathbf{j}_{\perp }\mathbf{=-}\dfrac{en_{s}}{m_{\perp
}c}\mathbf{B}_{\perp }\text{ and }\mathbf{\nabla }\times \mathbf{j}%
_{\parallel }\mathbf{=-}\dfrac{en_{s}}{m_{\parallel }c}\mathbf{B}_{\parallel
}.  \tag{6.27}
\end{equation}%
By combining Eq.s (6.25) and (6.27) and using results of our Sections 4.2.
and 4.4 we can rewrite these equations in the following suggestive
(London-type) form%
\begin{equation}
\mathbf{\tilde{\omega}}_{\perp }=e_{\perp }\mathbf{v}_{\perp }\text{ and }%
\mathbf{\tilde{\omega}}_{\parallel }=e_{\parallel }\mathbf{v}_{\parallel }. 
\tag{6.28}
\end{equation}%
This form allows us to make a connection with the inertial dynamics of a
nonrigid body. Following Kozlov, Ref.[65], we consider the motion of a
nonrigid body in which particles can move relative to each other due to
internal forces. Let the inertia axes of the body be the axes of the moving
frame. Let \textbf{K} be the angular momentum of the body relative to a
fixed point and $\mathbf{\omega }$ the angular velocity of the moving
trihedron while the inertia matrix \textbf{I} is diag ($I_{\perp },I_{\perp
},I_{\parallel })\footnote{%
For the sake of comparison with superconductors, we deliberately choose the
matrix in such form.}.$ The angular momentum and the angular velocity are
related by 
\begin{equation}
\mathbf{K}=\mathbf{I\omega }+\mathbf{\lambda ,}  \tag{6.29}
\end{equation}%
where $\mathbf{\lambda =}(\lambda _{\perp },\lambda _{\perp },\lambda
_{\parallel })$ is the gyroscopic torque originating from the motion of
particles inside the body. From here we obtain the Euler equation 
\begin{equation}
\mathbf{\dot{K}=\omega \times K=}0,  \tag{6.30}
\end{equation}%
which is a simple consequence of Eq.(6.29). In view of Eq.s(4.45) and (4.48)
we can identify Eq.s (6.28) with (6.29) thus formally making Eq.s (6.29), of
London type. The hydrodynamic analogy can be in fact extended so that the
hydrodynamically looking Lamb-type equation can be easily obtained and
analyzed. Details are given in Ref.[65], page 148.

\subsection{Dirac monopoles, dual Meissner effect, Abelian projection for
QCD and string models}

At this point our readers may have already noticed the following. 1.In our
derivation of Eq.(5.63) we made screening effects seemingly disappear while
the title of our work involves screening. 2.In Eq.(6.14) we introduced the
Nambu-Goto string normally used in hadron physics associated with non
Abelian Yang-Mills (Y-M) gauge fields. Quantum chromodynamics (QCD) of
hadrons and mesons is definitely not the same thing as scalar
electrodynamics (that is G-L model) discussed in our work. 3. Variation of
the action $S$ in Eq.(6.14) leading to the string equation of motion,
Eq.(6.16a), under some conditions reduces to the equation of motion for the
Heinsenberg (anti) ferromagnetic chain, which indeed describes the motion of
the vortex filaments [59]. \ From this reference it follows that such
equation of motion, in principle, can be obtained quite independently from
the Nambu-Goto string, QCD, etc. In this subsection we demonstrate that the
above loose ends are in fact indicative of the very deep underlying
mathematics and physics needed for a unified description of all of these
phenomena. \ 

\bigskip The formalism developed thus far in this work suffers from a kind
of asymmetry. On one hand, we started with a solution of hard spheres and
then we noticed that these hard speres in solution act as currents (if one
is using the magnetic analogy). The famous Biot-Savart law in magnetostatics
is causing two currents to be entangled with each other thus creating the
Gauss linking number, Eq.(4.8). Thus, \ it appears that in solution two
particles (currents) are always linked (entangled) with each other. That
this is indeed the case was noticed long ago as mentioned in the
Introduction, e.g. see Ref.[11]. We can treat the vortices causing such
linkages as independent objects. This is reflected in the fact that we
introduced the vorticity $\mathbf{\tilde{\omega}}\ \mathbf{(r})$ as $\mathbf{%
\tilde{\omega}}=k\oint\limits_{C}d\sigma \mathbf{v}(\sigma )\delta (\mathbf{r%
}-\mathbf{r}(\sigma ))$, e.g read comments after Eq.(5.38). In view of our
major equation $\mathbf{\tilde{\omega}}\ \mathbf{(r})=e\mathbf{v,}$ we can
think either about the velocity (or vorticity) of a particular hard sphere
or about the velocity of a particular vortex. Because of this, it is
possible to treat both particles and vortices on the same footing. In such a
picture (sketched in Appendix B) one can either eliminate vortices and think
about effective interactions between hard spheres or vice versa. In this
sense we can talk about the \textsl{duality} of descriptions and, hence,
about the \textsl{dual Meissner effect}-for loops instead of particles%
\footnote{%
It should be noted that in the case of usual superconductors one should
distinguish between the constant magnetic fields penetrating superconductors
and the fields made by vortices. In the case of colloidal suspensions it is
also possible to create some steady velocity current and to consider
velocity at a given point in the fluid as made of both steady and
fluctuating \ parts.}.

Before describing the emerging picture in more detail, we note the
following. Consider the expression for vorticity $\mathbf{\tilde{\omega}}%
=k\oint\limits_{C}d\sigma \mathbf{v}(\sigma )\delta (\mathbf{r}-\mathbf{r}%
(\sigma ))$ from the point of view of reparametrization invariance. In
particular, since we have a closed contour, we can always choose it as going
from infinity to minus infinity (it is easy topologically to wrap it onto a
closed contour of any size). For the function $y=exp(\sigma )$ we have
evidently $0\leq y\leq \infty $ when $\sigma $ varies from -$\infty $ to $%
\infty .$ This means that we can replace $\sigma $ by $\ln y$ \ in the
expression for the vorticity in order to obtain 
\begin{equation}
\mathbf{\tilde{\omega}}=k\int\limits_{-\infty }^{0}dz\mathbf{v}(z)\delta (%
\mathbf{r}-\mathbf{r}(z)),  \tag{6.31}
\end{equation}%
which in a nutshell is the same thing as a Dirac monopole, Ref.[70], with
charge strength $k$, so that the vortices can be treated as Dirac monopoles.
In Appendix C \ we provide some facts about Dirac monopoles in relation to
vortices. According to Dirac [70] the string attached to such a monopole can
either go to infinity (as in the present case) or to another monopole of
equal and opposite strength. \ In our case this means only that when two
hard spheres become hydrodynamically entangled, they cannot escape the
linkage they formed. This is the (topological) essence of quark confinement
in QCD known as \textsl{monopole condensation\footnote{%
That is the Bose-Einstein-type condensation in view of results of Section 5.
This is explained further in Appendix C}}. But we are not dealing with QCD
in this work! How then we can talk about the QCD? The rationale for this was
put forward first by Nambu, Ref.[71]\footnote{%
We discuss his work briefly in Appendix C}. In his work he superimposed the
G-L and Dirac monopole theories to demonstrate quark confinement for mesons
(these are made \ of just two quarks: quark and antiquark). For this
qualitative picture to make sense, there should be some way of reducing QCD
to G-L type theory. The feasibility of such an \textsl{abelian reduction
(projection)} was investigated first by 't Hooft in Ref.[72]. Recent
numerical studies have provided unmistakable evidence supporting the idea of
quark confinement through monopole condensation, Ref.s [73,74]. Theoretical
advancements made since the publication of 't Hooft's paper took place along
two different (opposite) directions. In one direction, recently, Faddeev and
Niemi found knot-like topological solitons using a Skyrme-type nonlinear
sigma model and conjectured that such a model can correctly represent QCD in
the low energy limit [75,76]. That this is indeed the case was established
in a series of papers by Cho [77,78] and, more recently, by Kondo, Ref.[79].
In another direction, in view of the fact that, while macroscopically the
Meissner effect is triggered by the effective mass of the vector field,
microscopically this mass is generated by Cooper pairs [25], it makes sense
to look at detection of the excited states of such Cooper-like pairs
experimentally. The famous variational BCS treatment of superconductivity
contains at its heart the gap equation responsible for the formation of
Cooper pairs. The BCS treatment\ of superconductivity was substantially
improved by Richardson, Ref.[80], who solved the microscopic model exactly.
His model is known in literature as the Richardson model. Closely related to
this model is a model proposed by Gaudin. It is also exactly solvable (by
Bethe anstatz methods) [81]. The Gaudin model(s) describes various
properties of one dimensional spin chains in the semiclassical limit. Energy
spectra of the Gaudin and Richardson models are very similar. In particular,
under some conditions they are equidistant, like those for bosonic string
models.\footnote{%
Also, for monopoles models discussed in Appendix C.}. Recently, we were able
to find new models associated \ with Veneziano amplitudes, e.g. see Ref.s
[82,83], \ describing meson-meson scattering processes. In particular, we
demonstrated that the Richardson-Gaudin spin chain model producing
equidistant spectra can be obtained directly from Veneziano amplitudes.
Since the Veneziano amplitudes \ describe extremely well the meson mass
spectrum, and since we demonstrated that the Richardson-Gaudin model
(originally used in superconductivity and nuclear physics) can be recovered
from combinatorial and analytical properties of these amplitudes, this means
that the Abelian reduction can be considered as confirmed (at least for
mesons) not only numerically but also experimentally.

\subsection{Miscellaneous}

\bigskip In Section 3.3. we demostrated that for colloidal suspensions it is
sufficient to use only the Abelian version of the Chern-Simons theory for
description of emerging entanglements. There could be other instances where
such an Abelian treatment might fail. Examples of more sophisticated
non-Abelian fluids were considered in several recent excellent reviews
[84,85]. These papers might serve as points of departure for the treatment
of more elaborate hydrodynamical problems involving non-Abelian
entanglements. Finally, the force-free equation $\mathbf{\tilde{\omega}%
=\alpha v}$ which is used in our work, is known to possesss interesting new
physical properties when, instead of treating $\alpha $ as a constant, one
treats $\alpha $ as some function of \ the coordinates. Such treatment can
be found in Ref.[86] and involves the use of conformal \ transformations and
invariants recently considered in our work on the Yamabe problem, Ref.[87],
and the Poincare$^{\prime }$ conjecture, Ref.[68].

\qquad

\bigskip \qquad

\bigskip \textbf{Acknowledgement}. Both authors gratefully acknowledge
useful technical correspondence and conversations with Dr. Jack Douglas
(NIST). This work would look very different or even would not be written\ at
all without his input and his impatience to see this work completed.

\bigskip \qquad

\bigskip \textbf{Appendix A. Some facts from the theory of Green's functions}

\bigskip Consider an equation 
\begin{equation}
\left( \frac{\partial }{\partial t}-H\right) \Phi =0.  \tag{A.1}
\end{equation}%
Such an equation can be written in the form of an integral equation as
follows%
\begin{equation}
\Phi (\mathbf{x},t)=\int G_{0}(\mathbf{x},t;\mathbf{x}^{\prime },t^{\prime
})\Phi _{0}(\mathbf{x}^{\prime },t^{\prime })d\mathbf{x}^{\prime }dt^{\prime
}  \tag{A.2}
\end{equation}%
so that 
\begin{equation}
\Phi (\mathbf{x},t\rightarrow t^{\prime })=\Phi _{0}(\mathbf{x},t^{\prime }).
\tag{A.3}
\end{equation}%
Under such conditions, the Green's function $G_{0}(\mathbf{x},t;\mathbf{x}%
^{\prime },t^{\prime })$ must obey the following equation 
\begin{equation}
\left( \frac{\partial }{\partial t}-H\right) G_{0}(\mathbf{x},t;\mathbf{x}%
^{\prime },t^{\prime })=\delta (\mathbf{x}-\mathbf{x}^{\prime })\delta
(t-t^{\prime })  \tag{A.4}
\end{equation}%
provided that $G_{0}=0$ for $t<t^{\prime }.$ In a more complicated
situation, when 
\begin{equation}
\left( \frac{\partial }{\partial t}-H-V\right) G(\mathbf{x},t;\mathbf{x}%
^{\prime },t^{\prime })=\delta (\mathbf{x}-\mathbf{x}^{\prime })\delta
(t-t^{\prime })  \tag{A.5}
\end{equation}%
we can write a formal solution for G in the form of the integral (Dyson's)
equation%
\begin{equation}
G(\mathbf{x},t;\mathbf{x}^{\prime },t^{\prime })=G_{0}(\mathbf{x},t;\mathbf{x%
}^{\prime },t^{\prime })+\int G_{0}(\mathbf{x},t;\mathbf{x}^{\prime
},t^{\prime })V(\mathbf{x}^{\prime },t^{\prime })G(\mathbf{x}^{\prime
},t^{\prime };\mathbf{x}^{\prime \prime },t^{\prime \prime })d\mathbf{x}%
^{\prime }dt^{\prime }  \tag{A.6}
\end{equation}%
or, symbolically, $G=G_{0}+G_{0}VG$ . In the case of Eq.(4.35) of the main
text, we have to replace Eq.(A.5) by 
\begin{equation}
\left( \frac{\partial }{\partial t}-H_{1}-H_{2}-V_{12}\right) G(\mathbf{x}%
_{1},\mathbf{x}_{2},t;\mathbf{x}_{1}^{\prime },\mathbf{x}_{2}^{\prime
},t^{\prime })=\delta (\mathbf{x}_{1}-\mathbf{x}_{1}^{\prime })\delta (%
\mathbf{x}_{2}-\mathbf{x}_{2}^{\prime })\delta (t-t^{\prime })  \tag{A.7}
\end{equation}%
and, accordingly, the Dyson type Eq.(A.6) is now replaced by the analogous
equation in which now we must have $G_{0}(\mathbf{x}_{1},\mathbf{x}_{2},t;%
\mathbf{x}_{1}^{\prime },\mathbf{x}_{2}^{\prime },t^{\prime })=G_{0}(\mathbf{%
x}_{1},t;\mathbf{x}_{1}^{\prime },t^{\prime })G_{0}(\mathbf{x}_{2},t;\mathbf{%
x}_{2}^{\prime },t^{\prime }).$ To check the correctness of such a
decomposition we note that for $\mathbf{x}\neq \mathbf{x}^{\prime }$ Eq.s
(A.1) and (A.4) coincide while for $t\rightarrow t^{\prime }$ integration of
Eq.(A.4) over a small domain around zero and taking into account that $%
G_{0}=0$ for $t<t^{\prime }$ produces $G_{0}(\mathbf{x},t\rightarrow
t^{\prime };\mathbf{x}^{\prime },t^{\prime })=\delta (\mathbf{x}-\mathbf{x}%
^{\prime }).$ Repeating these arguments for the two-particle Green's
function and using Eq.(A.7) (with $V_{12}=0$ ) provides the needed proof of
the decomposition of $G_{0}$ in the two-particle case.

Define now formally the renormalized potential $\mathcal{V}$ via 
\begin{equation}
G=G_{0}+G_{0}\mathcal{V}G_{0}.  \tag{A.8}
\end{equation}%
Then, by comparing this equation with the original Dyson's equation for G we
obtain%
\begin{equation}
G-G_{0}=G_{0}\mathcal{V}G_{0}=G_{0}VG=G_{0}V(G_{0}+G_{0}\mathcal{V}G_{0}) 
\tag{A.9}
\end{equation}%
This allows us to write the integral equation for the effective potential $%
\mathcal{V}$ as%
\begin{equation}
\mathcal{V=}V+VG_{0}\mathcal{V}.  \tag{A.10}
\end{equation}

\qquad

\textbf{Appendix B. Dual treatment of \ the dynamics of colloidal supensions
and hydrodynamic screening }

\qquad

We begin by first considering screening. The path integral for the
functional, Eq.(5.5), can be conveniently rewritten as follows 
\begin{eqnarray}
\mathcal{F[}\mathbf{A},\varphi ] &=&\frac{\rho }{2}\int d^{3}r\{\left( 
\mathbf{\nabla }\times \mathbf{A}\right) ^{2}+D_{0}\left\vert (\mathbf{%
\nabla }-i\frac{2\pi e}{D_{0}}\mathbf{A)}\varphi \right\vert ^{2}\} 
\nonumber \\
&=&\frac{\rho }{2}\int d^{3}r\{\left( \mathbf{\nabla }\times \mathbf{A}%
\right) ^{2}+\left( \frac{D_{0}}{\pi }\right) ^{2}(\mathbf{\nabla }\psi -%
\frac{2\pi e}{D_{0}}\mathbf{A)}^{2}\}  \TCItag{B.1}
\end{eqnarray}%
upon substitution of the ansatz $\varphi =\dfrac{\sqrt{2D_{0}}}{2\pi }\exp
(i\psi )$ into first line of Eq.(B.1). Such a substitution is consistent
with the current defined in Eq.(4.50).

Since $\mathbf{\nabla }\cdot \mathbf{A=}0\mathbf{,}$ we obtain 
\begin{equation}
(\mathbf{\nabla }\psi -\frac{2\pi e}{D_{0}}\mathbf{A)}^{2}=(\mathbf{\nabla }%
\psi )^{2}+\left( \frac{2\pi e}{D_{0}}\mathbf{A}\right) ^{2}-\frac{4\pi e}{%
D_{0}}\mathbf{A\cdot \nabla }\psi .  \tag{B.2}
\end{equation}%
Consider now the following path integral%
\begin{equation}
Z=\int D\{\psi \}\exp [-\frac{1}{2}\left( \frac{D_{0}}{\pi }\right) ^{2}\int
d^{3}\mathbf{r}((\mathbf{\nabla }\psi )^{2}-\frac{4\pi e}{D_{0}}\mathbf{%
A\cdot \nabla }\psi )].  \tag{B.3a}
\end{equation}%
Since it is of a Gaussian-type, it can be straightforwardly calculated with
the result%
\begin{equation}
Z=N\exp (-\frac{e^{2}}{2}A_{\mu }\frac{\partial _{\mu }\partial _{\nu }}{%
\nabla ^{2}}A_{\nu }).  \tag{B3b}
\end{equation}%
Here $N$ is some (normalization) constant. Using this result and Eq.(B.1) we
obtain the following final expression for the partition function for the
vector A-field with account of constraints 
\begin{equation}
\Xi =\int D[\mathbf{A}]\exp \{-\frac{\rho }{2k_{B}T}\int d^{3}\mathbf{r\{}%
A_{\mu }[-\delta _{\mu \nu }\mathbf{\nabla }^{2}-(1-\frac{1}{\tilde{\xi}}%
)\partial _{\mu }\partial _{\nu }]A_{\nu }+e^{2}A_{\mu }(\delta _{\mu \nu }-%
\frac{\partial _{\mu }\partial _{\nu }}{\nabla ^{2}})A_{\nu }\}.  \tag{B.4a}
\end{equation}%
This result is in complete accord with Eq.(4.51b) where for the mass $m$ of
the vector field \ \textbf{A} we obtained: $m=e$. The above derivation was
made \textsl{without the use of Higgs-type calculations}, Ref[88]. Surely,
it is in accord with these calculations. We would like now to rewrite the
obtained result in a somewhat formal \ (simplified) form as follows:%
\begin{equation}
\Xi =\int D[\mathbf{A}]\delta (\mathbf{\nabla }\cdot \mathbf{A)}\exp \{-%
\frac{\rho }{2k_{B}T}\int d^{3}\mathbf{r[}\left( \mathbf{\nabla }\times 
\mathbf{A}\right) ^{2}+e^{2}\mathbf{A}^{2}]\}.  \tag{B.4b}
\end{equation}%
This will be used below in such simplified form. To avoid extra notation, we
also set $\dfrac{\rho }{k_{B}T}=1.$ This factor can be restored if needed.

Now we are ready for the dual treatment, which can be done in several ways.
For instance, following the logic of Dirac's paper [70 ], we replace $\Xi $
by 
\begin{equation}
\Xi =\int D[\mathbf{A}]\delta (\mathbf{\nabla }\cdot \mathbf{A)}\exp \{-%
\frac{1}{2}\int d^{3}\mathbf{r[}\left( \mathbf{\nabla }\times \mathbf{A}%
\right) +\mathbf{v})^{2}+e^{2}\mathbf{A}^{2}]\}  \tag{B.5}
\end{equation}%
where $\mathbf{v}=\dfrac{\mathbf{\tilde{\omega}}}{e}=\oint\limits_{C}d\sigma 
\mathbf{v}(\sigma )\delta (\mathbf{r}-\mathbf{r}(\sigma )).$ Next, we use
the Hubbard-Stratonovich-type identity allowing us to make a linearization,
e.g.%
\begin{equation}
\exp \{-\frac{1}{2}\int d^{3}\mathbf{r[(}\left( \mathbf{\nabla }\times 
\mathbf{A}\right) +\mathbf{v)}^{2}\}=\int D[\mathbf{\Psi }]\exp [-\frac{1}{2}%
\int d^{3}\mathbf{r\Psi }^{2}+i\int d^{3}\mathbf{r(}\left( \mathbf{\nabla }%
\times \mathbf{A}\right) +\mathbf{v)\cdot \Psi ]}  \tag{B.6}
\end{equation}%
Then, we take advantage of the fact that $\left( \mathbf{\nabla }\times 
\mathbf{A}\right) \cdot \Psi =\left( \mathbf{\nabla }\times \mathbf{\Psi }%
\right) \cdot \mathbf{A+\nabla \cdot (A\times \Psi )}$.By ignoring surface
terms this allows us to rewrite the above result as follows%
\begin{eqnarray}
&&\int D[\mathbf{\Psi }]\exp [-\frac{1}{2}\int d^{3}\mathbf{r\Psi }%
^{2}+i\int d^{3}\mathbf{r(}\left( \mathbf{\nabla }\times \mathbf{A}\right) +%
\mathbf{v)\cdot \Psi ]}  \nonumber \\
&=&\int D[\mathbf{\Psi }]\exp [-\frac{1}{2}\int d^{3}\mathbf{r\Psi }%
^{2}+i\int d^{3}\mathbf{r(}\left( \mathbf{\nabla }\times \mathbf{\Psi }%
\right) \cdot \mathbf{A}+\mathbf{v\cdot \Psi )]}  \TCItag{B.7}
\end{eqnarray}%
Using this result in Eq.(B.5) and using the Hubbard-Stratonovich
transformation again we obtain: 
\begin{eqnarray}
\Xi &=&\int D[\mathbf{A}]\delta (\mathbf{\nabla }\cdot \mathbf{A)}\exp \{-%
\frac{1}{2}\int d^{3}\mathbf{r[}\left( \mathbf{\nabla }\times \mathbf{A}%
\right) +\mathbf{v})^{2}+e^{2}\mathbf{A}^{2}]\}  \nonumber \\
&=&\int D[\mathbf{\Psi }]\delta (\mathbf{\nabla }\cdot \mathbf{\Psi )}\exp [-%
\frac{1}{2}\int d^{3}\mathbf{r[\Psi }^{2}+\frac{1}{e^{2}}\left( \mathbf{%
\nabla }\times \mathbf{\Psi }\right) ^{2}]+i\int d^{3}\mathbf{r\Psi \cdot v].%
}  \TCItag{B.8}
\end{eqnarray}%
Since exp$(i\int d^{3}\mathbf{r\Psi \cdot v)=}\exp (i\oint\limits_{C}d\sigma 
\mathbf{v}(\sigma )\cdot \mathbf{\Psi }(\sigma ))$ we can use this
expresiion in Eq.(5.34) in order eventually to arrive at the functional of
G-L-type (analogous to Eq.(5.6) with obviously redefined constants). The
vector field $\mathbf{\Psi }$ is now massive. It is convenient to make a
replacement: $\mathbf{\Psi \rightleftarrows }e\mathbf{\Psi }$ to make the
functional for the $\Psi $ field look exactly as in Eq.(B.5) (with $\mathbf{v%
}=0$). The above transformations provide a manifestly dual formulation of
the colloidal suspension problem. These transformations can be made
differently nevertheless. Such an alternative treatment is useful since the
end result has relevance to string theory and to the problem of quark
confinement in QCD as was first noticed by Nambu, Ref.[71]. This topic is
discussed briefly in the next appendix.

\qquad

\bigskip\ \textbf{Appendix C \ Nambu string and colloidal suspensions: Some
unusual uses of Dirac monopoles.}

We begin with Eq.(B.5) but this time we treat it differently. In particular,
we have%
\begin{eqnarray}
\Xi &=&\int D[\mathbf{A}]\delta (\mathbf{\nabla }\cdot \mathbf{A)}\exp \{-%
\frac{1}{2}\int d^{3}\mathbf{r[}\left( \mathbf{\nabla }\times \mathbf{A}%
\right) +\mathbf{v})^{2}+e^{2}\mathbf{A}^{2}]\}  \nonumber \\
&=&\int D[\mathbf{A}]\delta (\mathbf{\nabla }\cdot \mathbf{A)}\exp \{-\frac{1%
}{2}\int d^{3}\mathbf{rv}^{2}-\int d^{3}\mathbf{r[}\left( \mathbf{\nabla }%
\times \mathbf{A}\right) \cdot \mathbf{v}-\frac{1}{2}\int d^{3}\mathbf{r}%
\left( \mathbf{\nabla }\times \mathbf{A}\right) ^{2}-\frac{e^{2}}{2}\int
d^{3}\mathbf{rA}^{2}\}  \nonumber \\
&=&\int D[\mathbf{A}]\delta (\mathbf{\nabla }\cdot \mathbf{A)}\exp \{-\frac{1%
}{2}\int d^{3}\mathbf{rv}^{2}-\int d^{3}\mathbf{r[}\left( \mathbf{\nabla }%
\times \mathbf{v}\right) \cdot \mathbf{A}-\frac{1}{2}\int d^{3}\mathbf{r}%
\left( \mathbf{\nabla }\times \mathbf{A}\right) ^{2}-\frac{e^{2}}{2}\int
d^{3}\mathbf{rA}^{2}\}  \nonumber \\
&=&\int D[\mathbf{A}]\delta (\mathbf{\nabla }\cdot \mathbf{A)}\exp \{-\frac{1%
}{2}\int d^{3}\mathbf{rv}^{2}+e^{2}\sum\limits_{i<j}\oint\limits_{C_{i}}%
\oint\limits_{C_{j}}\frac{d\mathbf{l(\sigma }_{i})\cdot d\mathbf{l(\sigma }%
_{j})}{\left\vert \mathbf{r}(\sigma _{i})-\mathbf{r}(\sigma _{j})\right\vert 
}\exp (-\frac{\left\vert \mathbf{r}(\sigma _{i})-\mathbf{r}(\sigma
_{j})\right\vert }{\xi _{H}})\}.  \TCItag{C.1}
\end{eqnarray}%
The exponent in Eq.(C.1) is useful for comparison with that given in
Eq.(4.40). Such a comparison suggests that\ while the second (linking) term
is \ essentially the same as in Eq.(4.40)\footnote{%
We have mentioned already that the screening is not affecting the
topological nature of this term.}, the first term in the exponent of
Eq.(C.1) might be analogous to the "kinetic" string-like term in Eq.(4.40).
This line of reasoning can be found in the paper by Nambu [71]. If one
ignores quark masses as is usually done in string-theoretic literature, then
Eq.(13) of Nambu's paper looks very much like our Eq.(C.1), provided that we
identify the first term with the stringy Nambu-Goto term\footnote{%
E.g. see Eq.(6.14).}. To do so, we formally need to use the results of our
Sections 5.5 and 6.2. This time, however, we have to allow for self-linking.
Also, we have to take into account that for this case the energy and the
helicity become the same (up to a constant). Thus, one can consider the
helicity instead of energy. A very detailed treatment of helicity was made
in the paper by Ricca and Moffatt, Ref.[89], from which it follows that the
helicity is ideally suited for the description of self-linking. In such a
case we have to deal with closed curves of finite thickness. In fact, it is
sufficient to have a closed tube instead of a closed infinitely thin curve.
On such a tube one can perform the Dehn surgery by cutting a tube at some
section, twisting the free ends through a relative angle $2\pi n_{0},$ where 
$n_{0}$ is some integer, and reconnecting the ends. This operation makes a
self-linking proportional to $n_{0}$. If we agree that the Dehn twists are
made only in increments of $\pm 2\pi ,$ we obtain the "spectrum" which is
equidistant and, hence, string-like.

This intuitive picture can be made more quantitative as follows. Taking into
account Eq.(5.48), the kinetic term in the exponent of Eq.(C.1) can be
tentatively written as follows%
\begin{equation}
\frac{1}{2}\int d^{3}\mathbf{rv}^{2}=\frac{e^{2}}{2}\sum\limits_{i}\oint%
\limits_{C_{i}}\oint\limits_{C_{i}}d\mathbf{\sigma }d\mathbf{\sigma }%
^{\prime }\frac{\mathbf{v}(\sigma )\cdot \mathbf{v}(\sigma ^{\prime })}{%
\left\vert \mathbf{r(\sigma )}-\mathbf{r}(\sigma ^{\prime })\right\vert }. 
\tag{C.2}
\end{equation}%
This expression suffers from two apparent deficiencies. First, while the
second term in the exponent of Eq.(C.1) accounts for screening effects,
Eq.(C.2) is written without such an account. Second, since energy and
helicity are proportional to each other and since the Dehn surgery can be
made only for surfaces, Eq.(C.2) should be modified by replacing infinitely
thin contours by tubes. To repair the first problem we follow the book by
Pismen, Ref.[90], where on page 186 we find the following information.
Consider our Eq.s(4.52a) or (4.52b) and take into account Eq.(4.15). Then,
we can write 
\begin{equation}
\nabla ^{2}\mathbf{A}-e^{2}\mathbf{A}=-e\oint\limits_{C}d\sigma \mathbf{v}%
(\sigma )\delta (\mathbf{r}-\mathbf{r}(\sigma )).  \tag{C.3}
\end{equation}
The solution of the equation for vector potential $\mathbf{A}$, Eq.(5.38),
should \ now be modified to account for screening and boundary effects. The
result for energy, Eq.(5.47), \ now will be changed accordingly so that the
screening exponent will emerge in Eq.(C.1). To account for surface effects
we recognize that the self-linking expression, Eq.(C.2) is reparametrization
invariant. If, instead of infinitely thin contours we consider fluctuating
tubes, the reparametrization invariance should survive. The surface analog
of the expression $\oint\limits_{C}d\sigma \mathbf{v}(\sigma )\delta (%
\mathbf{r}-\mathbf{r}(\sigma ))$ is given in Eq.(6.16b). By introducing the
notation 
\begin{equation}
S^{\alpha \beta }=\frac{\partial x^{\alpha }}{\partial \sigma }\frac{%
\partial x^{\beta }}{\partial \tau }-\frac{\partial x^{\alpha }}{\partial
\tau }\frac{\partial x^{\beta }}{\partial \sigma }  \tag{C.4}
\end{equation}%
the self-linking term can be brought into the following final form (for just
one loop for brevity)%
\begin{equation}
\frac{1}{2}\int d^{3}\mathbf{rv}^{2}=\frac{e^{2}}{2}\int d\sigma d\tau \int
d\sigma ^{\prime }d\tau ^{\prime }S^{\alpha \beta }(\sigma ,\tau )\frac{\exp
(-\frac{\left\vert \mathbf{r}(\sigma ,\tau )-\mathbf{r}(\sigma ^{\prime
},\tau ^{\prime })\right\vert }{\xi _{H}})}{\left\vert \mathbf{r}(\sigma
,\tau )-\mathbf{r}(\sigma ^{\prime },\tau ^{\prime })\right\vert }^{\prime
}S^{\alpha \beta }(\sigma ^{\prime },\tau ^{\prime }),  \tag{C.5}
\end{equation}%
which is just what Nambu obtained. \ He further demonstrated that such a
term can be transformed into $-m\int d\sigma d\tau \sqrt{-g}$ \ (e.g. see
our Eq.(6.14)) with the constant $m$ (the string tension) being related to
coupling constant(s) of the theory. Since in the limit of infinitely thin
tubes results just obtained match those discussed in our Section 5.5, we
would like to take advantage of this observation. In Section 5.5 we
considered fully flexible (Brownian) loops. From the theory of polymer
solutions it is known that such loops can be made of the so called
semiflexible polymers whose rigidity is rather weak. Following our work,
Ref.[91], the path integrals describing semiflexible polymer chains \ are
given by 
\begin{equation}
I=\int D[\mathbf{u}(\tau )]\exp (-S[\mathbf{u}(\tau )])  \tag{C.6}
\end{equation}%
with action $S[\mathbf{u}(\tau )]$ given by 
\begin{equation}
S[\mathbf{u}(\tau )]=\frac{\kappa }{2}\int\limits_{0}^{N}d\tau \left( \frac{d%
\mathbf{u}}{d\tau }\right) ^{2}+\int\limits_{0}^{N}d\tau \lambda (\tau )(%
\mathbf{u}^{2}(\tau )-1).  \tag{C.7}
\end{equation}%
The rigidity constant is $\kappa .$ For brevity it will be put equal to one.
The Lagrange multiplier $\lambda $ takes care of the fact that the "motion"
is taking place on the surface of a 2-sphere. Minimization of the action $S$
produces%
\begin{equation}
\frac{d^{2}}{d\tau ^{2}}\mathbf{u}=\lambda \mathbf{u}  \tag{C.8}
\end{equation}%
with Lagrange multiplyer being determined by the constraint $\frac{d}{d\tau }%
\mathbf{u}^{2}=0$ thus producing instead of Eq.(C.8) the following result:%
\begin{equation}
\mathbf{\ddot{u}=-(\dot{u}\cdot \dot{u})u,}  \tag{C.9}
\end{equation}%
where $\mathbf{\dot{u}=}\frac{d}{d\tau }\mathbf{u,}$ etc. In view of the
results of this subsection, consider now an \ immediate extension of the
obtained results known as the Neumann model\footnote{%
Some useful details related to Neumann's model can be found in our work,
Ref.[92].} 
\begin{equation}
\mathbf{\ddot{u}+Gu=}\lambda \mathbf{u,}\text{ }\mathbf{u}^{2}=1  \tag{C.10}
\end{equation}%
for some matrix \textbf{G} which always can be brought to the diagonal form.
By analogy with Eq.(6.8), we can rewrite Eq.(C.10) in the following
equivalent form 
\begin{equation}
\mathbf{u}\times \lbrack \mathbf{\ddot{u}+Gu]=}0  \tag{C.11}
\end{equation}%
since $\mathbf{u}\times \lambda \mathbf{u=}0$ . The above equation is just a
special case of the Landau-Lifshitz (L-L) equation describing dynamics of
Heisenberg (anti)ferromagnets. In one space and one time dimension the L-L
equation reads 
\begin{equation}
\frac{\partial }{\partial t}\mathbf{u}=\{\mathbf{u}\times \lbrack \mathbf{%
\ddot{u}+Gu]\},}  \tag{C.12}
\end{equation}%
where now $\mathbf{\dot{u}=}\frac{d}{dx}\mathbf{u,}$ etc. In Sections 6.3
and 6.5 we mentioned already that L-L equation describes the dynamics of
vortex filaments in fluids, plasmas, etc and is also obtainable from the
Lund-Regge theory. Following Veselov.[93], consider a special solution of
the L-L equation obtained by inserting the ansatz $\mathbf{u}(x,t)=\mathbf{u}%
(x-i\theta t)$ into Eq.(C.12). Such a substitution produces:%
\begin{equation}
-i\theta \mathbf{\dot{u}=}\{\mathbf{u}\times \lbrack \mathbf{\ddot{u}+Gu]\},u%
}^{2}=1.  \tag{C.13}
\end{equation}%
Taking a vector product of both sides of this equation produces%
\begin{equation}
\mathbf{\ddot{u}+Gu=}\lambda \mathbf{u+}i\theta \lbrack \mathbf{\dot{u}%
\times u].}  \tag{C.14}
\end{equation}%
This equation describes the classical motion of a charged particle in the
presence of a Dirac monopole. At the quantum level such a problem was
studied in detail by Dunne, Ref.[94], who demonstrated that in the limit $%
\theta \rightarrow \infty $ the monopole spectrum is equidistant. This
result is compatible with the result of Ricca and Moffat [89], and explains
the role of monopoles in quark confinement (in view of results of our
Section 5.5). Furthermore it corroborates the results of our recent work,
Ref.[83], briefly mentioned in Section 6.5., where the spectrum of the1-d
Heisenberg XXX spin chain was recovered directly from the combinatorics of
scattering data supplied by uses of Veneziano amplitudes in scattering
experiments.

\qquad \qquad

\bigskip \textbf{References}

\qquad

[1] \ \ A. Einstein, Ann. der Physik 17 (1905) 549.

[2] \ \ A. Einstein, Ann. der Physik 19 (1906) 289.

[3] \ \ M.S. Selim, M.A. Al-Naafa, M.C. Jones,

\ \ \ \ \ \ Apl.ChE Journal 39 (1993) 3.

[4] \ \ W. Russel, D. Saville, W. Schowalter, Colloidal

\ \ \ \ \ \ Dispersions, Cambridge University Press, Cambridge, 1989.

[5] \ \ W. Hoover, F. Ree, J. Chem. Phys. 40 (1964) 2048.

[6] \ \ R. Roscoe, British Journal of Appl. Phys. 3 (1952), 267.

[7] \ \ J. Rallison, J. Fluid Mech. 186 (1988) 471.

[8] \ \ T. Tadros, Adv. in Colloid and Interface Sci. 12 (1980), 141.

[9] \ \ H. Brenner, Int. J. Multiphase Flow 1 (1974), 195.

[10] \ M. Doi, S.F. Edwards, The Theory of Polymer Dynamics,

\ \ \ \ \ \ \ Oxford University Press, Oxford, 1986.

[11] \ P. Hawksley, British. J. Appl. Phys. 5 (1954) S1-S5.

[12] \ S. Edwards, M. Muthukumar, Macromolecules 17 (1984) 586.

[13] \ N.Van Kampen, Stochastic processes in Physics and Chemistry,

\ \ \ \ \ \ \ North-Holland, Amsterdam, 1981.

[14] \ S. Meeker, W. Poon, P. Pusey, Phys. Rev. E 55 (1997) 5718.

[15] \ P. Segre, S. Meeker, P. Pusey, W. Poon,

\ \ \ \ \ \ \ Phys. Rev. Lett. 75 (1995) 958.

[16] \ H. Hooper, J. Yu, A. Sassi, D. Soane,

\ \ \ \ \ \ \ J. Appl. Polym. Sci. 63 (1997) 1369.

[17] \ G. Phillies, J. Colloid Interface Sc. 248 (2002) 528.

[18] \ S. Phan, W. Russel, Z. Cheng, J. Zhu, P. Chaikin, J. Dunsmuir,

\ \ \ \ \ \ \ R.Ottewill, Phys. Rev. E 54 (1996) 6633.

[19] \ J.Brady, J.Chem.Phys. 99 (1993) 567.

[20] \ J.Bicerano, J.Douglas, D.Brune, Rev.Macromol.

\ \ \ \ \ \ \ Chem.Phys.C39 (1999) 561.

[21] \ R. Ferrell, Phys. Rev. Lett. 24, (1970) 1169.

[22] \ J.Chorin, Vorticity and Turbulence, Springer-Verlag, Berlin, 1994.

[23] \ A.Kholodenko, J.Douglas, Phys.Rev.E 51 (1995) 1081.

[24] \ F.London, H.London, proc.Roy.Soc.London, Ser.A 149 (1935) 71.

[25] \ E. Lifshitz, L. Pitaevskii, Statistical Physics Part 2, Landau and

\ \ \ \ \ \ \ Lifshitz, Course of Theoretical Physics, Volume 9,

\ \ \ \ \ \ \ Pergamon Press, \ London, 1980.

[26] \ V.Ginzburg, L.Landau, Zh.Exp.Theor.Phys 20 (1950) 1064.

[27] \ G.Batchelor, J. Fluid Mech. 74 (1976) 1.

[28] \ S.Lovesey, Theory of Neutron Scattering From Condensed Matter,

\ \ \ \ \ \ \ Volume 1, Oxford University Press, Oxford, 1984.

[29] \ L.Landau, E. Lifshitz, Fluid Mechanics, Landau and Lifshitz Course

\ \ \ \ \ \ \ of Theoretical Physics, Volume 6, Pergamon Press, London, 1959.

[30] \ L. Landau, E.Lifshitz, Statistical Physics Part 1.,Course in
Theoretical

\ \ \ \ \ \ \ Physics,Vol.5, Pergamon Press, London, 1982.

[31] \ J.Jost, Riemannian Geometry and Geometric Analysis,

\ \ \ \ \ \ \ Springer-Verlag, Berlin, 2005.

[32] \ E.Witten, Comm.Math.Phys.121 (1989) 351.

[33] \ A.Kholodenko, T.Vilgis, Phys.Reports 298 (1998) 251.

[34] \ F. Tanaka, Prog. Theor. Phys. 68 (1982) 148.

[35] \ F. Ferrari, I. Lazzizzera, Nucl. Phys. B 559 (1999) 673.

[36] \ L. Landau, E.Lifshitz, Electrodynamics of Continuous Media,

\ \ \ \ \ \ \ Landau and Lifshitz Course of Theoretical Physics, Volume 8,

\ \ \ \ \ \ \ Pergamon Press, London, 1984.

[37] \ G. Batchelor, An Introduction to Fluid Dynamics,

\ \ \ \ \ \ \ Cambridge University Press, Cambridge, 1967.

[38] \ M. Brereton, S. Shah, J. Phys. A: Math. Gen. 13 (1980) 2751.

[39] \ P.Ramond, Field Theory: A Modern Primer,

\ \ \ \ \ \ \ Addison-Wesley Publ.Co., New York, 1989.

[40] \ R.Feynman, Phys.Rev.80 (1950) 440.

[41] \ A. Kholodenko, A. Beyerlein, Phys. Rev. A34 (1986) 3309.

[42] \ H.Kleinert, Gauge Fields in Condensed Matter, Volume2,

\ \ \ \ \ \ \ World Scientific, Singapore, 1989.

[43] \ A.Polyakov, Gauge Fields and Strings,

\ \ \ \ \ \ \ Harwood Academic Publishers, New York, 1987.

[44] \ A.Kholodenko, A.Beyerleyin, Phys.Rev.E A34 (1986) 3309.

[45] \ F.Lund, T.Regge, Phys.Rev.D 14 (1976) 1524.

[46] \ B.Dubrovin, A.Fomenko, S.Novikov, Modern Geometry-

\ \ \ \ \ \ \ Methods and Applications, Volume 2, Springer-Verlag, Berlin,
1985.

[47] \ A.Kholodenko, D.Rolfsen, J.Phys.A 29 (1996) 5677.

[48]. A.Kholodenko, T.Vilgis, Phys.Reports 298 (1998) 251.

[49] \ A.Kholodenko, Landau's last paper and its impact on

\ \ \ \ \ \ \ mathematics, physics and other disciplines in new millenium,

\ \ \ \ \ \ \ arxiv: 0806.1064.

[50] \ R.Feynman, Statistical Mechanics, Addison Wesley Publ.Co,

\ \ \ \ \ \ \ New York, 1972.

[51] \ F.London, Superfluids, Volume 2,

\ \ \ \ \ \ \ J.Wiley \& Sons Co., New York, 1954.

[52] \ P-G. De Gennes, Solid State Communications 10 (1972) 753.

[53] \ K.Iida, G.Baym, Phys.Rev. D 63 (2001) 074018.

[54] \ A.Leggett, Quantum Liquids,

\ \ \ \ \ \ \ Oxford University Press, Oxford, 2006.

[55] M.Rasetti, T.Regge, Physica 80A (1975) 217.

[56] V.Berdichevsky, Phys.Rev.E 57 (1998) 2885.

[57] V.Berdichevsky, Continuum Mech.Thermodyn.19 (2007) 135.

[58] W.Schief, Phys.Plasmas 10 (2003) 2677.

[59] C.Rogers, W.Schief, J.Math.Phys.44 (2003) 3341.

[60] L.Garcia de Andrade, Phys.Scr.73 (2006) 484.

[61] N.Bogoliubov, D.Schirkov, Introduction to the Theory of

\ \ \ \ \ \ Quantized Fields, Wiley Intersicience, New York, 1959.

[62] H.Moffatt, J.Fluid Mech.\ 35 (1969) 117.

[63] V.Arnold, B.Khesin, Topological Methods in Hydrodynamics,

\ \ \ \ \ \ \ Springer-Verlag, Berlin, 1998.

[64] K.Brownstein, Phys.Rev. A 35 (1987) 4856.

[65] V.Kozlov, General Theory of Vortices,

\ \ \ \ \ \ Springer-Verlag, Berlin, \ 1998.

[66] H.Zaghloul, O.Barajas, Am.J.Phys.58 (1990) 783.

[67] F.Gonzales-Gascon, D.Peralta-Salas, Phys.Lett. A 292 (2001) 75.

[68] A.Kholodenko, J.Geom. Phys.58 (2008) 259.

[69] V.Ferraro, C.Plumpton, Magneto-Fluid Mechanics,

\ \ \ \ \ \ Oxford University Press, Oxford, 1961.

[70] P.Dirac, Phys.Rev.74 (1948) 817.

[71] Y.Nambu, Phys.Rev.D10 (1974) 4262.

[72] G. t'Hooft, Nucl.Phys B 190 (1981) 455.

[73] T.Suzuki, I. Yotsuyanagi, Phys.Rev. D 42 (1990) 4257.

[74] J.Stack, S.Neiman, R.Wensley, Phys rev.D50 (1994) 3399.

[75] L.Faddeev, A.Niemi, Spin-charge separation, conformal

\ \ \ \ \ \ covariance and the SU(2) Yang-Mills theory, arXiv: hep-th/0608111

[76] L.Faddeev, Knots as possible excitations of the quantum

\ \ \ \ \ \ Yang-Mills field, arXiv.0805.1624

[77] Y.Cho, Phys.Rev.D 21 (1980) 1080.

[78] Y.Cho, D.Pak, Phys.Rev.D 65 (2002) 074027.

[79] K-I. Kondo, Phys.Rev.D 74 (2006) 125003.

[80] R.Richardson, Journal of Math.Phys. 9 (1968) 1327.

[81] M.Gaudin, La Function d'Onde de Bethe\textit{,}

\ \ \ \ \ \ Masson, Paris, 1983.

[82] \ A.Kholodenko, \textit{\ }J.Geom.Phys.56 (2006)1387.

[83] \ A.Kholodenko, New strings for old Veneziano amplitudes IV.

\ \ \ \ \ \ \ Connections with spin chains, arxiv: 0805.0113.

[84] \ R.Jackiw, V.Nair,S-Y. Pi, A.Polychronakos, J.Phys.A 37 (2004) R327.

[85] \ A.Polychronakos, Noncommutative Fluids, arxiv: 0706.1095.

[86] \ I.Benn, J.Kress, J.Phys.A 29 (1996) 6295.

[87] \ A.Kholodenko, E.Ballard, Physica A 380 (2007 115.

[88] \ W.Cottingham, D.Greenwood, An Introduction to the Standard

\ \ \ \ \ \ \ Model of Paricle Physics, Cambridge U.Press, Cambridge, 2007.

[89] \ H.Moffatt, R.Ricca, Proc.R.Soc. London A 439 (1992) 411.

[90] \ L.Pismen, Vortices in Nonlinear Fields, Clarendon Press,

\ \ \ \ \ \ \ Oxford, 1999.

[91] \ A.Kholodenko, Th.Vilgis, Phys.Rev. E 52 (1995) 3973.

[92] \ A.Kholodenko, Quantum signatures of solar system dynamics,

\ \ \ \ \ \ \ arXiv: 0707.3992.

[93] \ A.Veselov, Sov.Phys.\ Dokl. 28 (1983) 458.

[94] \ G.Dunne, Ann.Phys. 215 (1992) 233.

\ \ \ \ \ \ \ 

\end{document}